\documentclass[12]{article}
\usepackage{times}

\usepackage{graphicx}
\title{Deep Underground Science and Engineering Lab \\
S1 Dark Matter Working Group}
\author{Contributors:\footnote{Working Group members and contributors
    to the report, and their affiliations, are listed at the end of
    the document}~~D.S.~Akerib (Co-chair), E.~Aprile (Co-chair), \\
E.A.~Baltz, M.R.~Dragowsky, P.~Gondolo, R.J.~Gaitskell, A.~Hime, 
C.J.~Martoff, \\ D.-M.~Mei, 
H.~Nelson, B.~Sadoulet, R.W.~Schnee, A.H.~Sonnenschein and L.E.~Strigari}
\date{}

\long\def\symbolfootnote[#1]#2{\begingroup%
\def\thefootnote{\fnsymbol{footnote}}\footnote[#1]{#2}\endgroup}
    \def\@fnsymbol#1{\ifcase#1\or *\or \dagger\or \ddagger\or
      \mathchar ``278\or \mathchar ``27B\or \|\or **\or \dagger\dagger
      \or \ddagger\ddagger \else\@ctrerr\fi\relax}

\newcommand{\RnD}{{\small R\&D}}

\newcommand{\CDMS}{{\small CDMS}}

\newcommand{\CDMSII}{{\small CDMS\,II}}
\newcommand{\WIMP}{{\small WIMP}}
\newcommand{\WIMPs}{{\small WIMP}s}
\newcommand{\LHC}{{\small LHC}}
\newcommand{\ILC}{{\small ILC}}
\newcommand{\SUSY}{{\small SUSY}}
\newcommand{\SuperCDMS}{Super{\small CDMS}}

\newcommand{\etal}{et al}	% et al in appropriate format, the following . 
\begin{document}

%\ifpdf
%\DeclareGraphicsExtensions{.pdf, .jpg, .tif}
%\else
%\DeclareGraphicsExtensions{.eps, .jpg}
%\fi

\maketitle

\begin{flushright}
\textit{FINAL VERSION: February 17, 2007 \\ 
Revised from February 19, 2006 vers. 1.1 \\
based on reviewers's and S1 PI's feedback}
\end{flushright}

\section{Overview}
\label{overview}

The discovery of dark matter is of fundamental importance to
cosmology, astrophysics, and elementary particle physics. A broad
range of observations from the rotation speed of stars in ordinary
galaxies to the gravitational lensing of superclusters tell us that
80--90\% of the matter in the universe is in some new form, different
from ordinary particles, that does not emit or absorb
light. Cosmological observations, especially the Wilkinson Microwave
Anisotropy Probe of the cosmic microwave background radiation, have
provided spectacular confirmation of the astrophysical evidence. The
resulting picture, the so-called ``Standard Cosmology,'' finds that a
quarter of the energy density of the universe is dark matter and most
of the remainder is dark energy. A basic foundation of the model, Big
Bang Nucleosynthesis ({\small BBN}), tells us that at most about 5\%
is made of ordinary matter, or baryons. The solution to this ``dark
matter problem'' may therefore lie in the existence of some new form
of non-baryonic matter. With ideas on these new forms coming from
elementary particle physics, the solution is likely to have broad and
profound implications for cosmology, astrophysics, and fundamental
interactions.  While non-baryonic dark matter is a key component of
the cosmos and the most abundant form of matter in the Universe, so
far it has revealed itself only through gravitational
effects---determining its nature is one of the greatest scientific
issues of our time.

Many potential new forms of matter that lie beyond the Standard
Model of strong and electroweak interactions have been suggested as
dark matter candidates, but none has yet been produced in the
laboratory.  One possibility is that the dark matter is comprised of
Weakly Interacting Massive Particles, or \WIMPs, that were produced
moments after the Big Bang from collisions of ordinary matter. \WIMPs\
refer to a general class of particles characterized primarily by a
mass and annihilation cross section that would allow them to fall out of
chemical and thermal equilibrium in the early universe at the dark
matter density. Several extensions to the Standard Model lead to
\WIMP\ candidates. One that has received much attention is
Supersymmetry (\SUSY), which extends the Standard Model to include a
new set of particles and interactions that solves the gauge hierarchy
problem, leads to a unification of the coupling constants, and is
required by string theory.  The lightest neutral \SUSY\ particle, or
neutralino, is thought to be stable and is a natural dark matter
candidate.  Intriguingly, when \SUSY\ was first developed it was in no
way motivated by the existence of dark matter. This connection could
be a mere coincidence---or a crucial hint that \SUSY\ is responsible
for dark matter.

% dsa - new paragraph - reviewer B.6

The possibility that a new class of fundamental particles could be
responsible for the dark matter makes the search for \WIMPs\ in the
galactic halo a very high scientific priority. In resolving this
puzzle, it is intrinsically important to carefully search the parameter
space defined by a \WIMP\ signal in the galactic halo using
terrestrial detectors. A direct detection of dark matter in the halo
would be the most definitive way to determine that \WIMPs\ make up the
the missing mass. As we discuss further below, the study of \WIMP\
candidates in accelerator experiments is critical in determining the
relic density of these particles. The indirect detection of
astrophysical signals due to \WIMP-\WIMP\ self-annihilation may also
provide important clues but in many cases may be difficult to
unambiguously separate from more mundane astrophysical sources. That
leaves direct detection as playing a central role in establishing the
presence of \WIMPs\ in the universe today. Given both the technical
challenge and fundamental importance of direct \WIMP\ detection, it is
vital to have the means to confirm a detection in more than
one type of detector. In addition to giving a critical cross check on
systematic errors that could fake a signal, detection of \WIMPs\ in
multiple nuclei will yield further information about the \WIMP\ mass
and couplings. Eventually, if \WIMPs\ are discovered, then the
ultimate cross check will be confirming their galactic origin by
observing secondary signatures related to the motion of the earth and
solar system.

Many accelerator-based experiments have searched for \SUSY's
predictions. Although no direct evidence has yet been found, the
unexplored landscape on the energy frontier is still rich with
candidates. Indeed, \SUSY\ particles are the prime quarry of the large
experiments at Fermilab's Tevatron and {\small CERN}'s Large Hadron
Collider (\LHC), and the laboratory production of \WIMP\ candidates
would be a great help in solving the dark matter problem. However,
accelerator experiments alone cannot offer a solution because it is
impossible to determine whether the particles are sufficiently
long-lived.  Ultimately, only astrophysical observations can determine
whether \WIMPs\ exist in nature.

Astroparticle physics experiments worldwide are actively searching for \WIMPs\
under the hypothesis that they make up the missing mass in the
Galaxy. The two main approaches are direct detection of \WIMP-nucleus
scattering in laboratory experiments and indirect detection through
the observation of \WIMP-\WIMP\ annihilation products. An
astrophysical discovery of \WIMPs\ would be a landmark event for
cosmology and give new information on fundamental particle physics
that may otherwise be inaccessible even to the \LHC. In fact, if \SUSY\ eludes
the \LHC, astrophysical evidence for new particles could provide the
key guidance for the planned International Linear Collider (\ILC). A
solution to the dark matter puzzle will also address lingering
questions about our understanding of gravity.

{\small US} scientists are in a world-leading position in direct
detection, by having pioneered the development and deployment of several
of the best technologies, and by engaging in an active \RnD\ program that
promises to continue this leadership. The detection of dark matter is
an experimental challenge that requires the development of
sophisticated detectors, suppression of radioactive contamination, 
and---most relevant to this report---siting in deep underground
laboratories to shield from cosmic-ray-induced backgrounds. By
building the world's premier deep laboratory, together with bringing
the ongoing \RnD\ efforts to fruition, the {\small US} will be in a very
advantageous position internationally to attract and lead the major
experiments in this field.

A number of recent reports have highlighted the importance of dark
matter searches. Two {\small NRC} reports, ``Connecting Quarks with
the Cosmos,'' chaired by M.~Turner~\cite{turner}, and ``Neutrinos and
Beyond,'' chaired by B.~Barish~\cite{barish}; and the {\small HEPAP}
report ``The Quantum Universe,'' chaired by P.~Drell~\cite{drell},
have pointed out the high scientific priority of this enterprise.  In
reviewing these findings, the {\small OSTP} Interagency Working
Group's ``Physics of the Universe'' report directed that in the area
of dark matter ``{\small NSF} and {\small DOE} will work together to
identify a core suite of physics experiments. As stated in the report,
this work will include research and development needs for specific
experiments, associated technology needs, physical specifications, and
preliminary cost estimates''~\cite{ostp}.  The central role that dark
matter plays at the intersection of cosmology and fundamental physics
was highlighted most recently in the NRC report ``Revealing the Hidden
Nature of Space and Time: Charting the Course for Elementary Particle
Physics,'' chaired by H.~Schapiro and S.~Dawson~\cite{epp2010}. In
that report, the search for dark matter was included among the top
three priorities of the long range {\small US} {\small HEP} program.
These stated priorities and directives in these various reports are
well-aligned with the {\small DUSEL} science program.

In Section~\ref{motivation} of this report we describe the
cosmological and astrophysical evidence for dark matter and the deep
connection to particle physics suggested by the \WIMP\ hypothesis.  A
concordant picture of dark matter will require information from both
astrophysical observations and laboratory measurements. On the one
hand, accelerator experiments alone cannot establish a solution to the
dark matter problem because of the question of particle stability. On
the other hand, an astrophysical detection does not constrain the
particle physics parameters well enough to determine their relic
density.

In Section~\ref{detection} we define the basic challenge of the direct
detection of \WIMPs, which is based on elastic scattering between
\WIMPs\ in the halo and atomic nuclei in a terrestrial detector. We
examine there the general strategy, including the need to mitigate
background sources by enhancing the sensitivity for nuclear recoil
events relative to the predominant electromagnetic backgrounds. We
also discuss methods for comparing the sensitivity of different
approaches, interpreting results within the framework of particle
physics models, and conclude that section with a brief description of
the present status of \WIMP\ searches.

Following the general discussion of direct searches, we give an
overview in Section~\ref{indirect} of indirect astrophysical methods
to detect \WIMPs. These methods are based on looking for \WIMP\
annihilation products including gamma rays, neutrinos, and other
particle species. The sources are based on enhanced concentrations of
\WIMPs\ that arise from scattering and clumping of the particles in
astrophysical objects. Depending on the nature of the source and the
annihilation channel, the sources range from broad-band to
monoenergetic and include both directional and diffuse sources. While
there are several interesting hints, the challenge remains to
establish that these are not due to other astrophysical
processes.

In Section~\ref{candidates}, we take up in more detail the particle
physics behind \WIMP\ candidates. With the anticipated turn-on of the
\LHC\ at {\small CERN} and the high priority of the \ILC\ to the
{\small US} High Energy Physics program, it is particularly timely to
examine the complementarity of the accelerator- and
non-accelerator-based approaches to answering questions about the
nature of both dark matter and fundamental interactions. Clearly, the
laboratory production of \WIMP\ candidates could provide important
guidance to the astrophysical searches. By the same token,
astrophysical measurements of \WIMP\ properties can help to constrain
physics beyond the Standard Model and provide knowledge that would
otherwise be difficult to obtain at accelerators during a similar
time frame.

The multi-decade scientific program at {\small DUSEL} will include dark matter
searches at the ton-scale and beyond, with ultra-clean ultra-sensitive
detectors. In this context, we examine in Section~\ref{synergies} some of
the other physics that could also be pursued in combination with dark
matter instruments. Specifically, we look at the related requirements
of double-beta-decay experiments, which have in common the need for
low-radioactive background and excellent energy resolution. Germanium,
xenon and tellurium all have double-beta-decay nuclides and are being
used as detection media in both fields. In Section~\ref{synergies}, we
also examine the prospects for detecting neutrinos from
supernovae. These neutrinos, with energy in the 10-MeV range,
would exhibit scattering rates in nuclei that are enhanced by coherent
scattering and will give recoils above dark matter detection
thresholds.

In the latter part of the report we focus on the specific experiments
and technical requirements in Sections~\ref{experiments}
and~\ref{infrastructure}, respectively. The status of several
candidate experiments for {\small DUSEL} are briefly reviewed, looking
at both current status of progenitors and plans for the future. Infrastructure
requirements are treated in subsections on depth, materials handling,
and space and facility needs.

The report concludes with sections on the international context
regarding experiments and laboratories outside the {\small US}, and
our assessment of the long-range dark matter road map.

\section{WIMP Dark Matter: Cosmology, Astrophysics, and Particle Physics}
\label{motivation}

The evidence for dark matter is overwhelming, and arises from a wide
range of self-consistent astrophysical and cosmological data. Dark
matter appears to be ubiquitous in spiral galaxies, dating to
observations from the 1970's by Rubin and Ford~\cite{rubin}. It is
evident there in so-called ``flat'' rotation curves, in
which the rotation speed of stars and gas as a function of galactic radius is
larger than can be attributed to the centripetal force exerted by the
luminous mass.

\begin{figure}[htb]
\begin{center}
\includegraphics[width=2.38in]{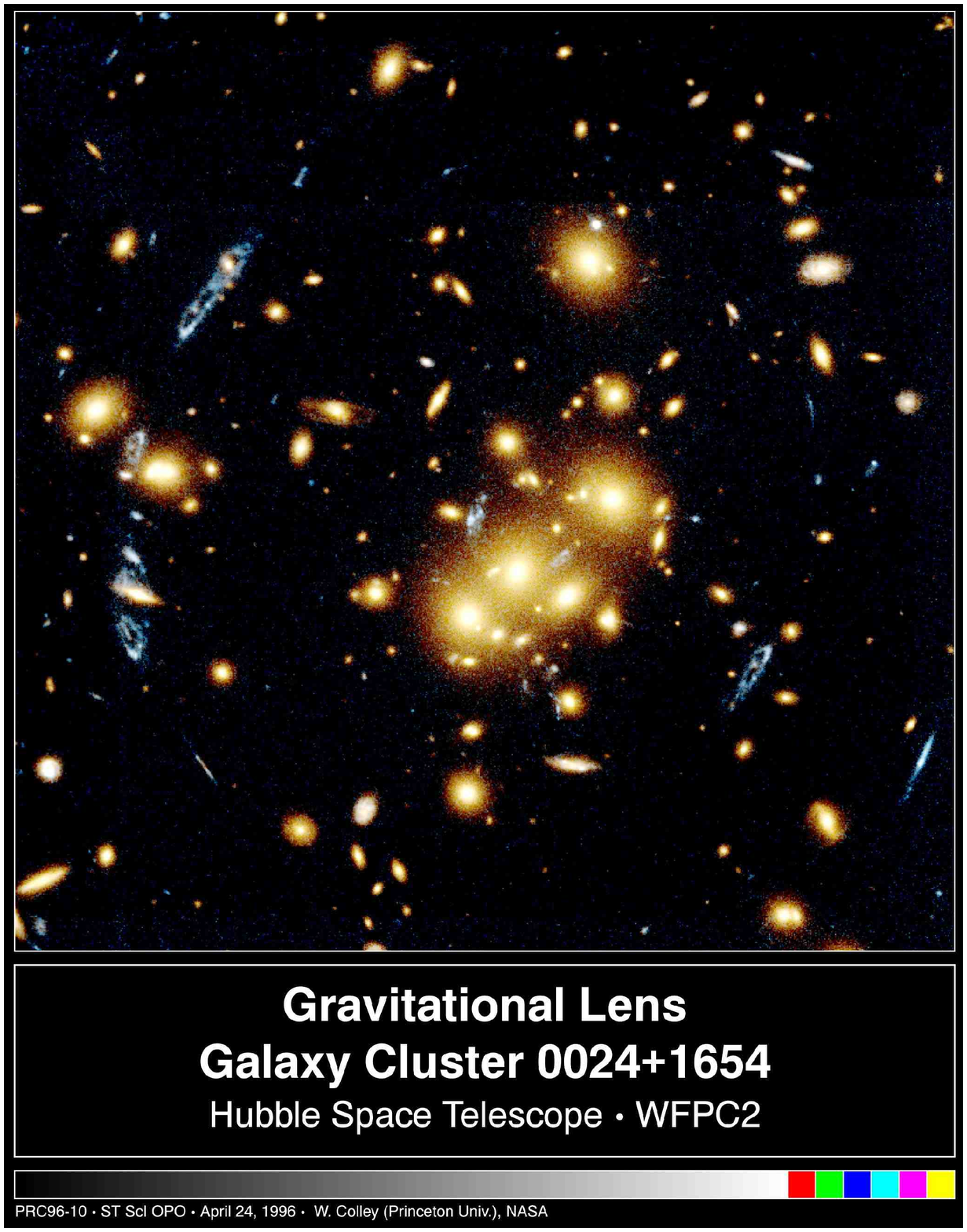}
\includegraphics[width=3.82in]{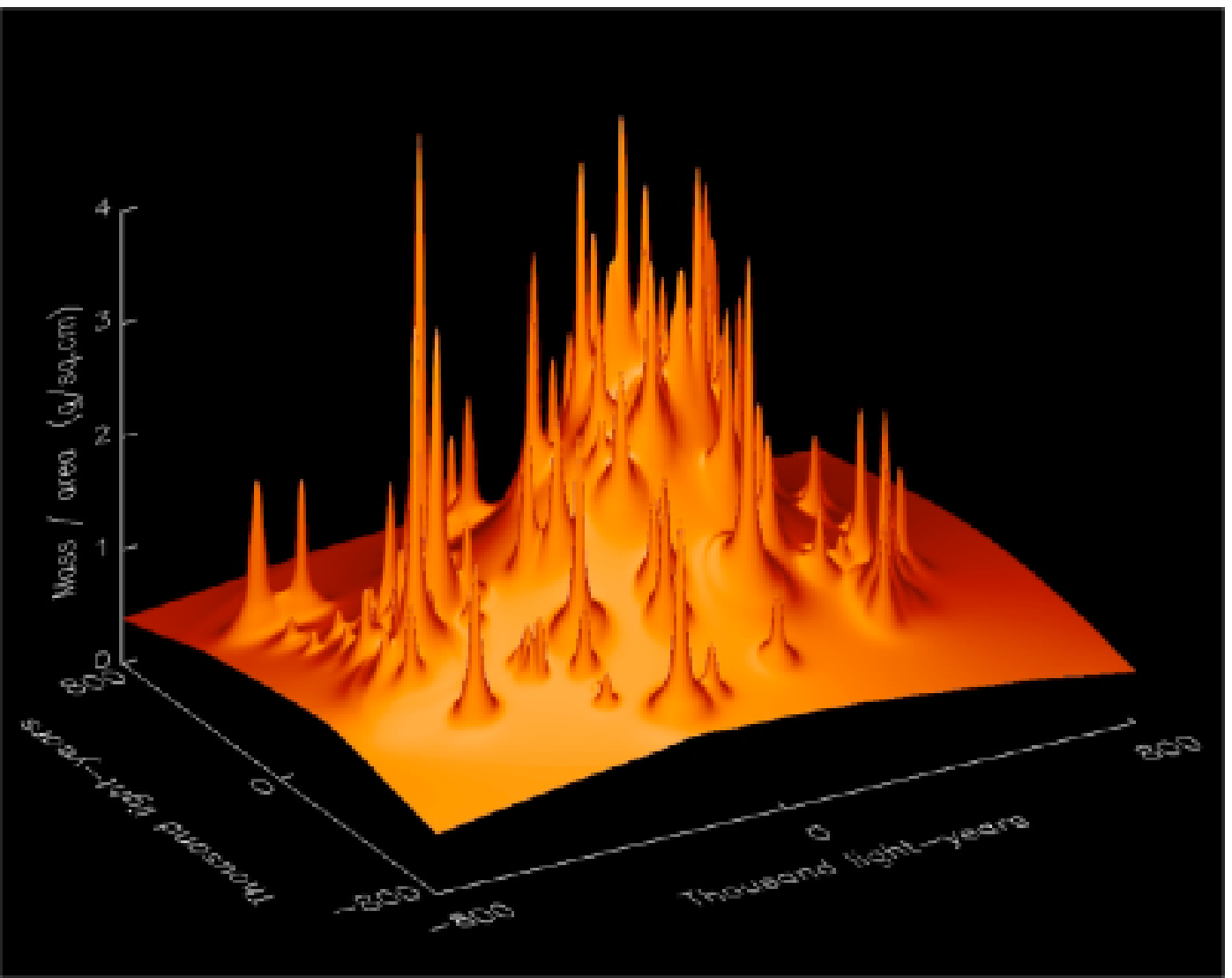}
\end{center}
\caption{\small Left: The foreground cluster of galaxies
  gravitationally lenses the blue background galaxy into multiple
  images~\cite{colley}.  Right: A parametric inversion for the
  strength and shape of the lens shows a smooth background component
  not accounted for by the mass of the luminous objects~\cite{tyson}.}
\label{fig:lens}
\end{figure}

On larger scales, studies of clusters of galaxies show even larger
fractions of dark matter relative to stars and intergalactic
gas. Indeed, it was Zwicky's observation of the Coma cluster in 1933
that provided the first evidence for a missing-mass problem~\cite{zwicky}.
Independent studies of virial speeds of cluster-bound galaxies,
gravitational-bound X-ray-emitting gas, and gravitational lensing 
of background objects by
the cluster all reveal similar amounts of
dark matter.  A striking example, shown in Figure~\ref{fig:lens}, is
the gravitational lensing of a background galaxy into multiple images
by a cluster in the foreground~\cite{colley}. The mass of the cluster
derived from its luminosity is insufficient to account for the
strength of the lens, as indicated by the fit to a smooth distribution
of matter underlying the galaxies, themselves~\cite{tyson}. 
The mean matter density of the Universe attributed to
measurements of X-rays emitted from the gas finds that {\small
$\Omega_m=0.26^{+0.06}_{-0.04}$}~\cite{allen2004} in units of critical
density.

On cosmological scales, detailed mapping of the anisotropy of the
cosmic microwave background ({\small CMB}) and observations of
high-redshift supernovae and the large-scale distribution of galaxies
have led to a concordance model of cosmology, which is consistent with
the mean matter density of the universe inferred from clusters and the
primordial light-element abundances from {\small BBN}. In this very successful
model, the universe is made of $\sim$4\% baryons which constitute the
ordinary matter, $\sim$23\% nonbaryonic dark matter and $\sim$73\%
dark energy (see, for example Refs.~\cite{spergel} or~\cite{tegmark}, for
fits to cosmological parameters). From these observations we also know
that the dark matter must have been nonrelativistic at the time that
the energy density of radiation and matter were equal, and is thus
generically referred to as ``cold'' dark matter.  Modifications to
gravity that can explain all the observations that indicate dark
matter seem unlikely because the observations cover such a large
distance scale.

% http://www.journals.uchicago.edu/cgi-bin/resolve?id=doi:10.1086/378560

Thermal production of \WIMPs\ in the early Universe, followed by
freeze-out, provides a natural mechanism for creation of dark matter
with the observed relic density~\cite{lee1977, jungman}. This is a
straight-forward extension of the standard {\small BBN}
scenario, which successfully predicts the abundance of the light
elements. The time and temperature at which the \WIMPs\ decouple from
ordinary matter is determined by their cross section for annihilation
with themselves or other particles in the hot plasma, and the
abundance at the decoupling time is determined by the temperature and
the mass. Therefore, the relic density of dark matter today would
depend principally on the \WIMP's mass and annihilation cross
section. Also, simple dimensional arguments relating the mass and
cross section, along with the constraint that the relic density be
equal to the dark matter density, naturally satisfy the criterion that
the particles are non-relativistic at decoupling.  Purely cosmological
considerations lead to the conclusion that \WIMPs\ should interact
with a cross section similar to that of the Weak Interaction.

Other scenarios for particle cold dark matter exist, such as the very
light axions~\cite{axions} or the very heavy \WIMP
zillas~\cite{wimpzillas}.  Axions arise as a solution to the strong CP
problem and arise in the early universe as a Goldstone boson from the
{\small QCD} phase transition. Like \WIMPs, they are also the subject
of active dark matter search experiments but require a different set
of experimental techniques (which are carried out in surface
laboratories).  Unlike axions and \WIMP zillas, \WIMPs are produced
thermally and represent a generic class of Big Bang relic particles
that are particularly interesting because of the convergence of
independent arguments from cosmology and particle physics. That is,
with the required ranges of mass and cross section characteristic of
the Weak scale, \WIMPs\ occur precisely where we expect to find
physics beyond the Standard Model. Specifically, new physics appears
to be needed at the Weak scale to solve the mass hierarchy problem.
Namely, that precision electroweak data constrain the Higgs mass in
the Standard Model to be less than 193\,GeV/c$^2$ at
95\%\,C.L.~\cite{abbiendi2003} in spite of the radiative corrections
that tend to drive it to the much higher scale of Grand Unified
Theories ({\small GUT}s).  These corrections tend to be cancelled in
Supersymmetry, which naturally predicts that the Lightest
Supersymmetric Partner ({\small LSP}) is stable and interacts at
roughly the Weak-Interaction rate. More recently, compact extra
dimensions have been proposed to solve this problem, also leading to
\WIMP\ candidates.  The theoretical parameter space is very
unconstrained so any empirical data that restrict it, both from dark
matter detection and accelerator-based methods, are extremely valuable

If \WIMPs\ are indeed the dark matter, their local density in the
galactic halo inferred from the Milky Way's gravitational potential
may allow them to be detected via elastic scattering from atomic
nuclei in a suitable terrestrial target~\cite{goodman}.  Owing to the
\WIMP-nucleus kinematics for halo-bound particles, the energy
transferred to the recoiling nucleus is on the order of 10\,keV.  The
expected rate of \WIMP\ interactions, which is currently limited by
observations to less than 0.1 events/kg/day~\cite{cdms119}, tends to
be exceeded in this energy range by ambient radiation from
radioisotopes and cosmic rays, and so sensitive high-radiopurity
detectors and deep underground sites are required.

\begin{figure}[!h]
\begin{center}
\includegraphics[width=3.2in]{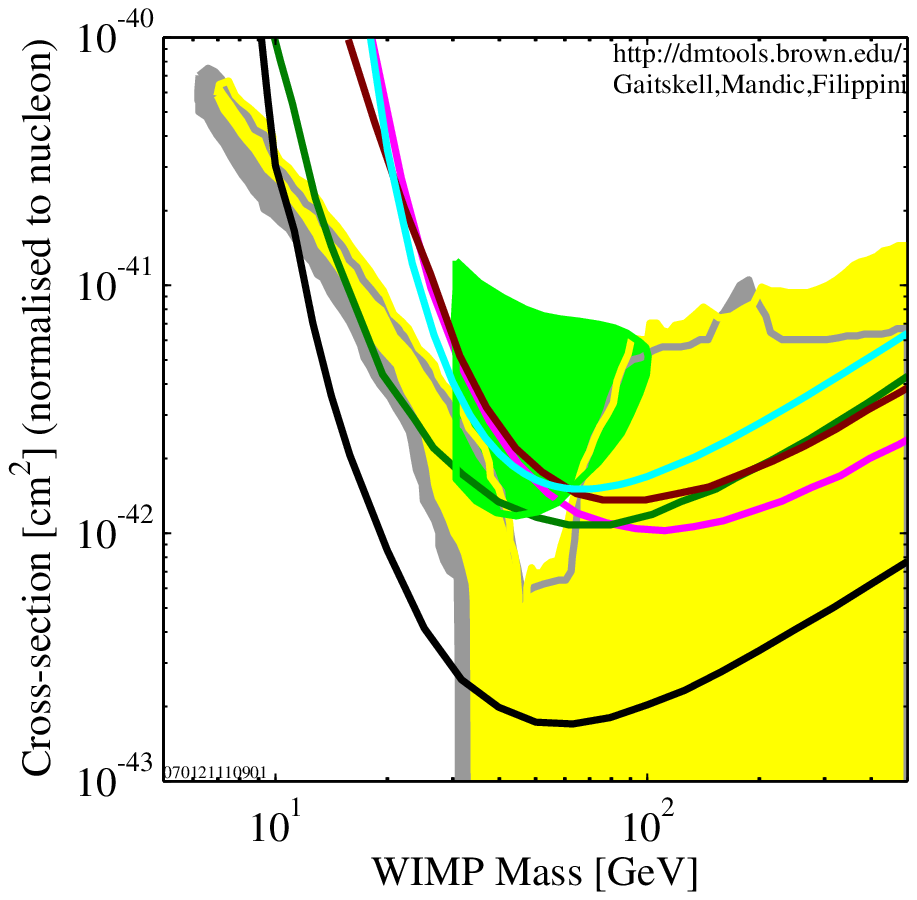}
\includegraphics[width=3.2in]{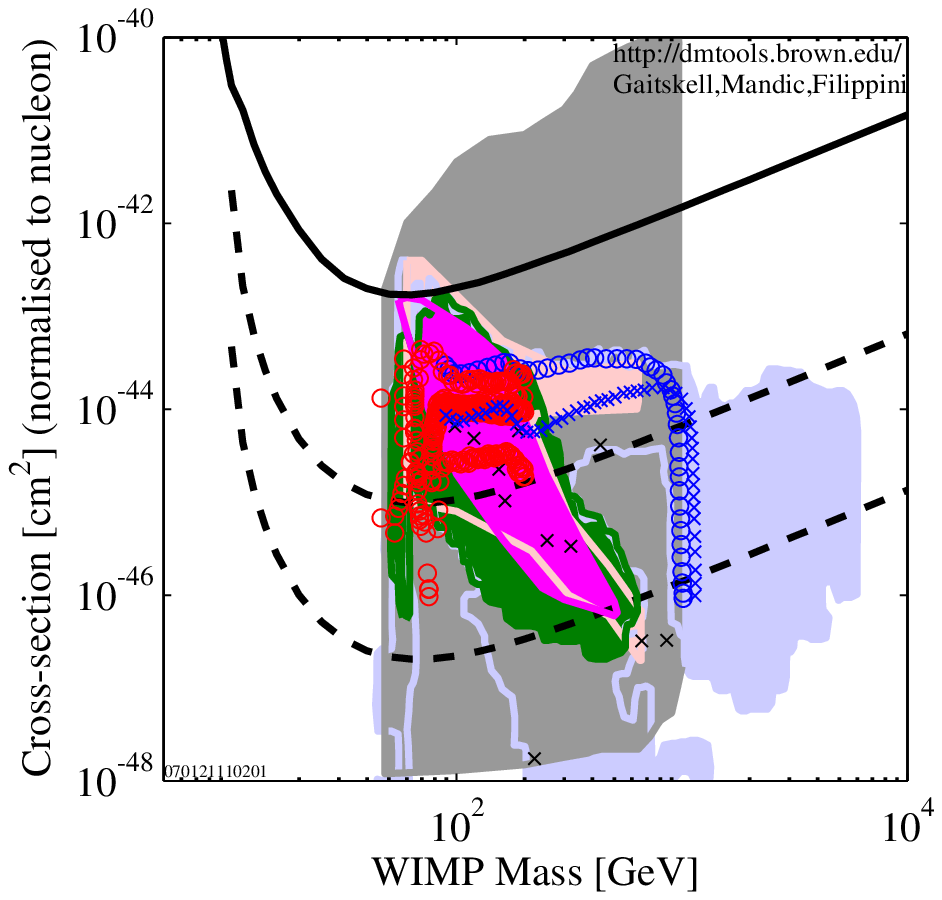}
\end{center}
\caption{\small Plots of the elastic scattering cross section for
spin-independent couplings versus WIMP mass. (a) The left panel shows
the leading experimental results in which the solid curves represent
experimental upper limits from the ZEPLIN-I single-phase liquid xenon
detector (dark green)~\cite{zeplin}, the EDELWEISS thermal and
ionization cryogenic detectors (dark red)~\cite{edelweiss}, the CRESST
thermal and scintillation cryogenic detectors
(cyan)~\cite{cresst2005}, the WARP two-phase liquid argon detector
(magenta; 55 keV threshold)~\cite{warp}, and the CDMS-II
athermal-phonons and ionization detectors (black)~\cite{cdms119}, in
all cases assuming the standard halo model~\cite{lewin96} and
nuclear-mass-number $A^2$ scaling. The contested DAMA
annual-modulation claim~\cite{dama} is shown by the green region.  The
yellow and grey regions represent unconstrained Minimal Supersymmetric
Standard Model (MSSM) predictions for low-mass WIMPs that result from
relaxing the GUT-scale unification of gaugino masses~\cite{Bottino03}.
(b) The right panel displays a broad range of models. The most
unconstrained MSSM predictions are indicated by the large grey
region~\cite{Kim02}.  Models within the minimal supergravity ({\small
mSUGRA}) framework are shown in dark green~\cite{Baltz04},
magenta~\cite{Chattopadhyay04} and pink~\cite{Baer03}; the light blue
region is from relaxing the constraint imposed by the SUSY
interpretation of the muon ``$g-2$'' anomalous magnetic moment
measurement~\cite{Baltz04}. More specific predictions are given by
split Supersymmetry models shown by blue circles and crosses (for
positive and negative values of the $\mu$ parameter,
respectively)~\cite{Giudice04} and red circles~\cite{Pierce04}. The
set of representative post-LEP LHC-benchmark models are shown by black
crosses~\cite{Battaglia03}. Experimental projections are shown as
black-dashed curves and are representative of the sensitivity of next- and
next-next-generation searches with detector masses in the range from
several tens of kilograms up to the ton scale. The CDMS-II limit is
shown again as a solid black curve for reference with the left panel.}
\label{fig:models_limits}
\end{figure}

Figure~\ref{fig:models_limits} illustrates the common landscape of
direct \WIMP\ searches and the theoretical parameter space in plots
of the \WIMP-nucleon cross section versus \WIMP\ mass. The experimental
bounds are represented by the U-shaped curves, above which cross
sections are ruled out for a given experiment. Theoretical models
that are constrained by the required relic density and all known
accelerator bounds are differentiated by alternative 
theoretical ideas. In some of the theoretical cases, bounded regions
indicate the remaining allowed parameter space. In other cases, the
model is fully specified and serves as a so-called benchmark,
which is typical of a given class of models and provides for a more
direct sensitivity comparison among different approaches, e.g.,
astrophysical searches or accelerator experiments. As is evident in
the figure, some models of interest are already being constrained by
direct detection methods.

In addition to the direct searches for \WIMP\ dark
matter, \WIMPs\ are being sought by indirect means by looking for
\WIMP\ annihilation products. Since \WIMPs\ would scatter and lose energy
in astrophysical objects such as the Earth, Sun and in the galactic
center, the number density could become high enough to detect
annihilation or subsequent decay products such as neutrinos, gamma
rays or charged particles. We describe the influence of these
methods in Section~\ref{indirect}.

\section{Direct Detection of WIMPs}
\label{detection}

\WIMP-search experiments seek to measure the interactions of dark
matter particles bound in our galactic halo with the atomic nuclei of
detector target materials on Earth.  The calculation of the 
rate in terrestrial detectors depends on 
the \WIMPs\ velocity distribution and density in the galaxy,
the \WIMP\ mass and elastic-scattering
cross section on nucleons, and the nuclear structure of the target
nuclei.  Unfortunately, the relic density of \WIMPs\ provides only a
loose constraint on the scattering cross section because the relic
density depends on the processes through which the particles
annihilate. In the well-studied example of \SUSY, model-dependent
co-annihilation channels prevent the making of a simple
crossing-symmetry argument to relate the density and the scattering
rate~\cite{jungman}. Indeed, \SUSY-based calculations for neutralinos
show at least five orders of magnitude variation in the nucleon
coupling, and can even vanish in finely-tuned cases.  Covering the
bulk of the \SUSY\ parameter space for \WIMPs\ will require an
increase in sensitivity from the current rate limits of
$\sim$0.1\,event/kg/d to less than 1\,event/ton/year, demanding
increases in detector mass and exposure, and reduction in and/or
improved rejection of radioactive and cosmogenic backgrounds.

Given the measured properties of the Milky Way galaxy and fairly
generic assumptions about the spatial and velocity distribution of its
dark matter halo, the spectrum of \WIMP\ energy depositions in a
detector can be calculated in terms of the unknown \WIMP\ mass and total
cross section for nuclear scattering. Because the \WIMPs\ must be highly
non-relativistic if they are to remain trapped in our galaxy, the
shape of the spectrum has almost no dependence on the detailed
particle physics model. As discussed in~\cite{lewin96}, the differential
rate takes the form

\begin{equation}\label{spectrum}
{dN \over dE_r} = {\sigma_0 \rho_\chi \over 2 \mu^2 m_\chi} F^2(q)
\int^{v_{esc}}_{v_{min}} {f(v) \over v} dv,
\end{equation}

\medskip \noindent where $\rho_\chi$ is the local \WIMP\ density,
$\mu$ is the \WIMP-nucleus reduced mass $m_\chi m_N/(m_\chi + m_N)$
(assuming a target nucleus mass $m_N$, and \WIMP\ mass $m_\chi$), and
the integral takes account of the velocity distribution $f(v)$ of
\WIMPs\ in the halo. The term $v_{min}$ is the minimum \WIMP\ velocity
able to generate a recoil energy of $E_r$, and $v_{esc}$ is the
galactic escape velocity. $F^2(q)$ is the nuclear form factor as a
function of the momentum transfer $q$ and $\sigma_0$ is the total
\WIMP-nucleus interaction cross section. 

The astrophysical uncertainties in Equation~\ref{spectrum} are
contained in $\rho_\chi$ and $f(v)$, which can be estimated by
comparing our galaxy's measured rotation curve to dark matter halo
models. Simple models of galactic structure indicate that the
particles in the halo should be relaxed into a Maxwell-Boltzmann
distribution with an {\small RMS} velocity related to the maximum
velocity in the rotation curve. In the Milky Way, the value of the
{\small RMS} velocity is estimated to be $v_{RMS} \sim$220\,km/s. This
will be assumed below in comparing direct detection
experiments. However, the velocity structure of the dark halo may be
more complex than a simple Maxwell-Boltzmann distribution. Current
understanding of structure formation posits that a dark halo of
galactic dimensions is built hierarchically from the merging of
smaller dark halos. In the central parts of the galaxy, where many
small halos have converged, the short orbital times and the strong
tidal forces blend the halos together into a relaxed structure with a
Maxwell-Boltzmann distribution. In the outer parts of the galaxy,
where the orbital times are long (several Gyr), small halos are still
falling into the galaxy at present. Matter is pulled out of the small
halos by tidal forces and is redistributed into long arms in front and
behind them. The capturing of a multitude of small halos renders the
outer galaxy a crisscross of tidal streams. The crucial question for
estimating direct detection rates is if the solar system lies in the
outer or the inner part of the galactic halo, i.e. where the halo
velocity distribution is Maxwell-Boltzmann or where it is not. Recent
data on the Sagittarius dwarf galaxy show that it may be one the
``small halos'' still falling into our Milky Way, and one of its tidal
arms may pass very close, if not even across, the solar
system~\cite{majewski}. The amount of matter in the Sagittarius arm
seems however to be only a fraction of the local dark matter
density~\cite{freese04,freese05}. Theoretical calculations using
N-body simulations and semi-analytic evaluations of the small halo
tidal disruption rate are still too controversial to answer the
question of how far from the galactic center small halos are well
merged~\cite{diemand,zhao}. The discovery of \WIMPs\ in direct
detection experiments and the subsequent measurement of the \WIMP\
velocity distribution by means of directional detectors seems to be a
primary route to provide an answer and inspire theoretical
investigations.

The local dark matter density $\rho_\chi$ has significant uncertainty,
since the local rotation velocity only constrains the total mass
inside our solar system's radius while the matter distribution in the
halo is not well-constrained by empirical data~\cite{kamion1998}. For
example, deviations of the dark halo from a spherical shape to a
flattened spheroid can produce an increase of the estimated local dark
matter density from 0.3 to 2 GeV/c$^2$/cm$^3$~\cite{gates95}. For the
purpose of comparing the sensitivity of different dark matter
experiments, it is conventional to use the value quoted by the
Particle Data Group, $\rho_\chi$=0.3 GeV/c$^2$/cm$^3$~\cite{pdg04},
which is known to within a factor of two.

Integrating equation~\ref{spectrum} over the sensitive recoil-energy
range of the detector gives the expected rate for a calorimetric
detector. Since the spectrum is featureless, using secondary
signatures to distinguish a \WIMP\ signal from the ambient backgrounds
is desirable, such as the following two well-studied possibilities.
First, the direction
of the recoiling nucleus is correlated with the motion of the
laboratory through the galactic rest frame. This manifests itself as a
diurnal modulation in a terrestrial detector owing to the Earth's
rotation. Second, the recoil-energy spectrum undergoes a seasonal
kinematic variation owing to the Earth's orbit around the
Sun.\symbolfootnote[2]{This feature, which is the basis of a
controversial detection claim by the {\small DAMA} Collaboration's NaI
array, appears as an annual modulation in the scattering rate over a
fixed recoil-energy range.  Aside from the {\small DAMA} claim, which
has been largely ruled by other experiments under standard
assumptions, no other detections have been reported. See
Section~\ref{experiments} for further details.} 

If the \WIMP\ is a neutralino it couples to nucleons via neutral
current reactions mediated by exchange of Z$^0$'s, Higgs particles and
squarks. In the non-relativistic limit, the most general possible
interaction between a Majorana fermion such as the neutralino and the
nucleon reduces to a simple form with only two terms, one of which is
a spin-independent coupling and the other a coupling between the
neutralino spin and the nucleon spin. Because the de Broglie
wavelength of the momentum transfer is of nuclear dimensions (i.e.,
$\lambda=h/q\sim=1$\,fm for $m_\chi=100$\,GeV/c$^2$ and
$v=220$\,km/s), the interaction is at least partially coherent over
the target nucleus. For the spin-independent coupling, full coherence
results in a cross section $\sigma_0 \propto A^2$ for a nucleus of
atomic number $A$. For spin-dependent scattering, the coupling is
dominated by the net nuclear spin, since the contribution from paired
opposite-spin nucleons cancels. Corrections based on nuclear spin
structure functions can spoil this
cancellation, rendering odd-proton nuclei sensitive to
\WIMP-neutron couplings~\cite{jungman, tovey2000}, or vice versa.

In general, for a \WIMP\ with equal spin-dependent and spin-independent
couplings, detection via spin-independent scattering on a large-$A$
nucleus is favored owing to the coherent enhancement. The majority of
experiments performed to date have used heavy target materials to
maximize sensitivity to this scattering mode. However, models
exist (e.g., neutralinos that are pure gaugino or pure higgsino
states) in which the spin-independent coupling is highly suppressed;
a few experiments targeted at these models have been done with
low-$A$ high-spin target materials. Since the specific composition of
\WIMPs\ is not known, the long-term program should address this range
of possibilities.

\begin{figure}[htb]
\begin{center}
\includegraphics[width=4.0in]{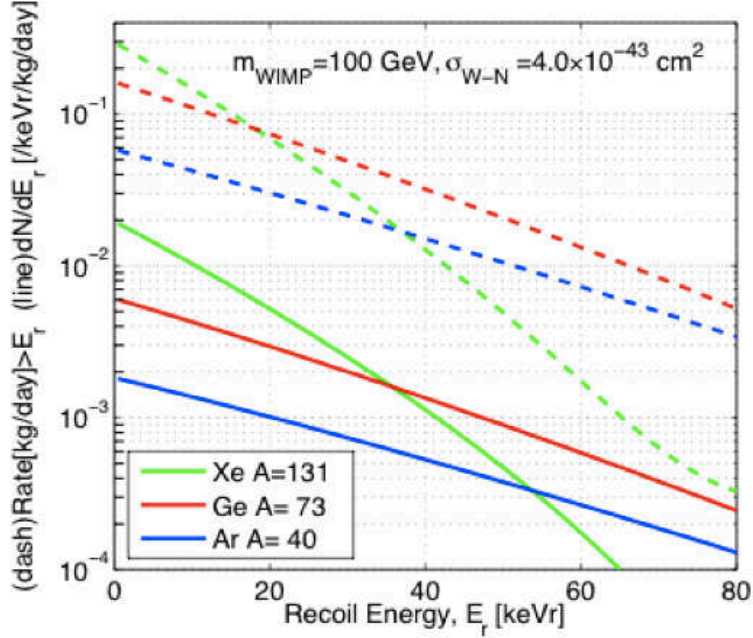}
\end{center}
\caption{\small Calculated differential spectrum in
evts/keV/kg/day (solid lines), and the integrated rate evts/kg/day
(dashed lines) above a given threshold energy (keV$_r$) for Xe, Ge and
Ar targets. A 100-GeV/c$^2$ WIMP with a
WIMP-nucleon spin-independent cross section of $\sigma = 4 \times
10^{-43}$\,cm$^2$ has been used. The plot shows that the total event rate
per mass in a Xe detector with a threshold of 16\,keV$_r$ is identical to
that of a Ge detector with a lower threshold of 10\,keV$_r$. An Ar
detector with a threshold of 16 (30)\,keV$_r$ would have a total event
rate per mass which is 1/3 (1/5) of Ge and Xe at the stated
thresholds.}
\label{fig:recoil_energy}
\end{figure}

For large target nuclei and large recoil energies, the interaction
between the \WIMP\ and a nucleus loses coherence and the
spin-independent cross section is suppressed for large
$q^2$. Therefore, there are important tradeoffs to be made when
designing experiments between the choice of target nucleus and the
obtainable energy threshold. This point can be best illustrated by
showing the results (Figure~\ref{fig:recoil_energy}) of a full
calculation assuming the spin-independent coupling dominates, using
standard halo parameters and the formalism discussed in~\cite{lewin96}. A
\WIMP\ mass of 100\,GeV/c$^2$ is chosen with a cross section normalized to
the nucleon, which is representative of the best current
limits in direct detection experiments. Figure~\ref{fig:recoil_energy}
shows both the differential and integrated (above the indicated
threshold) \WIMP\ event rate in keV$_r$ (which is the recoil energy
imparted to the nucleus) expected for single isotope targets of
$^{131}$Xe (similar for $^{129}$I), $^{73}$Ge, and $^{40}$Ar.

It can be seen that for a given elastic scattering cross section for
\WIMP-nucleon interactions, the smaller nuclei are penalized owing to
a combination of smaller coherence enhancement ($\sim A^2$) and the
less effective transfer of recoil energy to a target that is lighter
than the \WIMP. The recoil spectrum for the heavier Xe nucleus is
significantly suppressed by the loss of coherence for higher $q^2$
scattering events (form factor suppression). For a 100-GeV/c$^2$
\WIMP, the integrated event rate drops by a factor of two for a
threshold recoil energy increase of 13, 20, and 22 keV$_r$ for Xe, Ge,
and Ar, respectively. 

The scattering of \WIMPs\ on nuclei would produce signals in many types
of conventional radiation detectors. For example, scintillation
counters, semiconductor detectors and gas counters are all capable of
detecting nuclear recoils of a few keV. Unfortunately, these
instruments are also efficient detectors of environmental radiation,
such as cosmic rays and gamma rays from trace radioisotopes present in
construction materials. Detectors exposed to environmental radiation
in an unshielded room typically register about 10$^7$ events a day
per kg of detector mass, while \WIMPs\ are known to produce less
than 0.1 count per kg-day (approximately the current limit) and
significant \SUSY\ parameter space exists down to $10^{-6}$ per
kg-day. Exploring this parameter space will require ton-scale
detectors with nearly vanishing backgrounds.

Background reduction in \WIMP\ detectors can be approached with two
basic strategies: reduction in the background radiation level by
careful screening and purification of shielding and detector
components, and development of detectors that can discriminate between
signal and background events. Most experiments employ a combination of
the two strategies.

Technologies for producing ultra-pure low-radioactivity materials have
been pursued by many groups, and materials have been produced or
identified with levels of radioisotope contamination three to six
orders of magnitude below typical environmental contamination
levels. For example, $^{238}$U
is typically present at $10^{-6}$-g/g 
but levels in the $10^{-9}$--$10^{-12}$\,g/g range 
have been achieved~\cite{heusser}. 
This progress has gone hand-in-hand with the
development of ever more sensitive instruments for the detection of
contamination. After the common environmental radioisotopes have been
minimized by purification, the dominant source of radioactivity in
materials is often ``cosmogenic,'' i.e., originating from isotopes
that are produced in the material by cosmic-ray-induced secondaries
during the above-ground manufacturing process. Since purification and
assembly of all but a few types of detector materials in an
underground cosmic-ray-free environment poses a logistical challenge,
additional work is required to reduce activity below the lower limit
imposed by cosmogenic activation (see Section~\ref{handling} for
further discussion). Most recent experiments emphasize
improving the sensitivity using a background discrimination mechanism.

The development of new background discrimination mechanisms is a very
active field and many innovative techniques are being studied at the
kilogram scale. The most sensitive proven technique, which is used by
the \CDMS\ and {\small EDELWEISS} experiments, exploits the difference
in charge yield in semiconducting crystal targets for
nuclear-scattering compared to background electron-scattering
processes. The target crystals, which are made of silicon or
germanium, are cooled below 100 mK and instrumented with both charge
and phonon sensors. The ratio of the amplitudes of the charge and
phonon pulses (or simply the ratio of charge to temperature rise in
some versions of the technology) is used as a discrimination
parameter. The detailed shape and timing of the pulses contains
additional information which can be exploited to separate
nuclear-scattering events from events with imperfectly-collected
charge.

Other background discrimination technologies include low-temperature
detectors with phonon and scintillation readout, cryogenic noble
liquid targets with pulse shape discrimination and/or simultaneous
measurement of charge and light, low-pressure gas with sensitivity to
the energy density and direction of nuclear recoils, and heavy-liquid
bubble chambers with bubble nucleation conditions fine tuned to avoid
backgrounds.  Demonstrated discrimination factors for gamma rays from
these techniques (i.e., fraction of unrejected background events)
range down to $10^{-9}$ and some of them are plausibly scalable to
greater than one ton of target mass. Details and status reports on
several of the technologies are given in Section~\ref{experiments}.

The problem of reducing the background from neutron scattering in dark
matter detectors requires special mention. Neutrons interact
exclusively by nuclear scattering, so the discrimination methods
mentioned above are not effective. Discrimination is possible based on
the propensity for neutrons to multiple-scatter in a detector, while
\WIMPs\ would only scatter once. However, discrimination mechanisms
based on multiple scattering only become effective for very large
detectors with excellent spatial resolution or high granularity.
Neutrons up to 8\,MeV are produced by ($\alpha$,n) reactions and
spontaneous fission in
rock and common construction materials, and cosmic rays produce
non-negligible fluxes up to 1\,GeV, even at substantial depths
underground. The low-energy neutrons can be shielded by practical
thicknesses of hydrogenous moderating materials and minimized by
material selection, but the high-energy
neutrons are very difficult to attenuate by shielding. This problem,
which is discussed in detail in Section~\ref{depth}, strongly
influences the choice of appropriate underground laboratory depth.

We conclude this section with a description of the current results for
both spin-independent and spin-dependent scattering. The limits on the
spin-independent cross section, in particular from
\CDMSII~\cite{cdms119}, {\small EDELWEISS}~\cite{edelweiss}, 
{\small WARP}~\cite{warp} and
{\small ZEPLIN~I}~\cite{zeplin}, are beginning to significantly
constrain the \WIMP-nucleon scattering cross section in some \SUSY\
models, as shown earlier in Figure~\ref{fig:models_limits}.  As that
figure shows, it is difficult to reconcile the {\small DAMA} claim of
a \WIMP\ signal based on annual modulation of the rate of iodine
recoils~\cite{dama} with the current limits on the spin-independent
scattering cross section from other experiments. A recent
reinterpretation of {\small DAMA}'s data as an annual modulation on
the rate of the lighter sodium nuclei~\cite{Gondolo05} is sensitive to
lower-mass \WIMPs. Such \WIMPs\ are possible in \SUSY\ if gauge unification
is relaxed~\cite{Bottino03}, although these investigators predict
a cross section ten times lower than the corresponding {\small DAMA}
interpretation. The recent result from \CDMSII\ rules
out most of this parameter space~\cite{cdms119}. 

\begin{figure}[htb]
\begin{center}
\includegraphics[width=3.2in]{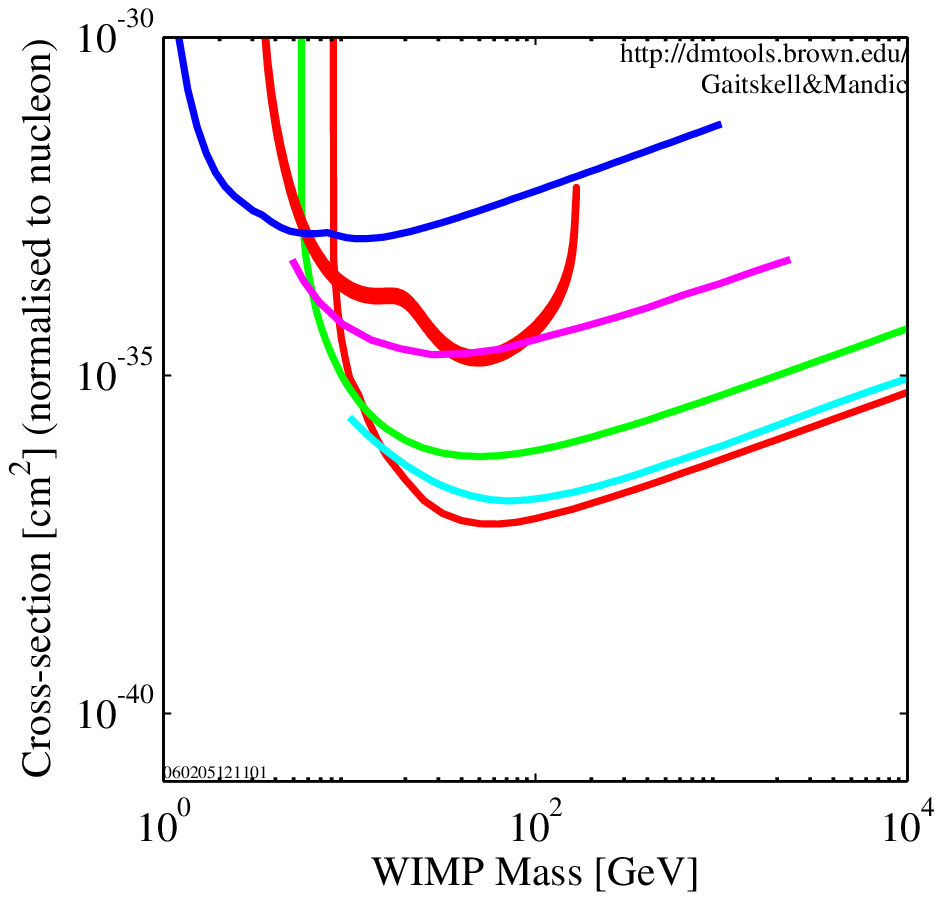}
\includegraphics[width=3.2in]{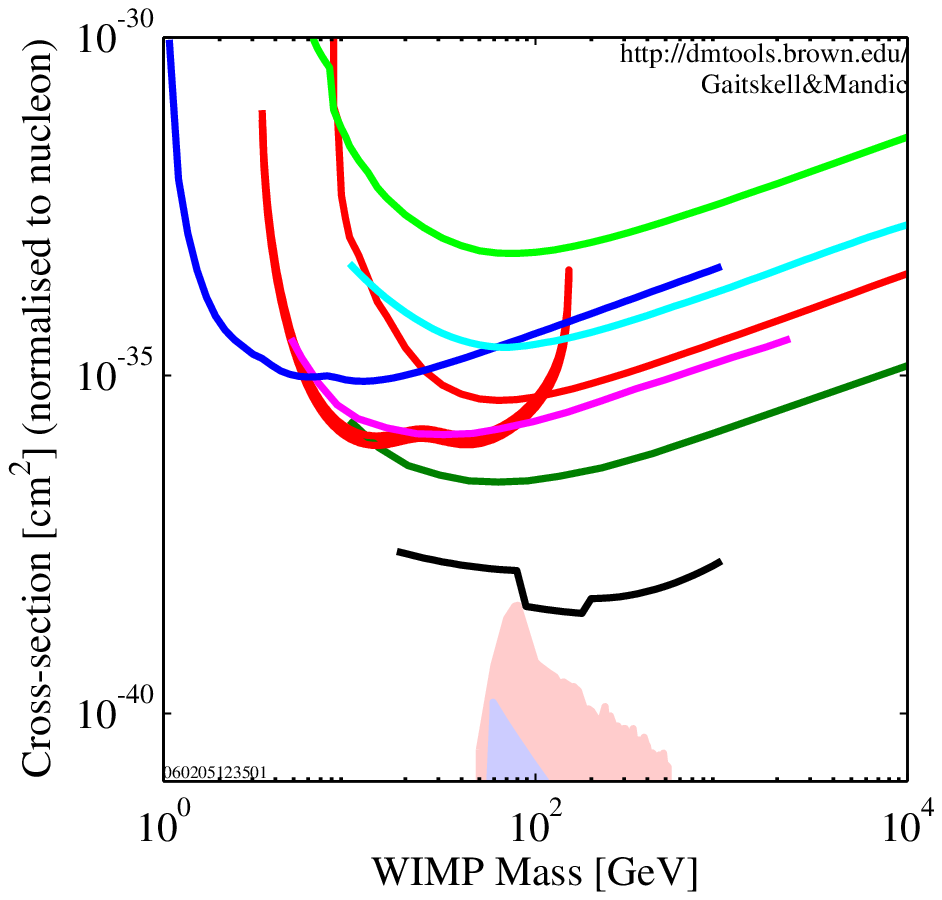}
\end{center}
\caption{\small Current limits on the WIMP-nucleon cross section for
spin-dependent neutron (left) and proton (right) scattering. The
filled red contours are from DAMA~\cite{Bernabei03, Savage04}. The
limit curves are from CRESST-I (blue)~\cite{CRESSTI, Savage04},
PICASSO (magenta)~\cite{Picasso05}, CDMS-II silicon (green) and
germanium (red)~\cite{cdms_sd}, and ZEPLIN-I
(cyan)~\cite{zeplin_sd}. The proton case also includes limits from
NAIAD (dark green)~\cite{naiad} and the indirect detection limit from
Super-Kamiokande (black)~\cite{SuperK}, as well as theoretical
predictions from the Minimal Supersymmetric Standard Model (MSSM)
(pink)~\cite{Ellis01} and the constrained MSSM (light
blue)~\cite{Ellis00}.}
\label{fig:sd_limits}
\end{figure}

Limits on the spin-dependent cross section on neutrons and protons are
shown in Figure~\ref{fig:sd_limits}. Since sensitivity is dominated by
the unpaired nucleon, different experiments tend to provide the best
limits in only one or the other case. The {\small DAMA} signal on the
odd-proton sodium and iodine nuclei for neutron couplings has been
ruled out by other experiments under standard halo
assumptions~\cite{lewin96}. The corresponding case for proton
couplings has been ruled out by other experiments except in the mass
range from about 5--15\,GeV/c$^2$. In addition to the limit curves in
the proton graph, a spin-dependent proton interpretation is excluded
above a \WIMP\ mass of 18\,GeV/c$^2$ by the SuperKamiokande
search~\cite{SuperK} for high-energy neutrinos resulting from \WIMP\
annihilation in the sun.

\section{Indirect Detection of WIMPs}
\label{indirect}

\WIMP\ dark matter can be searched for not only directly with
low-background detectors, but also indirectly through \WIMP\
annihilation products. For example, Supersymmetric \WIMPs\ like the
neutralino are identical to their antiparticles, and can mutually
annihilate and produce standard model particles. If the neutralino
density is large enough, the annihilation rate may be sufficient to
produce detectable signals in the form of neutrinos, gamma-rays, and
cosmic rays.  Detectable neutrinos would be produced in the center of
the Sun~\cite{Silk:1985} or of the
Earth~\cite{Freese:1986,Krauss:1986}; gamma-rays, positrons,
anti-protons, and anti-deuterons would be produced in the galactic
halo \cite{Gunn:1978,Stecker:1978,Silk:1984,Donato:2000}, at the
galactic center \cite{Gondolo:1999}, or in the halos of external
galaxies \cite{Gondolo:1994}.

Typically, the annihilation products have
an energy equal to a fraction of the neutralino mass, with the notable
exception of gamma-ray lines at an energy equal to the mass of the
neutralino. The number density of \WIMPs\ is higher 
inside compact celestial bodies such as
the Sun, the Earth, and the 
Galactic center. In addition
to a mono-energetic energy signal in some annihilation channels, a
directional signal pointing back to a celestial body may also provide
a distinguishing observable.

\begin{figure}[!h]
\begin{center}
\includegraphics[width=3.0in]{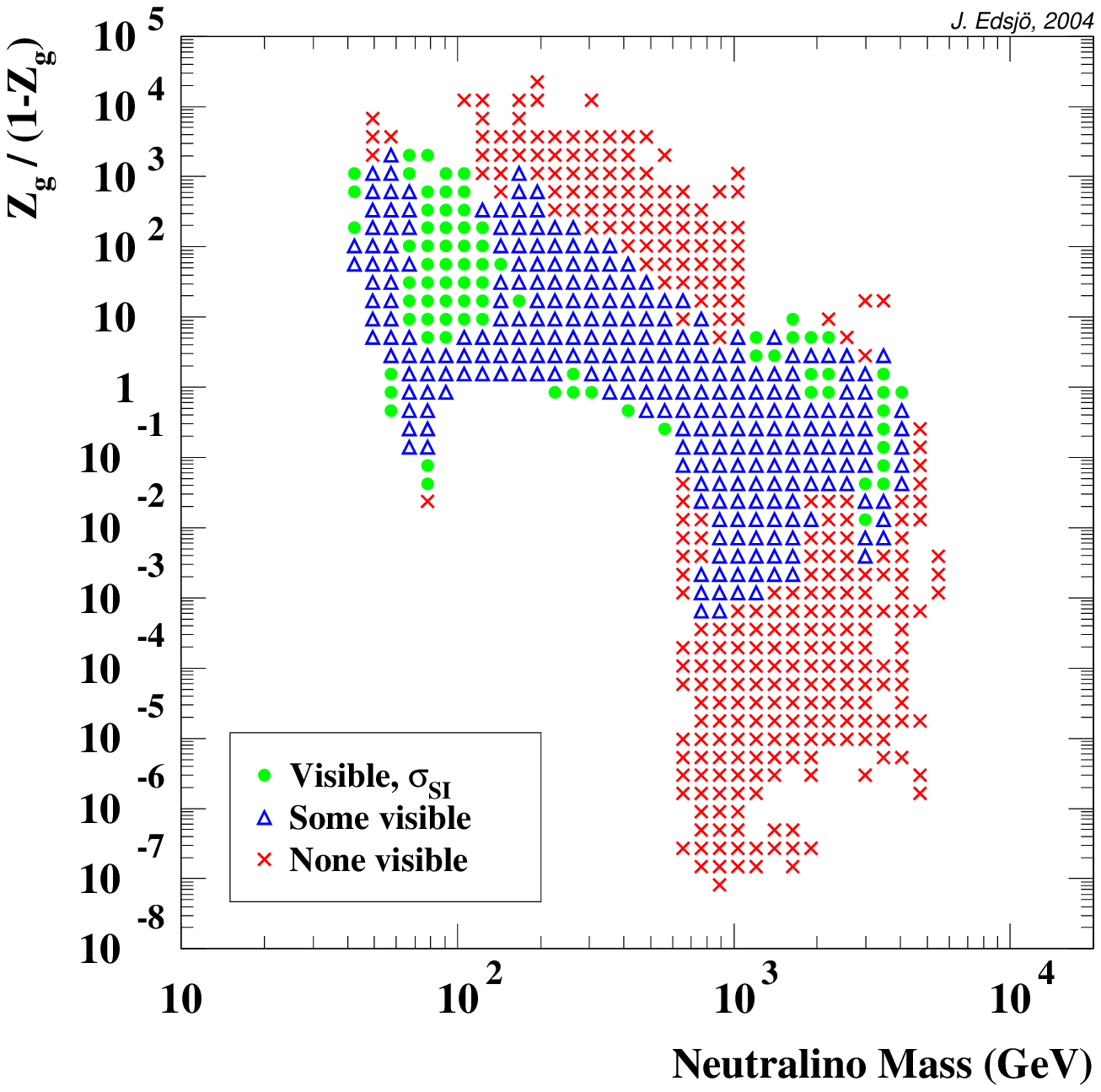}
\includegraphics[width=3.0in]{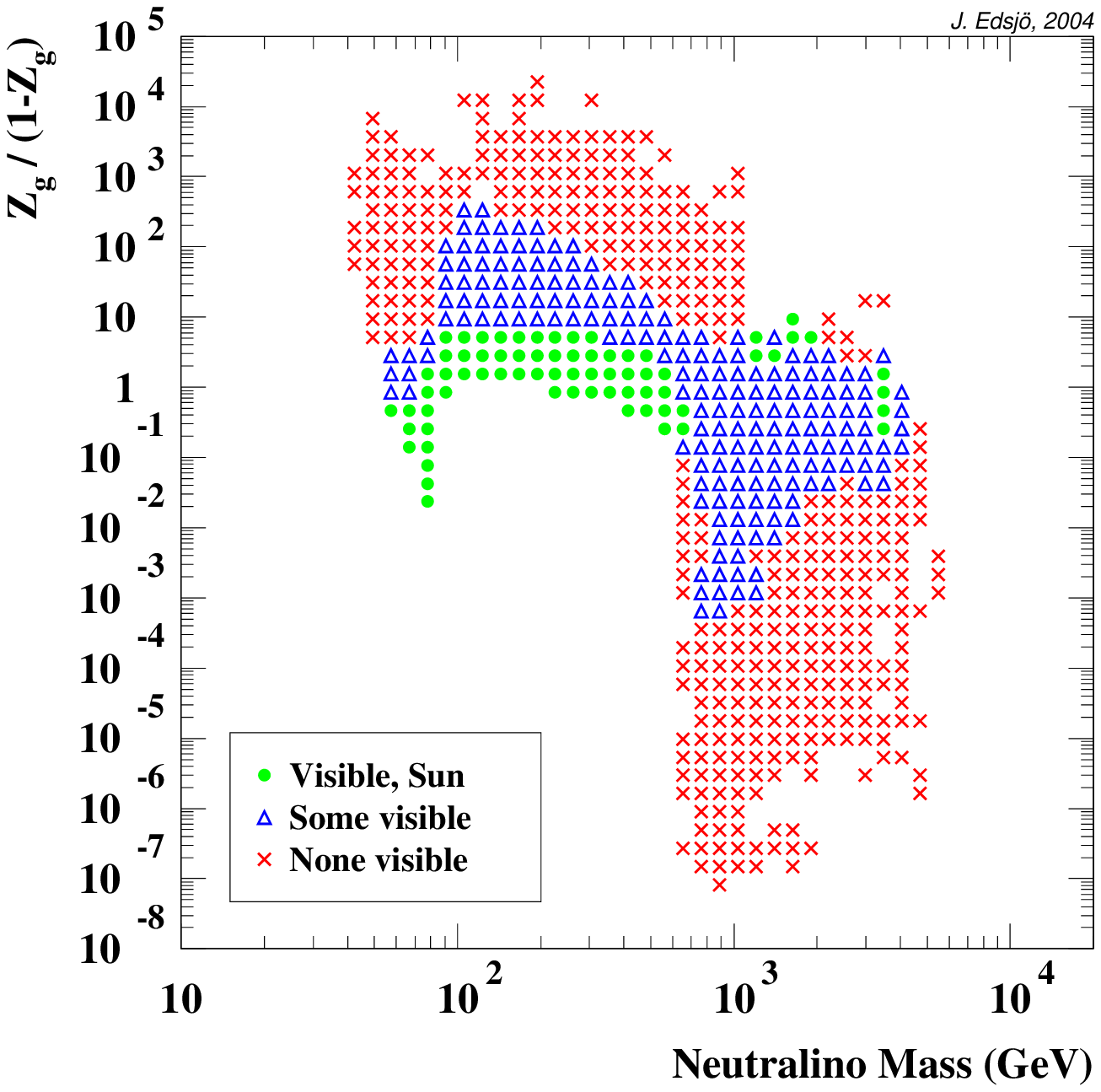}
\end{center}
\begin{center}
\includegraphics[width=3.0in]{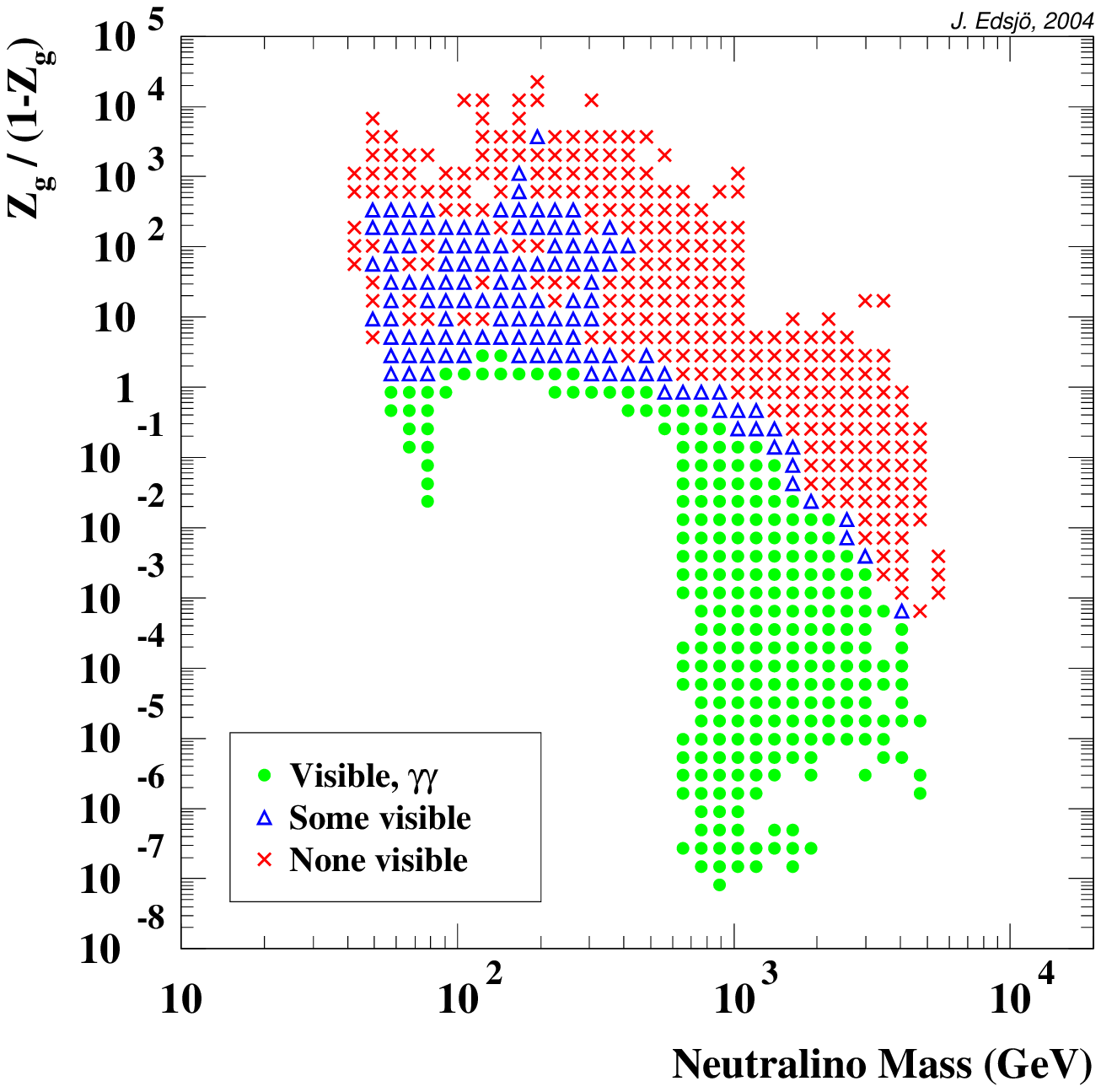}
\includegraphics[width=3.0in]{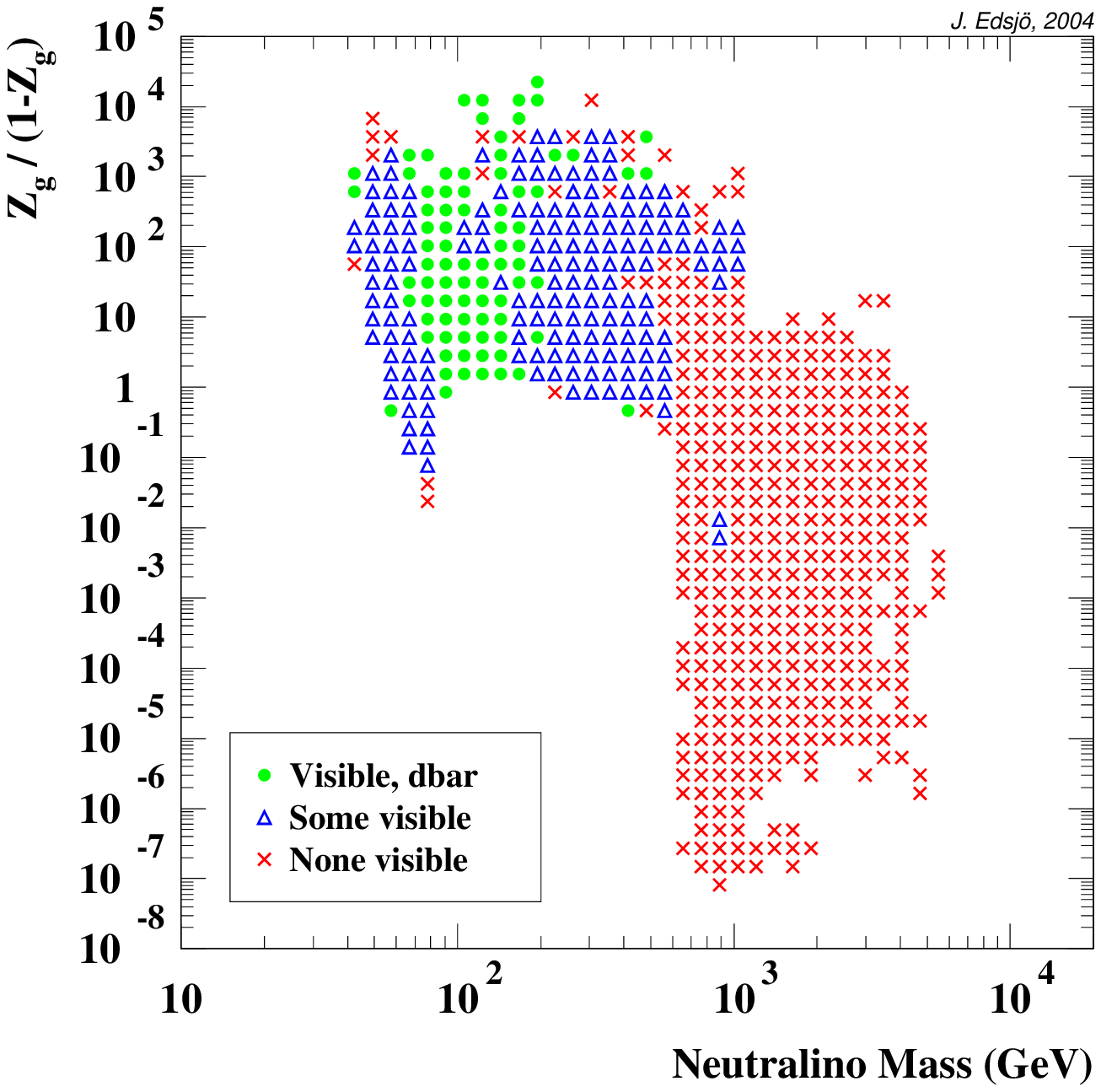}
\end{center}
\caption{\small Reach of next generation searches for neutralino dark
matter: (a) Top left: direct searches, (b) Top right: indirect
searches via high-energy neutrinos from the Sun, (c) Bottom left:
searches for a gamma-ray line in the Galactic halo, (d) Bottom right:
searches for antideuteron in cosmic rays. The figures show a
projection of a 7-parameter space onto the neutralino mass-composition
diagram. The neutralino composition is represented by
$Z_g/(1-Z_g)=Z_g/Z_h$, i.e., the ratio of gaugino and higgsino
fractions. Green full dots indicate that all points in the projection
can be reached in the respective search, blue triangles that some but
not all projected points can be reached, red open circles that none
can be reached. (All from~\cite{edsjo}.)}
\label{fig:indirect}
\end{figure}

We examine here the relationship of indirect searches to direct
searches in deep underground laboratories. The extent of
Supersymmetric parameter space that can be probed by the next
generation of direct detectors (one order of magnitude improvement in
the current limits) is illustrated in Figure~\ref{fig:indirect}(a) for
spin-independent scattering cross sections. The figure represents a
projection of a 7-parameter Supersymmetric space onto the neutralino
mass-composition diagram~\cite{edsjo}. Green full dots indicate that
all models projected onto that point can be reached by direct
detection searches, blue triangles indicate that some but not all
models can be reached, and red open circles that no model can be
reached. The horizontal axis is the neutralino mass $m_\chi$ in GeV/c$^2$,
while the vertical axis is the ratio of gaugino-to-higgsino fractions
$Z_g/Z_h$: neutralinos which are prevalently gauginos are at the top
of the diagram, and neutralinos which are prevalently higgsinos at the
bottom.  Neutralinos of mixed type fall around $Z_g/Z_h \sim 1$. Since
the spin-independent neutralino-nucleon scattering cross section is
chirality-suppressed for pure gauginos and pure higgsinos,
spin-independent direct searches are mainly sensitive to the
mixed-type region.  However, the sensitivity of direct searches has
been greatly improving, and the next generation detector will be able
to probe higgsino purity of the order of 1\%, and gaugino purity
of the order of 0.1\% (the latter at relatively low neutralino masses,
below 100 GeV/c$^2$, say). We will see that indirect searches by means of
gamma-rays and cosmic antideuterons are sensitive to the complementary
higgsino and gaugino regions.

Searches for \WIMP\ annihilation products from the centers of the Sun
and Earth are sensitive to regions of parameter space similar to those
reached by direct searches. Of the annihilation products produced
inside the Sun and the Earth, only the neutrinos interact weakly
enough to escape. These high-energy neutrinos from \WIMPs\ have been
searched for in deep underground neutrino detectors originally built
to search for proton decay. Currently the best upper bounds on
neutrinos from \WIMPs\ come from the SuperKamiokande and Baksan
detectors; the currently-running {\small AMANDA} experiment and the
future IceCube detector will be able to improve on these limits. The
\WIMP\ annihilation rate in the Sun and the Earth generally depends on
the scattering cross section of neutralinos with nuclei, which is the
same cross section probed by direct detection. This is so because for
all but very massive \WIMPs\ the annihilation rate in the Sun and
Earth equals the rate at which they capture neutralinos. Capture
occurs when repeated scattering off nuclei makes the neutralino lose
so much of its kinetic energy that it becomes gravitationally bound
and sinks to the center of the object. Thus the annihilation rate is
governed by the capture rate, in turn proportional to the scattering
cross section. For this reason, direct searches and indirect searches
via neutrinos from the Sun or Earth probe similar regions in
Supersymmetric parameter space. The region reached by indirect
searches for neutrinos from the Sun is shown by dots and triangles in
Figure~\ref{fig:indirect}(b). The reach of this kind of indirect
search is similar to direct searches, i.e., they probe mixed-type
neutralinos, although they do not extend so much into the pure gaugino
and higgsino regions. (One must also distinguish the cases of
spin-dependent and spin-independent scattering cross sections: the
spin-dependent cross section contributes to scattering from hydrogen
in the Sun while several, but not all, direct detection experiments
are mainly sensitive to the spin-independent cross section.)

A clear signature of \WIMP\ annihilation would be the detection
of a gamma-ray line at an energy equal to the \WIMP\ mass. The line is
produced by the two-body annihilation of a pair of \WIMPs\ into a pair
of photons, which can occur in the galactic halo, in
nearby external galaxies, and in the large scale structure at high
redshift. A similar annihilation into a Z boson and a photon produces
a gamma-ray line which for heavy \WIMPs\ is close in energy to the
$\gamma\gamma$ line. For neutralinos, the highest cross section for
annihilation into $\gamma\gamma$ occurs in the higgsino region, due to
the similar masses of (heavy) higgsino-like neutralinos and charginos,
and the possibility of forming short-lived bound states among
them. Searches for gamma-ray lines from neutralino annihilation are
thus mainly sensitive to the higgsino region, as shown in
Figure~\ref{fig:indirect}(c). Thus they probe a region of
Supersymmetric parameter space different from that probed in direct
searches. In this sense, direct searches and searches for a gamma-ray
line are complementary.

Antideuterons ($\overline{\rm d}$) are a very rare product of
astrophysical processes, and detection of antideuterons could
constitute evidence for \WIMP\ annihilation in our Galactic
halo~\cite{antid1}. Current searches are still an order of magnitude
away from the theoretical predictions for
neutralinos~\cite{BESSlimit}, but proposals such as the General
Anti-Particle Spectrometer ({\small GAPS})~\cite{GAPS} should be able
to reach into the Supersymmetric parameter space. Antiparticle
detection is achieved by collecting low-energy $\overline{\rm d}$ and
observing the de-excitation ladder x-rays and pion burst following
annihilation of the exotic atom.  While considerable work remains to
clarify near-Earth $\overline{\rm d}$ backgrounds~\cite{antid2}, these
backgrounds are far more favorable than the case of anti-protons,
which are relatively more abundant in the cosmic rays compared with
the signal rate from annihilation of Supersymmetric \WIMPs.  The
expected antideuteron flux decreases rapidly with increasing
neutralino mass and is sensitive to relatively light neutralinos. In
addition, due to the specifics of antideuteron production,
$\overline{\rm d}$ searches are mainly sensitive to gaugino-like
neutralinos, more so than indirect searches through gamma-ray lines,
as shown by comparing Figures~\ref{fig:indirect}(c) and~(d). They also
provide better sensitivity than direct searches in some of the
gaugino-higgsino parameter space.
 
There are other indirect searches for \WIMP\ annihilation products
beyond those illustrated above, such as searches for antiprotons,
positrons, and continuum gamma-rays from pion decay. Detection of
\WIMP\ signals has been claimed using these probes. For example, an
excess in the cosmic ray positron flux observed in the {\small HEAT}
detector~\cite{heat} has been attributed to \WIMP\ annihilation in the
halo~\cite{Baltz:2002,Kane:2002,Cumberbatch:2007}; an excess of
gamma-ray flux in the {\small EGRET} data has been construed as
emission from an unusually-shaped \WIMP\ halo~\cite{deboer}; the
detection of a TeV gamma-ray source at the Galactic Center by the
{\small HESS} collaboration~\cite{hess} (after tentative detections by
{\small VERITAS}~\cite{veritasgc} and {\small
CANGAROO}~\cite{cangaroogc}) has been interpreted as due to heavy
\WIMPs~\cite{horns,bergstrom05,profumo,hall}. Unfortunately, these
kinds of searches share a problem of interpretation when ascertaining
the origin of a claimed signal: the energy spectrum from \WIMP\
annihilation does not show unequivocal signatures of its origin, and
anomalies can be difficult to isolate from modifications to other
astrophysical inputs, such as cosmic ray fluxes. Whether these
modifications are reasonable is largely still a matter of taste,
although further data (for example from {\small GLAST}, {\small HESS},
{\small VERITAS}, etc.) may be able to shed some light on the issue.

In summary, indirect searches provide better sensitivity than direct
searches in some regions of parameter space, and comparable
sensitivity in others.  While there may be the challenge of systematic
errors from assessing background of non-\WIMP\ astrophysical sources
for some modes, indirect searches could play an equally important role
as direct searches in establishing the presence of particle dark
matter. On the one hand, laboratory approach offers more control over
systematic effects, but it is also possible that the actual physical
parameters of \WIMPs\ may lead to stronger evidence through the
observation of annihilation products, e.g., in the case of a 
directional mono-energetic feature.

\section{Dark Matter Candidates and New Physics in the Laboratory}
\label{candidates}

A longstanding goal of the particle-physics community has been to find
evidence of Supersymmetry. As noted in Section~\ref{motivation}, this
extension to the Standard Model promises to stabilize the Higgs mass
and unify the coupling constants at the {\small GUT} scale, and is
required by string theory.  The currently best-motivated \WIMP\
candidate is the Lightest Super Partner ({\small LSP}), which would be
stable in \SUSY\ models with R-parity conservation.  \SUSY\ particles,
in addition to the Higgs boson itself, are the prime quarry for
discovery at the \LHC\ and precision studies at the \ILC. A detection
of new particles at the \LHC\ beyond the Higgs would indicate new
physics at the electroweak scale, possibly related to dark matter.
While laboratory results cannot independently establish a solution to
the dark matter problem, the production of a compelling dark matter
candidate is of critical importance.  To identify such a particle as
the dark matter, a consistent astrophysical observation would be
needed to demonstrate that the same particle is found in the cosmos
and that it is stable compared with the age of the Universe.  The
precision in particle properties necessary to determine the relic
density with precision comparable to astrophysical and cosmological
measurements would likely require the \ILC.

Supersymmetry has over a hundred free parameters, and so the study of
its phenomenology spans a broad continuum of approaches. At one end of
the spectrum, constraints may be imposed that favor a particular
theoretical idea, such as minimal Supergravity ({\small mSUGRA}) or
Split Supersymmetry, in order to gain predictive power. In some cases,
fully-specified models are defined that are representative of a
characteristic region of parameter space, which serve as useful
benchmarks for scoping experiments or comparing the sensitivity of
different approaches. At the other end of the spectrum is the
exploration of the available parameter space imposing only known
empirical constraints and minimum theoretical bias. This approach
serves to define the range of experiments necessary to fully explore
the \SUSY/\WIMP\ hypothesis. Interestingly, a wide variety of models,
even including the maximally constrained ones, can lead to interesting
dark matter candidates, as illustrated earlier in
Figure~\ref{fig:models_limits}.

Examples of unconstrained models include those calculated by Baltz and
Gondolo~\cite{Baltz01}, and by Kim \etal.~\cite{Kim02}. These
modelers only use minimal unification assumptions, empirical
constraints from accelerator experiments and the \WIMP\
relic-density requirement.  Bottino \etal.~\cite{Bottino03} may
represent the most extreme form of such a model, in which unification
assumptions are relaxed and result in candidates with masses down to a
few GeV/c$^2$.\symbolfootnote[3]{As shown in
Figure~\ref{fig:models_limits}, \CDMSII\ has ruled out the low-mass
branch of this model down to 10\,GeV/c$^2$. The region from
5--10\,GeV/c$^2$ could accommodate the {\small DAMA} result for sodium
recoils, however the halo velocity distribution would need to be
altered to enhance the modulation amplitude since the predicted cross
section is too low by an order of magnitude under standard assumptions
(see also Figure~4 of Ref.~\cite{cdms119}).}

Examples of theoretically-constrained models, which serve as a useful
benchmark for \LHC\ searches, also show the complementarity of \WIMP\
and accelerator experiments.  For example, consider the Constrained
Minimal Supersymmetric model~\cite{Battaglia03, Ellis03} or the
{\small mSUGRA} model~\cite{Baltz04}, which impose unification of the
\SUSY\ parameters.  Some of the regions within the parameter space,
such as the focus-point region, which tends to large scalar masses,
are inaccessible at the \LHC\ but could be easily seen in the next
generation of dark matter experiments in the 10--100\,kg scale at a
\WIMP-nucleon cross section of 10$^{-44}$~cm$^2$. On the other hand,
in the ``co-annihilation'' region, the \LHC\ has a good chance of
seeing the 5 Higgs particles, as well as gaugino decays into
identifiable squarks and neutralinos. However, the \WIMP-nucleon cross
section could be anywhere in the range 10$^{-46}$--10$^{-44}$~cm$^2$,
which requires a ton-scale detector for a full exploration.

More highly-specified models, such as the recently-proposed Split
Supersymmetry model of Arkani-Hamed and Dimopolous~\cite{Arkani04},
lead to the \WIMP-nucleon cross-section range of
10$^{-45}$--10$^{-44}$~cm$^2$ ~\cite{Giudice04, Pierce04}. This model,
inspired by the large number of vacua in string theory and the
difficulty of accounting for the apparent small value of the
cosmological constant, further decreases the effective number of
parameters by allowing the scalars to be very massive.  An anthropic
argument is used to pay the price of fine-tuning the Higgs mass to
account for the dark matter---otherwise galaxies (and observers)
would not have formed.  In addition to the narrow range of predicted
elastic cross sections, there are also testable consequences in
accelerator experiments.  The \LHC\ would detect only one Higgs, the
neutralino (if it is light enough), and a long-lived
gluino~\cite{hewett}. Meanwhile, direct searches would check that
indeed the neutralino constitutes the dark matter, providing the
justification of the Split-Supersymmetry scheme, and fixing the
remaining parameters of the model. According to~\cite{Arkani04}, the
long life of the gluino would then be a signature for fine tuning.

In recent years, there has been much excitement about compact
extra dimensions of order 1\,TeV$^{-1}$ as an alternative explanation
of the hierarchy problem. In a broad class of such models, the
lightest Kaluza Klein excitation is stable and could be the dark
matter~\cite{KKdarkmatter}, with a mass of order
1\,TeV/c$^2$. Interestingly, the elastic scattering cross section
covers the range from 10$^{-46}$ to 10$^{-42}$\,cm$^2$ per nucleon.

A new benchmark study is well underway to look at connections between
cosmology and the \ILC. Because the actual parameters
of \SUSY\ are numerous and their values unknown, these benchmarks help
to define measurement scenarios for specific experimental signatures
at the \LHC, \ILC, and direct searches, as well as how a set of actual
measurements could, in turn, constrain the theoretical parameter
space.

\begin{figure}[!h]
\begin{center}
\includegraphics[width=3.0in]{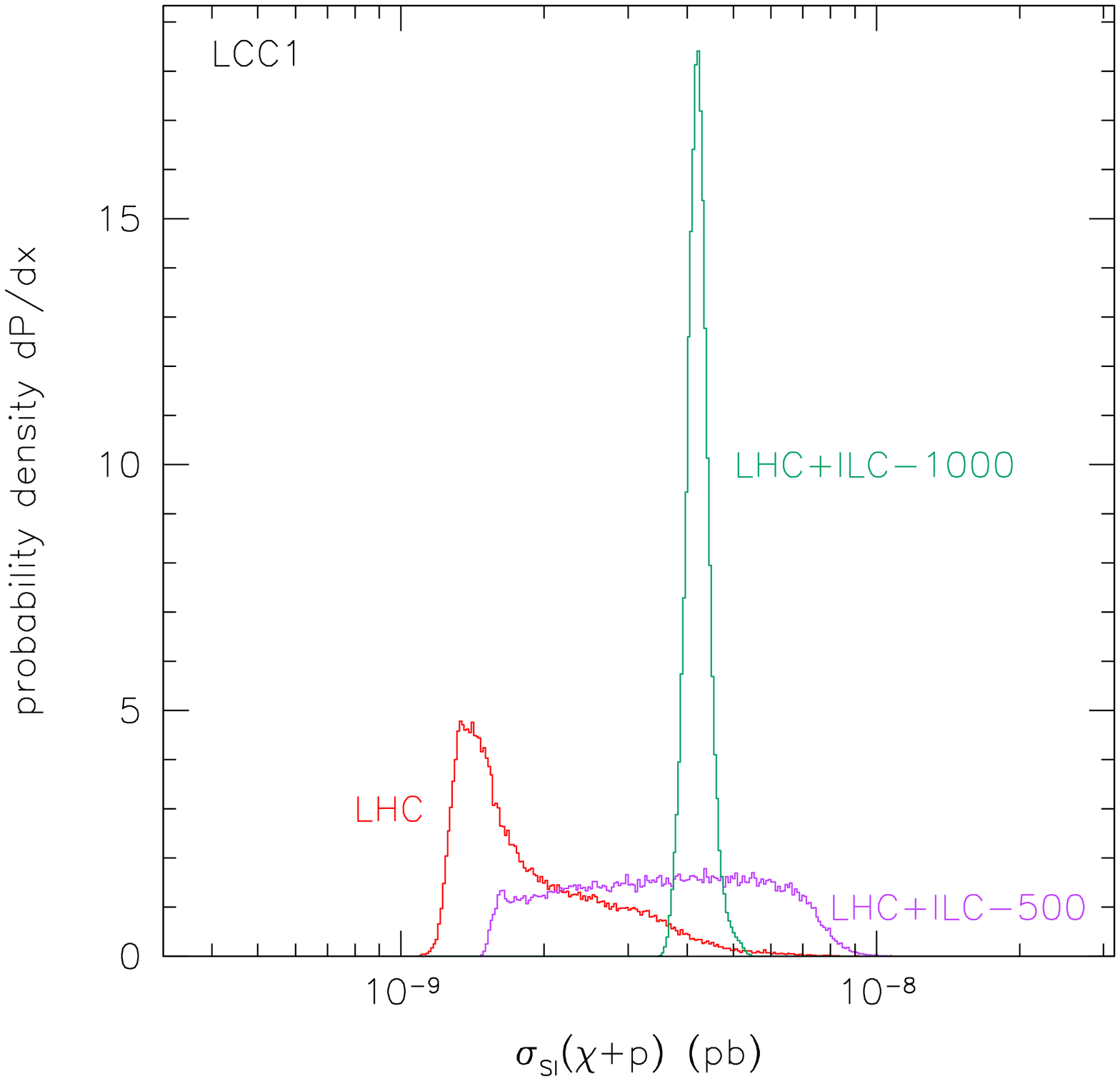}
\includegraphics[width=3.0in]{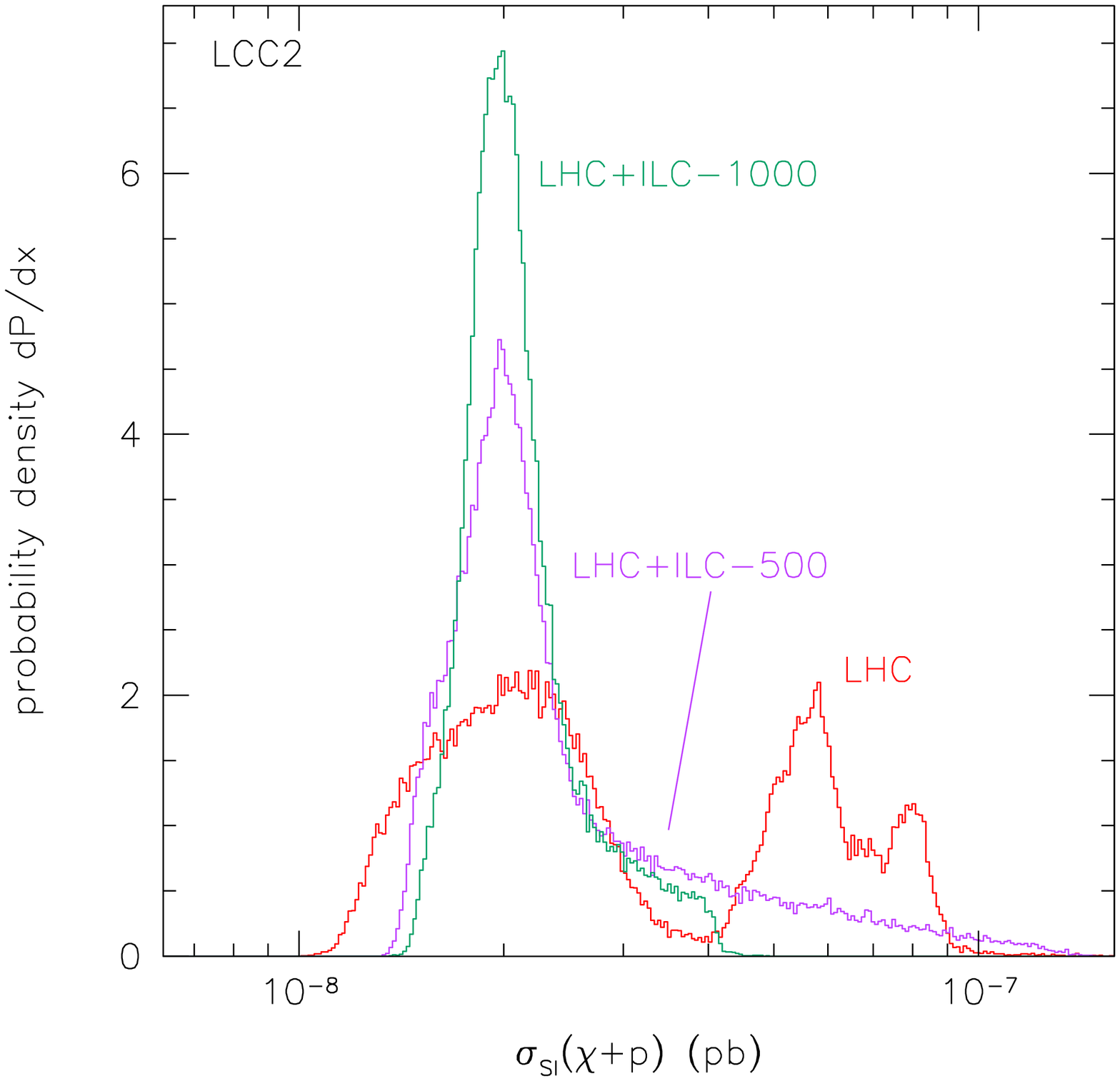}
\end{center}
\begin{center}
\includegraphics[width=3.0in]{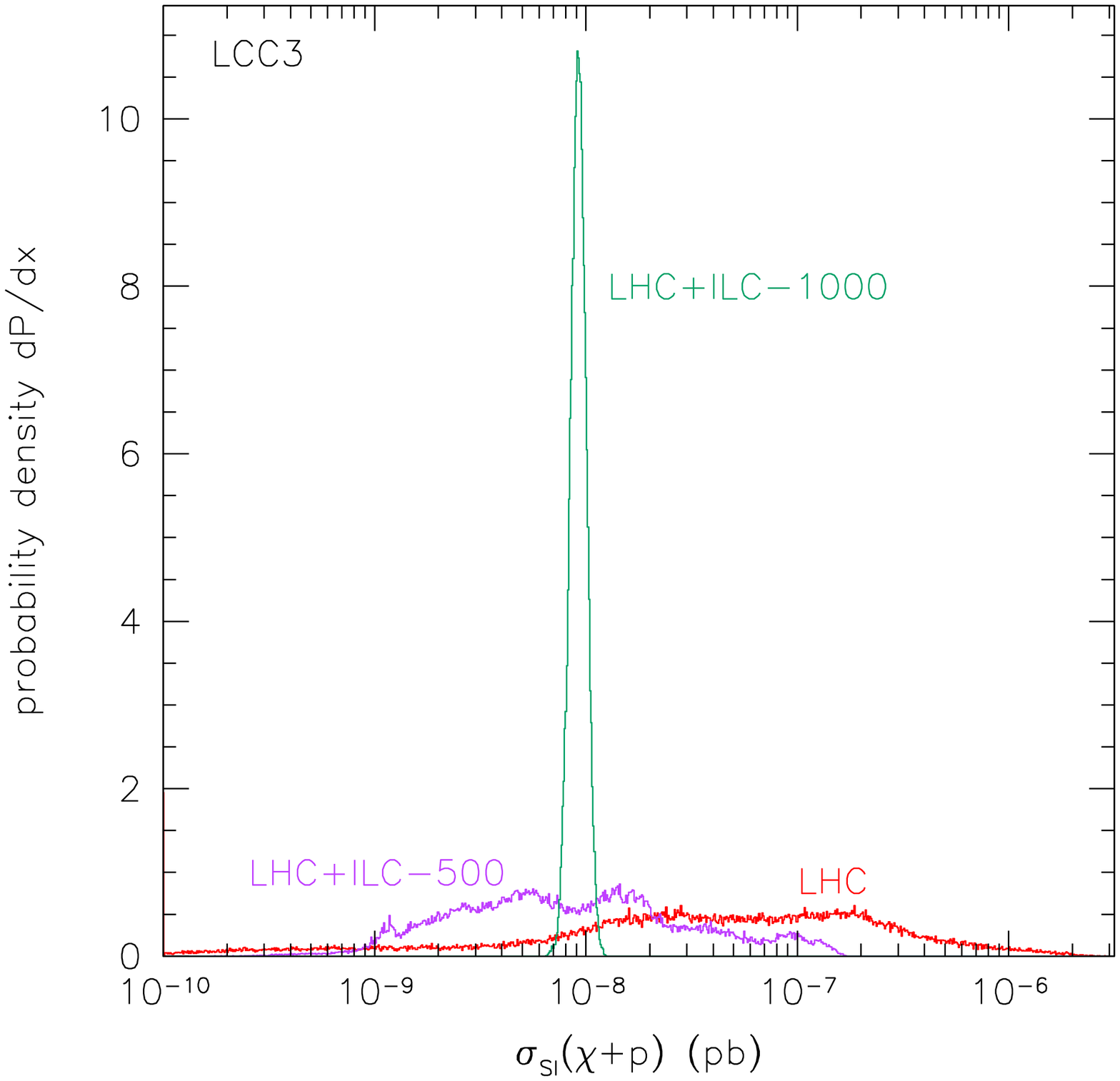}
\includegraphics[width=3.0in]{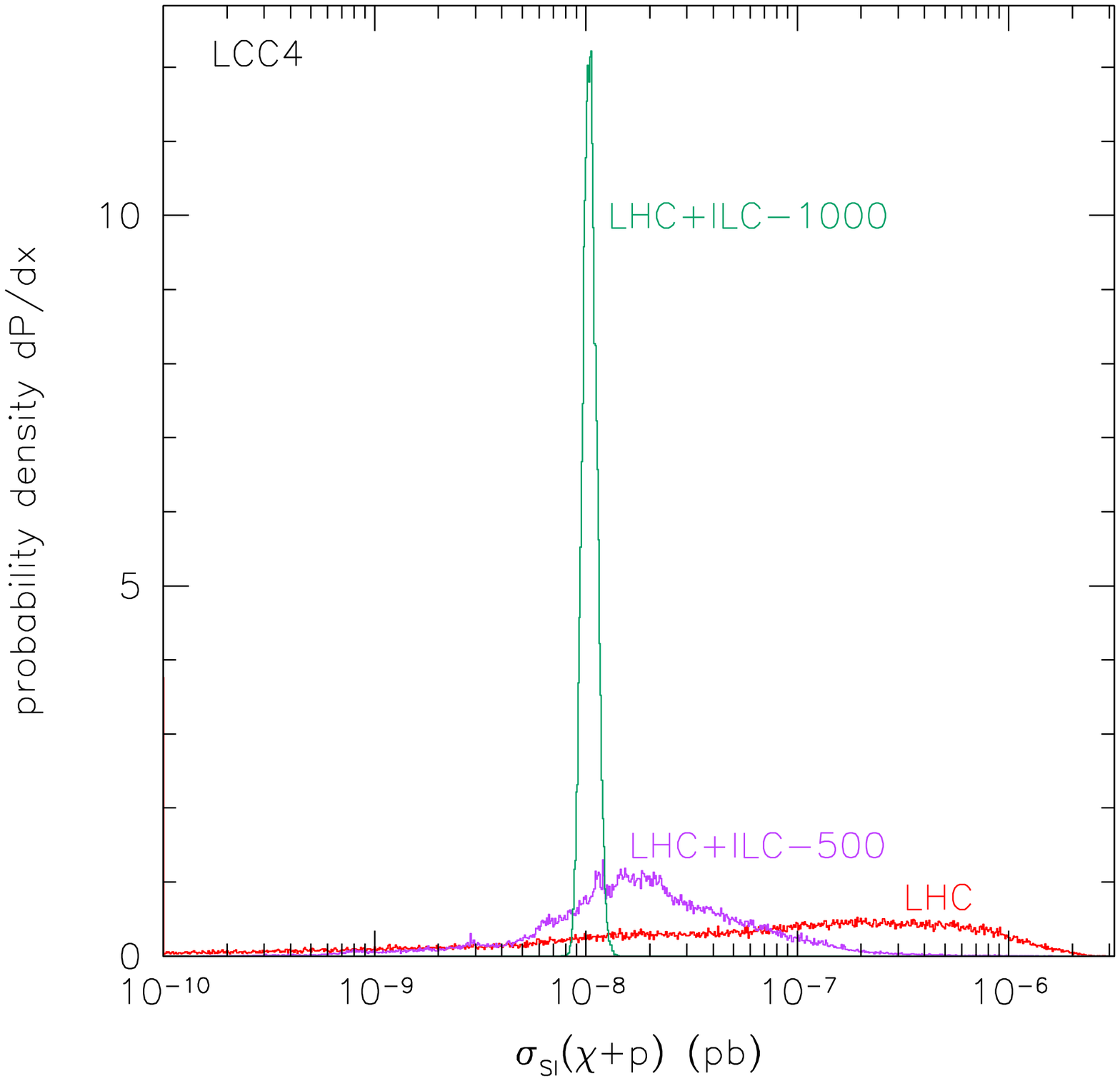}
\end{center}
\caption{\small Allowed range of the WIMP-proton elastic cross section
for LHC and ILC measurements of the benchmark models LCC1 (top left),
LCC2 (top right), LCC3 (bottom left), and LCC4 (bottom right). The
three curves in each graph correspond to measurements of these models
at the LHC, ILC 500-on-500\,GeV and ILC 1000-on-1000\,GeV. See 
text for details~\cite{lcc}.}
\label{fig:lcc_sigsi}
\end{figure}

Detailed calculations are being performed for each benchmark model by
computing a set of measured physical quantities, including
experimental uncertainties, and then examining the full range of
\SUSY\ parameter space that would be allowed by these
measurements. From this range of allowed parameters, specific derived
quantities of interest can be constrained. For example, to inform dark
matter searches the mass of the {\small LSP} and its elastic cross
section on nucleons could be constrained.  In particular, a set of
benchmark points accessible to both the \LHC\ and \ILC, and
interesting for cosmology, has been proposed.  Labeled {\small
LCC1--4}, each has a relic density broadly similar to that required to
explain the dark matter. While each is accessible to direct dark
matter searches, the aim of this study is to the examine and compare
what can be learned about the benchmark models from the different
approaches.

\begin{figure}[!h]
\begin{center}
\includegraphics[width=3.0in]{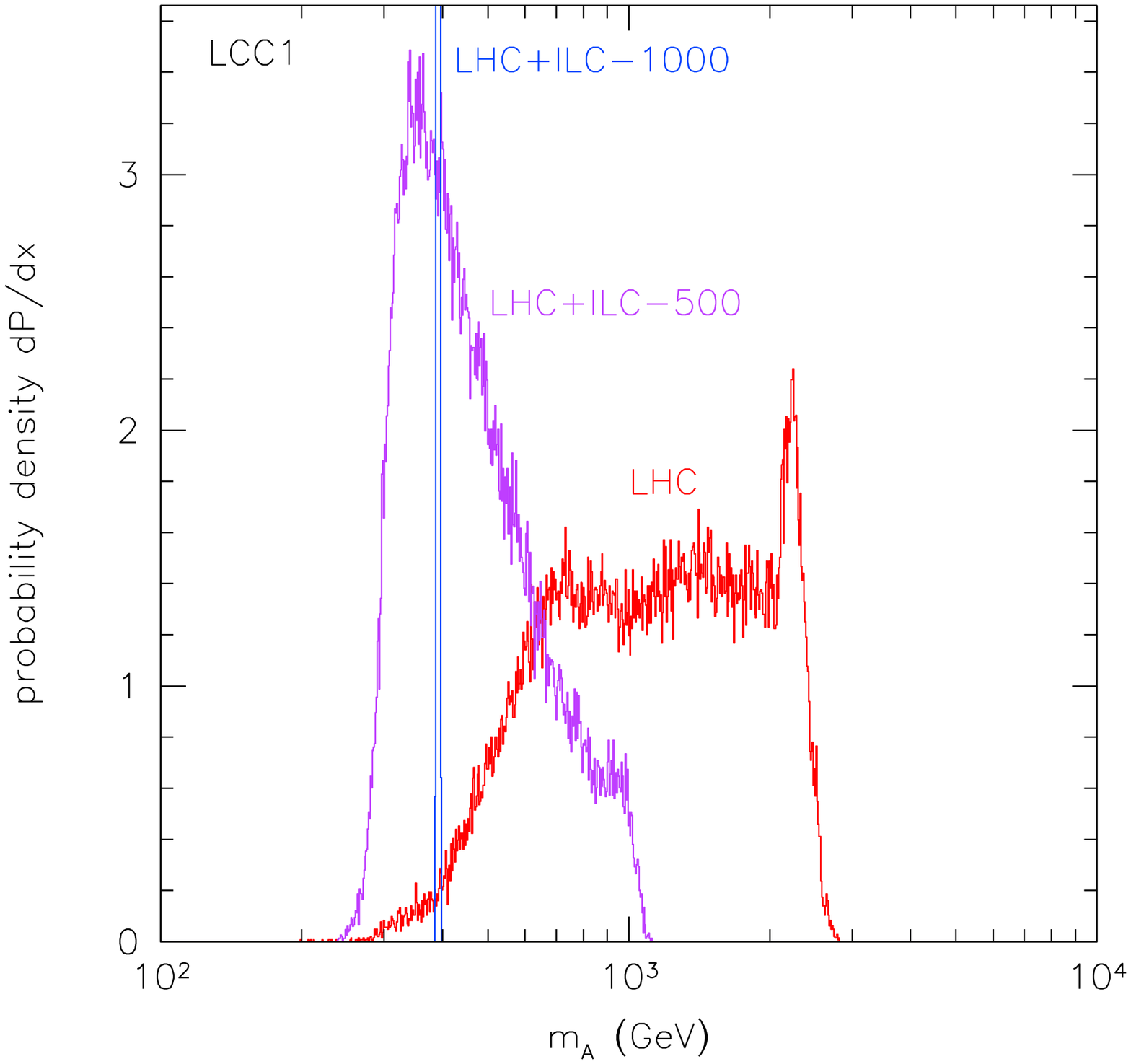}
\includegraphics[width=3.0in]{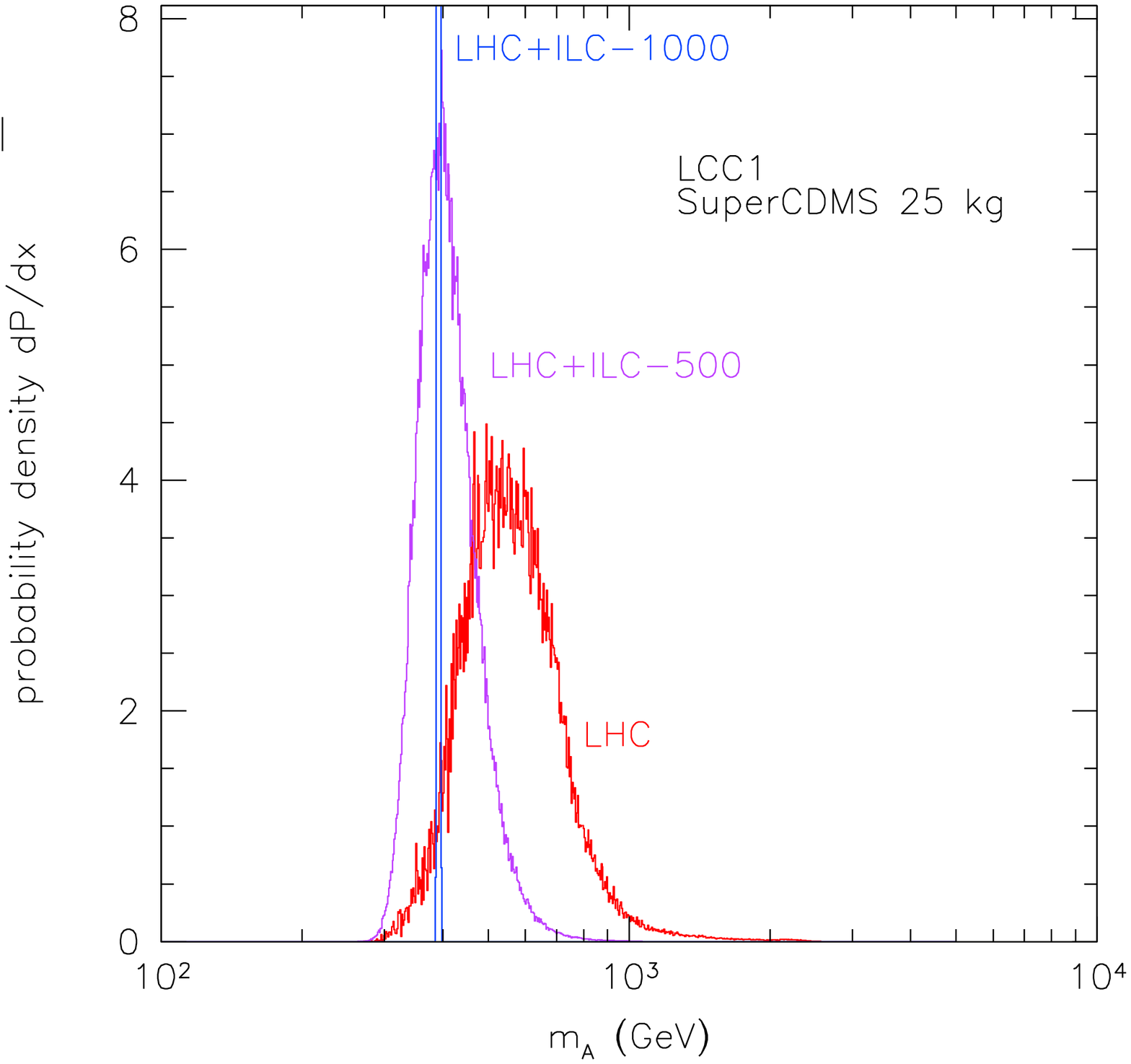}
\end{center}
\caption{\small Allowed range of the mass of the heavy CP-odd Higgs in
  the LCC1 model is shown for accelerator-based measurements (left)
  and combined accelerator and direct detection (right) with a
  per-nucleon cross-section sensitivity of 10$^{-45}$cm$^2$, such as
  the SuperCDMS 25\,kg experiment~\cite{lcc}.}
\label{fig:lcc1_ma}
\end{figure}

For benchmark point {\small LCC1} (in the bulk region where
coannihilations are not necessary), Baltz, Battaglia, Peskin and
Wizansky~\cite{lcc} determined that if that model is the actual model
in nature, and the \LHC\ measured accessible masses (three of the
neutralinos, sleptons except the heavy stau, squarks except for the
stops, the gluino, and the light Higgs), then the range of models
consistent with those measurements would allow the neutralino-proton
cross section to vary over an order of magnitude, as illustrated in
Figure~\ref{fig:lcc_sigsi}.  In fact, the central value is off by more
than a factor of two.  Also shown is the allowed range for what would
be measured for \ILC\ center-of-mass energies of both 500\,GeV and
1\,TeV.  The 500-GeV \ILC\ offers some improvement, but the TeV \ILC\
is required for a solid measurement.  This is because the important
quantity for direct detection in this model is the mass of the {\em
heavy} Higgs boson, which at 395\,GeV is accessible only to the TeV
\ILC.  The non-detection at \LHC\ or \ILC-500 gives the wide range of
possibilities, limited only by the artificial upper limit (taken to be
5\,TeV) put on this mass.  Figure~\ref{fig:lcc1_ma} illustrates
specifically what a direct-detection constraint can do for particle
physics.  In particular, the mass of the heavy Higgs bosons is usually
unconstrained until a TeV \ILC, unless $\tan\beta$ is large.  For
point {\small LCC1}, the mass distribution for the {\small CP}-odd
Higgs improves significantly for \LHC\ and even \ILC-500 measurements.
This shows that direct detection would strongly constrain this part of
the Higgs sector before a TeV \ILC\ was available, and again relates
dark matter and electroweak symmetry breaking!

For the benchmark model {\small LCC2}, (in the focus-point region
where neutralinos can annihilate to gauge bosons, and coannihilations
with charginos can be important), colliders can do a better job of
constraining the direct-detection cross section.  This model has a
``mixed'' neutralino giving larger elastic cross section, and it
depends on the light Higgs, easily constrained by the \LHC.  The
usually dominant heavy Higgs is more than 3\,TeV/c$^2$ in mass, and
thus irrelevant to direct detection.  At the \LHC, there is a
complication in that the data can not distinguish between the discrete
possibilities that the lightest neutralino is Bino, Wino, or Higgsino.
Assuming a standard halo, a few dozen events in a direct detection
experiment would serve to completely eliminate the incorrect
solutions.  The \ILC\ would be required to eliminate the incorrect
solutions without astrophysical input.

For the benchmark model {\small LCC3} (with stau coannihilations
important), again colliders give little information before a TeV \ILC,
though in this case, the heavy Higgs bosons are visible even to the
\LHC.  Here, the important uncertainty is the gaugino-higgsino mixing
angle that can only be readily constrained by a TeV \ILC, which can
observe most of the neutralinos and charginos.  Furthermore, the TeV
\ILC\ could also measure the width of the pseudoscalar Higgs, which
constrains $\tan\beta$ and thus the neutralino mass matrix.

The point {\small LCC4} is an interesting case.  This model has a
large resonant annihilation cross section (through the pseudoscalar
Higgs, which is closely tied in mass to the heavy Higgs).  Again,
these heavy Higgs particles are visible to the \LHC, but the
neutralino mixing angles can only be measured at a TeV \ILC.  As with
{\small LCC3}, measuring the width of the pseudoscalar Higgs at a TeV \ILC\
gives $\tan\beta$, but for this point, the measurement is crucial
for determining the relic density.

So we see that for all of these benchmarks, which have been chosen to
illustrate the power that colliders would have, direct detection would
make significant contributions to fundamental physics, in advance of a
TeV linear collider.  While such \LHC/\ILC\ measurements could help to
constrain the astrophysical searches, a \WIMP\ detection in the Galaxy
leading to a determination of the elastic cross section would place a
significant constraint on \SUSY, otherwise absent from the accelerator
experiments. Since a \WIMP\ detection could constrain the elastic
cross section to better than a factor of a few, the neutralino mass
matrix becomes better defined because the mixing angles are
constrained by including the elastic cross section measurement. The
uncertainties are dominated by halo uncertainties and the strange
content of the nucleon for {\small SI} interactions. {\small SD}
interactions suffer, in addition, from uncertainty in the spin
structure of the nucleus. From the direct-search viewpoint, it is
important and attainable that the long-term program address the
sensitivity requirements for both {\small SI} and {\small SD}
interactions. Sorting out this physics will require a variety of
target nuclei with different masses and spins, and the present \RnD\
program is headed in that direction (see Section~\ref{experiments} for
further details).

As for the \LHC/\ILC\ informing the solution to the dark matter
problem in the absence of halo searches, the best they can do is refine
the candidates but can say nothing relevant about the stability on the
Hubble time scale.  In fact, if \LHC/\ILC\ measurements result in a
robust prediction that astrophysical searches should have seen
something {\it but didn't} then we may arrive at the very interesting
conclusion that the particles are unstable and that R-parity is not
conserved.  This result could not be obtained from accelerator
experiments, alone. Thus the question of particle stability, the
province of dark matter searches, is of fundamental importance to both
cosmology and particle physics. Moreover, the {\small WMAP}
measurement of the relic density is broadly recognized as a powerful,
though hypothetical constraint on \SUSY\ parameters---that constraint
is only realized once it is demonstrated that the neutralino is stable
on the Hubble time scale and the particle mass is consistent between
the halo and accelerator measurements. It remains that an
astrophysical detection of \WIMPs\ is the essential link between
early-universe physics and fundamental particle physics in this sector.

% dsa - new paragraph - reviewer B.6

\begin{center}

* \qquad * \qquad * 

\end{center}

In summarizing sections \ref{detection}, \ref{indirect} and
\ref{candidates}, we see that detecting \WIMP\ scatters in terrestrial
detectors, observing astrophysical sources of \WIMP\ annihilation
products, and producing \WIMPs\ and related particles in the
laboratory can each play an important role in resolving the dark
matter problem and elucidating the new fundamental physics that could
be behind it. Direct detection remains the clearest and most promising
means for establishing that \WIMPs\ exist in the galaxy. If \WIMPs\
are comparable in mass that that of detector target nucleus, then the
\WIMP\ mass can be determined well enough to cross check against
accelerator-produced candidates, and even if \WIMPs\ are heavier, the
target mass serves as a strong lower-bound. With detection in multiple
targets, the nature of the \WIMP\ coupling to nuclei can be
constrained. A measurement of the elastic cross section also has the
potential to provide information about the underlying particle model,
e.g., the neutralino mixing angles. Indirect searches have the
challenge of discriminating against astrophysical backgrounds in most
channels, but the possibility of a ``smoking gun'' signal remains, for
example, from a mono-energetic and/or directional signal. Together,
either the direct or indirect techniques are needed to establish
particle stability comparable to the age of the universe---and could
also inform us on the nature of R-parity. While indirect means will
inform the annihilation cross section, it remains with the
accelerator-based experiments to work out the full phenomenology of
the \WIMP\ sector and to permit a sufficiently robust calculation of
the relic density. Naturally, we stand to learn the most if \WIMPs\
are observed in all three regimes, in terms of both overall confidence
and consistency, and through reducing uncertainties of both the
particle properties and the astrophysical parameters (such as halo
density and velocity distribution). If we are unfortunate in that the
observation of \WIMPs\ proves elusive, then it is equally important to
push on all fronts since there are substantial regions of parameter
space that are searched uniquely by only one method. Unless and until
new information arrives, say, a detection of dark matter axions,
\WIMPs\ remain an excellent hunting ground.

\section{Synergies with Other Sub-Fields}
\label{synergies}

% dsa & ea: re reviewer comment A.3:
%  - modified end of 1st paragraph
%  - modified xenon paragraph in dbl-beta subsection

Dark matter searches push the frontier on low-background low-threshold
high-energy-resolution particle detection. These capabilities are
similar to some of the requirements for double-beta decay experiments,
and not surprisingly there has historically been a strong connection
between these fields.  Despite the great differences in the
characteristics of the rare decays sought, background concerns for the
two experiments are very similar and result in many of the same
shielding and cleanliness requirements.  However, the demands on both
types of experiments have become so strict that it is extremely
difficult for a single experiment to be competitive in both
sub-fields.  For double-beta decay, discrimination of electron recoils
from nuclear recoils is not important, but excellent energy resolution
at relatively high energies ($\sim$MeV) and rejection of multiple-site
events are critical.  While it is very unlikely that a single
experiment can pursue both dark matter and double-beta decay without
compromising the sensitivity for either process, technological
advances driven by a dark matter experiment could benefit a
double-beta decay experiment based on the same target material. This
synergy will become especially relevant as experiments move to larger
scale and cost. We discuss in more detail below the requirements and
opportunities for a multi-purpose experiment, in some specific target
material.

Following that, we examine the prospects for the detection of
supernova neutrinos in dark matter experiments. As the experiments
reach the ton scale, supernova neutrinos become interesting quarry
because, like \WIMPs, the scattering rate from nuclei is enhanced by
coherence effects.

\subsection{Double-Beta Decay}

Dark matter experiments using Ge detectors could also, if enriched in
$^{76}$Ge, be useful for searching for neutrinoless double-beta decay.
Cryogenic detectors based on {\small NTD} thermistors or
athermal-phonon sensors could result in better energy resolution than
the $\sim4$\,keV of Ge ionization detectors alone, resulting in
improved background rejection, especially of the important
$\beta\beta2\nu$ background.  The intrinsic energy resolution of
athermal-phonon sensors can be better than 200\,eV ({\small FWHM}),
but better correction for the position dependence of the collected
energy will be necessary in order to achieve resolution better than
5\,keV at energies of 2\,MeV.  Furthermore, athermal-phonon sensors
may provide sufficient position reconstruction to allow improved
rejection of multiple-site events from Compton backgrounds and
inelastic neutron interactions.  At a scale of 500\,kg to a ton, a
single experiment combining the technologies of current dark matter
and double-beta decay experiments may be able to achieve both goals,
with sensitivity to effective Majorana neutrino mass of 20\,meV.

Xenon is another element that is being used for both dark matter
detection and double-beta decay, where the latter exploits the isotope
$^{136}$Xe. However, the operating principle and the readout of a
liquid xenon TPC for dark matter and for double-beta decay are quite
different and, at present, exclude the possibility of a single
experiment for both processes. The poor energy resolution of a
dual-phase xenon TPC as currently implemented for dark matter
detection, where the emphasis is rather very low energy threshold and
recoil discrimination, is incompatible with a double-beta decay search
which has to emphasize energy resolution to distinguish the
two-neutrino process at the continuum endpoint from the mono-energetic
zero-neutrino process. In the long term, the goal of building a
directional gaseous {\small TPC} for dark matter may offer a unique
opportunity for double-beta decay detection if xenon enriched in
$^{136}$Xe is a component of the chamber. Preliminary studies are
underway to assess the background-rejection capability and the physics
content that would result from a measure of the opening angle of the
two mono-energetic electrons emitted by the zero-neutrino decay
process.

Also notable is the element tellurium, which has a double-beta decay
nucleus in $^{130}$Te. This element is being pursued primarily for
double-beta-decay studies by instrumenting TeO$_2$ with {\small
NTD} thermistors in a cryogenic array known as Cuore. Owing to the
very low radioactive background and an energy threshold of 10\,keV,
this experiment will also have sensitivity to dark matter signals.

Thus, while the emphasis of the dark matter and double-beta decay
studies are different in several respects, these differences tend not
to be mutually exclusive, especially in the case of cryogenic crystals
with very fine spectroscopy. For these, as well as for other materials
such as liquid xenon, many technology developments, such as
radio-purification, low-background assays, ways to improve energy
resolution, ways to fabricate large arrays, etc., are common to both
efforts, and there has been much positive exchange between the fields.
It is possible that a particular experiment could be well-suited to
performing frontier studies on both counts.  However, as the needs for
each become more restrictive and the experiments become more
complicated, it is critical to optimize each experiment's design to
ensure obtaining its primary measurement of interest.

\subsection{Supernova Neutrino Detection}

The core-collapse of a massive star, and the subsequent Type-{\small
II} supernova event, releases a burst of $\sim 10^{58}$ neutrinos of
all flavors at of order 10\,MeV energy. Neutrinos with these
energies have an enhanced rate for scattering from nuclei due to
coherent interactions, producing nuclear recoils of order
10\,keV---above dark matter detection thresholds. As detectors
reach the ton scale and beyond within the dark matter program at
{\small DUSEL}, the detection of these neutrinos, in particular from a
supernova in our galaxy, becomes possible. Here we summarize the rates
and prospects for detecting supernova neutrinos.

The condition for coherent nuclear scattering is $\Delta R_A < 1$,
where $\Delta$ is the three-momentum transferred to the nucleus, and
$R_A$ is the radius of the nucleus. For all nuclei of interest up to
germanium, this condition is satisfied for neutrinos up to energies of
about 50\,MeV. Supernova neutrinos emerge with nearly thermal spectra,
with mean energies of 13, 15, and 24\,MeV for electron-neutrino,
electron anti-neutrino, and muon/tau neutrino flavors,
respectively. The average energy of the recoiling nucleus is $2/(3A)
(E_{\nu}/{\rm MeV})^2$, where $A$ is the mass number of the nucleus,
making the coherent-scattering channel sensitive to the high-energy
tail of the thermal spectrum. Heavier target nuclei have an $A^2$
enhancement of the total cross section, but the energy of the recoils
is smaller than for lighter nuclei.

As an example of expected yields from a Galactic supernova at 10\,kpc,
there will be 18~events in 1~ton of germanium from muon and tau
neutrinos, with 10~events above 10\,keV recoil
energy~\cite{horowitz03}. Electron neutrinos contribute a total yield
of 4~events. For lighter nuclei, in particular for the case of a neon
target, there will be a total yield of about 4~events, including all
neutrino flavors. For lighter nuclei there is a larger probability
that all of the recoils will be above 10\,keV. In some cases, the
intrinsic threshold for neutrino scatters may be substantially lower
than for dark matter. Ordinarily, the threshold for dark matter
detectors that rely on a dual measurement for gamma and beta
rejection is determined by optimizing
efficiency versus background rejection. Since the neutrino events
occur in a burst coincident with an externally-defined time window,
background rejection is not needed and so the intrinsic lower
energy-threshold of the device can be used.  In the case of cryogenic
detectors, their intrinsic 1\,keV threshold allows the full
recoil-energy spectrum to be detected. With burst-detection thresholds of
a few keV, nearly the full spectrum would be accessible to
scintillation detection, e.g., in noble liquids.

The yield from a Galactic burst is clearly small compared with the
expected yields in much larger neutrino detectors, such as
Super-Kamiokande, {\small KamLAND}, and {\small
SNO}~\cite{beacom01}. However, even a few events in a 1-ton
dark matter detector could be the first detection of neutrino-nucleus
coherent scattering. This is very promising if we consider that larger
neutrino detectors will accurately determine the average energies of
the different flavors. From just a few recoil events, we can make the
first measurement of the coherent cross section, and look for
deviations from the expected standard model result.

In addition to the yield from a Galactic supernova, potential 100-ton
detectors may have sensitivity to the Diffuse Supernova Neutrino
Background ({\small DSNB}). Modern predictions indicate that the
{\small DSNB} flux is approximately 6\,cm$^{-2}$s$^{-1}$, including
contributions from all neutrino flavors~\cite{strigari05}. In a
100-ton germanium detector, this corresponds to about 2~events per
year. In addition to this low event rate, a hindrance for {\small
DSNB} detection is that the mean {\small DSNB} neutrino energy is
lower relative to the spectrum of neutrinos from a galactic supernova
burst, due to cosmological redshifting of neutrino energies. However,
this flux may be detectable if the \WIMP-nucleus cross section is at
the lower end of the theoretical models.

\section{Direct Detection Experiments: Status and Future Prospects}
\label{experiments}

The initial goal of direct dark matter searches is to detect 
a signal that can be confidently attributed to elastic scatters of galactic
halo \WIMPs\ from the nuclei in an earthbound detector. It would be essential 
to follow up an initial detection by confirming the signal and its
galactic-halo origin in several ways:

\begin{itemize}

\item Detect again using a different target nucleus, to confirm that
the cross section scales with nuclear mass or spin as expected for
scattering of \WIMPs\ rather than neutrons or other background.

\item Constrain the \WIMP\ mass to the greatest extent possible by
measuring the recoil spectrum in a well-matched target nucleus to test
whether it is compatible with candidates resulting from
accelerator-based experiments.

\item Confirm the galactic origin of the signal by detecting the
expected annual modulation effect, which would require increased
statistics, and the diurnal modulation effect, which would require a
directionally-sensitive detector.

\item Ultimately, refine our understanding of the galactic halo by
accumulating larger statistics in both recoil energy and directional
measurements to exploit information in spectrum and modulation
signals.

\end{itemize}

With these broad goals in mind, we review in this section the present
status of the direct detection experiments and of the \RnD\ programs
underway for future experiments.

The following subsections, listed alphabetically, detail the status
and prospects of specific experiments, with an emphasis on those that
are included in the infrastructure matrix assembled in the context of
the ``Solicitation 1'' process. (For a more comprehensive review of
dark matter searches, see, for example, the recent review by
Gaitskell~\cite{gaitskell04}.)

\subsection{CLEAN and DEAP} 

% http://mckinseygroup.physics.yale.edu/Publications/SNOLAB%20workshop.pdf    

The {\small CLEAN} experiment, or Cryogenic Low Energy Astrophysics
with Neon, is a large cryogenic scintillation detector with 130\,tons
of liquid neon as the active material.  The central dewar is to be
approximately 6\,m on a side, located within a 10-m-diameter by
12-m-tall water shielding tank.  The combination of water shielding,
self-shielding by a layer of neon outside the central fiducial volume,
and the radiopurity of neon, results in very low backgrounds.  The
scintillation light output has a time structure that allows
discrimination between electron recoils and nuclear recoils. The
design sensitivity is 10$^{-46}$\,cm$^2$ per nucleon for WIMP-nucleus
spin-independent scattering.  Other physics goals include a 1\%
measurement of $p$-$p$ solar neutrinos, supernova neutrino detection,
and 10$^{-11} \mu_b$ neutrino magnetic moment detection.  The
collaborators describe {\small CLEAN} as an \RnD\ collaboration at
this time with emphasis on the development of smaller-scale prototypes
to realize the technical requirements of a large LNe detector. The
program has evolved to enable such smaller-scale prototypes to be
exploited in a dedicated \WIMP\ search using LAr in place of LNe based
upon recent developments to exploit LAr in a detector dubbed {\small
DEAP} (Dark matter Experiment with Argon and Pulse shape
discrimination).

\subsection{Chicagoland Observatory For Underground Particle Physics (COUPP)}

%http://collargroup.uchicago.edu/projects/wimp/#[1]

A heavy liquid bubble chamber (CF$_3$I) sensitive to nuclear recoil
events has been operated in a thermodynamic regime in which it is
insensitive to minimum-ionizing radiation and electron recoils from
gamma and beta backgrounds. It thus affords a high degree of
background discrimination. The critical challenge of long-term stable
operation of the chamber has been met, which requires a glass-walled
vessel to minimize surface-induced nucleation events.  The high
content of nonzero spin nuclei is noteworthy for this target
material. The present 2-kg room-temperature prototype, which is expected
to have a sensitivity of approximately of 10$^{-42}$ and 10$^{-40}$\,cm$^2$
per nucleon for spin-independent and spin-dependent \WIMP\ scattering,
respectively, would be scaled up in phases to 250\,kg, 1000\,kg, then
several tons.  This prototype is now running in a 300\,m.w.e.-depth
site at Fermilab.

\subsection{Time Projection Chambers (Directional-TPC group and DRIFT III)}

Two collaborations, Directional-{\small TPC} group and the {\small
DRIFT-III} (Directional Recoil Identification from Tracking)
collaboration, are proposing Time Projection Chambers ({\small TPC's})
with total mass in the hundreds of kilograms, to initiate the study of
\WIMP\ astronomy by tracking nuclear recoils in low-pressure gas at
room temperature.  The solar system orbits around the galactic center
with tangential speed similar to that expected for halo \WIMPs, and
presently directed toward the constellation Cygnus.  Combined with the
Earth's rotation on its axis, this will result in a sidereal-day
modulation of the {\it direction} of \WIMPs\ in an earthbound lab, as
the apparent source of the \WIMP\ wind rises and sets.  A large
low-pressure negative-ion {\small TPC} can measure the orientation of
\WIMP-recoil tracks, as was shown by the 0.2 kg {\small DRIFT-I} and
{\small DRIFT-II} prototypes in the Boulby Mine. While such detectors
also have very good discrimination properties, work is ongoing to
demonstrate this at sufficiently low thershold. These devices
presently rely on use of a toxic flammable chemical dopant (CS$_2$)
for the negative-ion drift. Although {\small DRIFT} underwent a
thorough safety review to operate in Boulby and has had no accidents
of any sort in roughly five years of operation, \RnD\ is underway to
identify alternative dopants.

\subsection{European Underground Rare Event Calorimeter Array (EURECA)}

%astro-ph/0504241

The {\small CRESST} and {\small EDELWEISS} collaborations have
developed extremely sensitive cryogenic solid detectors operating at
milliKelvin temperatures. Electron- versus nuclear-recoil
discrimination is achieved by combining heat detection with a
secondary signal that has a different response for the two recoil
types. The {\small CRESST} detectors augment the thermal signal with a
scintillation signal in CaWO$_4$ targets which is detected thermally
in a second adjacent device. Two 300-g detectors were successfully
operated in Gran Sasso and were limited by nuclear recoils consistent
with that expected from neutron-background (and the limited shielding
deployed in this early run). The {\small EDELWEISS} detectors augment
the thermal signal in germanium targets with an ionization signal. Their
2002 and 2003 data sets from an array of three 300-g detectors were a
then world-best upper limit. The sensitivity was limited by a
low-energy beta background that suffers ionization-signal loss near the
detector surface. Improved designs are being developed, some of which
include a highly-resistive metal-film readout that should offer
additional information for discriminating surface events from bulk
nuclear recoils.  The {\small
EURECA} program is the joint effort of these two collaborations to
plan a ton-scale recoil-discriminating cryogenic array to achieve sensitivity
in the 10$^{-46}$\,cm$^2$/nucleon range.

\subsection{Scintillation and Ionization in Gaseous Neon (SIGN)}

% http://www.physics.tamu.edu/research/list-high_energy.html
% email 8 jan 06 from jim white

The {\small SIGN} (Scintillation and Ionization in Gaseous Neon)
concept is a modular room-temperature nuclear-recoil-discriminating
pressurized neon target for dark matter and neutrino detection. A
detector module would consist of a cylindrical vessel with a diameter
of $\sim$50\,cm and a length of $\sim$5\,m. An array of 100 modules at
100\,atm would result in a 10-ton target mass and occupy a detector
space of less than 1000\,m$^3$, including a 2-m-thick shield. The
signal would be read out from both ends of the cylinder either as
charge or as light via wavelength-shifting fibers, and would provide
position information along the cylinder.  Timing between the
ionization and scintillation signals would provide radial
positioning. The primary scintillation signal would be detected using
a photocathode on the inside diameter of the cylinder. Nuclear-recoil
discrimination is achieved by the ratio of ionization to
scintillation. Measurements have recently been completed that confirm
excellent discrimination between gammas and nuclear recoils.  The
physics reach for \WIMPs\ is 
$10^{-45}$\,cm$^2$/nucleon per ton of detector assuming zero
background. Such a detector would also see $\sim$2.5 nuclear-recoil
events per ton from a supernova at a distance of 10\,kpc (center of
galaxy). Low energy nuclear recoils ($<$3\,keV) would also be
observable from coherent scattering with solar neutrinos with energies
above 10\,MeV.

\subsection{SuperCDMS}

The \CDMS\ collaboration has developed semiconductor detectors of both
silicon and germanium which operate at milliKelvin temperatures.  Recoil
discrimination is achieved by combining ionization and athermal-phonon
readout. While the ionization signal suffers the same tendency with
regard to low-energy betas as the {\small EDELWEISS} detectors, the
athermal technology provides phonon-pulse shape information that
allows event localization. This information permits surface-versus-bulk
discrimination to reject surface betas, which would otherwise be the
dominant internal background. \CDMS\ has set world-leading
limits in the mid-10$^{-43}$\,cm$^2$/nucleon range using several 0.25\,kg
detectors in the Soudan Mine, and expects to improve the sensitivity
by a factor of 10 in 2006--2007.  \SuperCDMS\ would further develop this
technology to larger detector modules with improved performance to
further reject surface betas, and to engage industrial partners to
eventually reach an array of total mass of about one
ton. Sensitivity is anticipated to reach 10$^{-46}$\,cm$^2$/nucleon.

\subsection{XENON}

% http://gaitskell.brown.edu/physics/talks/0307_EPSHEP_Aachen/XENON/Gaitskell_XENON_v07.pdf

% http://meetings.aps.org/Meeting/APR05/Event/29211

The {\small XENON} collaboration is currently developing a
discriminating liquid-xenon 15-kg prototype detector with an expected
sensitivity of mid-10$^{-44}$\,cm$^2$/nucleon, coupled with advanced
\RnD\ to develop a ton-scale experiment. Signals from primary
scintillation in the liquid and proportional scintillation by
electrons extracted into the gas phase are being shown to allow both
electron-nuclear recoil discrimination and 3-D localization of
events. The prototype is currently being installed in a shielded setup
at Gran Sasso. Based on what is learned there, along with detailed
laboratory studies of various readout schemes to maximize
light-collection, the collaboration anticipates fielding a set of
100-kg modules to reach one ton of active mass and sensitivity to a
\WIMP-nucleon cross section of 10$^{-46}$\,cm$^2$/nucleon.

\subsection{ZEPLIN IV-Max}

The {\small ZEPLIN} collaboration has used several kilograms of liquid
xenon in a scintillation detector with pulse-shape discrimination
({\small ZEPLIN-I}) to set limits vying with the world's best.  In
addition to self shielding, the detectors have external
liquid-scintillator vetoes.  Currently the group is commissioning
detectors with tens of kilograms of active mass, which derive signals
from both primary scintillation and ionization in the liquid to
discriminate electron from nuclear recoils and to localize events
within the sensitive volume.  These systems use liquid-plus-gas
``two-phase" detection employing two different readout schemes in the
{\small ZEPLIN-II} and {\small ZEPLIN-III} detectors.  {\small
ZEPLIN-IV} (also referred to as {\small ZEPLIN-Max}) would scale up
the two-phase technique to a ton of active mass with sensitivity in
the range of 10$^{-46}$\,cm$^2$/nucleon.

\subsection{Summary}

The direct detection of dark matter is a vital, growing field with a
number of well-developed plans for developing ton-scale
experiments. The goal of all such experiments is to reach deep into
the \SUSY-predicted \WIMP-nucleon cross-section range of $\sim$
10$^{-46}$\,cm$^2$/nucleon for spin-independent couplings. While most
\SUSY\ models have stronger spin-independent couplings, the
sensitivity of these same experiments to spin-dependent couplings,
e.g., through the presence of isotopes such as $^{17}$F, $^{73}$Ge,
and $^{129}$Xe, is also of great interest.  Achieving this sensitivity
level will require sufficient depth and local shielding to bring
unvetoed neutron interactions in the detector material to well below
$3\times 10^{-5}$\,events/kg/day.  A combination of very effective
local shielding and muon vetoes, high radiopurity, event 
discrimination, and accessible well-planned deep laboratory space are
needed to reach this physics-driven goal.

\section{Infrastructure}
\label{infrastructure}

In this section we describe infrastructure and technical support that
is generally required for dark matter experiments aimed at ton-scale
detectors. Depth, which is a particularly important requirement
because it is the primary method of suppressing the cosmogenic neutron
background, is treated in Section~\ref{depth}. Following that
discussion, we describe in Section~\ref{handling} the materials
handling methods that are needed. Section~\ref{infrastructure} concludes
with an itemized list of the space and facilities needs defined by the
envelope of possible experiments, and a second list of
experiment-specific items that the lab needs to be prepared to
provide depending on which experiments are staged there.

Since a separate working group is devoted to low-background counting,
we do not discuss that here. Of course, such facilities are critical to
the preparation of dark matter experiments, and access to
state-of-the-art low-background assaying and cleaning methods
developed at the laboratory will be vital to the success of virtually
the full range of dark matter detection strategies.

\subsection{Depth Requirements} 
\label{depth}

The most important cosmic-ray-muon background for direct dark matter
detection experiments is due to fast neutrons (20--500\,MeV) produced
outside the detector shielding. These high-energy ``punch-through''
neutrons are difficult to tag with a conventional local muon-veto
system, as they can originate several meters within the cavern's rock
walls. They create secondaries in the surrounding shielding materials,
which can scatter in the detector target and generate signals quite
similar to those of \WIMPs. While a variety of countermeasures are
possible, such as thick active shields and wide umbrella vetoes
deployed in the cavern, an extensive earthen overburden remains
the most reliable means to reduce the muon flux and its accompanying
cosmogenic-neutron background. Moreover, as the mass and sensitivity
of dark matter searches grows to the ton-scale and beyond, as
expected on the time-scale of the 30-plus year {\small DUSEL} program,
the combination of depth and active shields may be called for.
Therefore siting the laboratory at the greatest depth possible is
critical to a successful long-term program.

Estimates of the size of the punch-through neutron background depend
on the muon flux, material composition of a specific underground site,
neutron-production cross sections, and the details of the subsequent
hadronic cascades. The muon flux is the best measured of these, and at
the earth's surface is about 170/m$^2$/s with a mean energy of about
4\,GeV. The attenuation at 4500\,m.w.e. is nearly a factor of $10^7$
with a flux of 800/m$^2$/y and a much harder spectrum with mean energy
of about 350\,GeV. For comparison, the flux at 1700\,m.w.e. is 100
times higher, while at 6000\,m.w.e. it is 15 times lower.  The neutron
flux in the energy range of interest does not decrease as quickly with
increasing depth because the neutron production rate and yield (per
muon) increase with muon energy.

The neutron cross sections are far more uncertain than the muon flux
and spectrum. Energetic muons produce neutrons in rock through
quasielastic scattering, evaporation of neutrons following nuclear
excitation, photonuclear reactions associated with the electromagnetic
showers generated by muons, muon capture, and secondary neutron
production in the subsequent electromagnetic showers and hadronic
cascades. The neutron yield as a function of the mean muon energy is
approximately a power law, $N\propto\langle E_\mu\rangle^{0.75}$.
While various theoretical estimates of the high-energy neutron
spectrum at depth have been made, few experiments have been done.

A recent and comprehensive study of the cosmic-ray muon flux and the
activity induced as a function of overburden has been made for a suite
of underground laboratories ranging in depth from $\sim$1000 to
8000\,m.w.e.~\cite{mei_hime05} derived from the Monte Carlo
muon-shower propagation code {\small FLUKA}.  The {\small FLUKA} code
works reasonably well, with deviations from measurements being about
50\%.  This study also develops a Depth-Sensitivity-Relation ({\small
DSR}) in an attempt to characterize the depth requirements of
next-generation experiments searching for neutrinoless double-beta
decay and \WIMP\ dark matter.

\begin{figure}[htb]
\begin{center}
\bigskip
\includegraphics[width=3.00in]{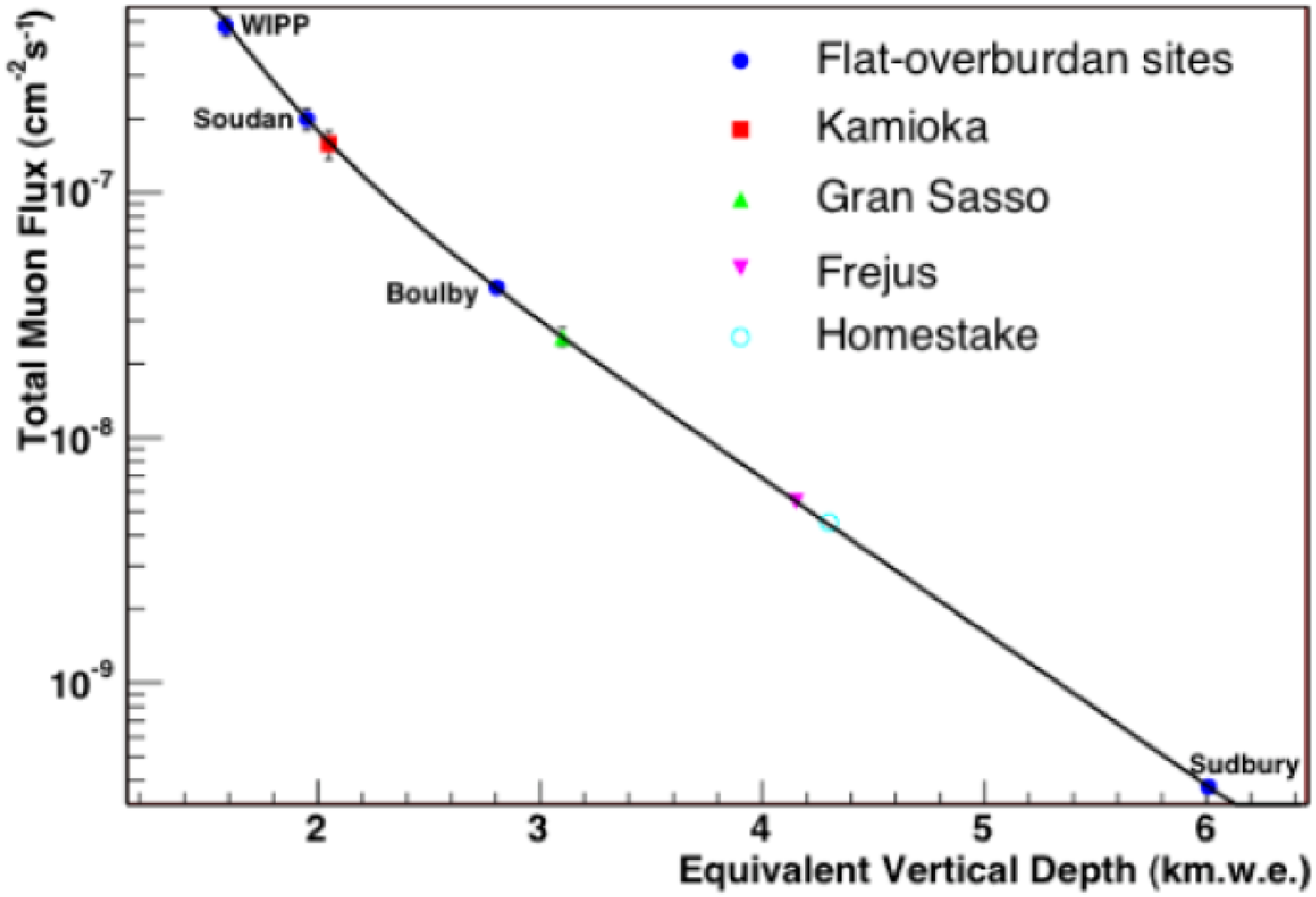}
\includegraphics[width=3.05in]{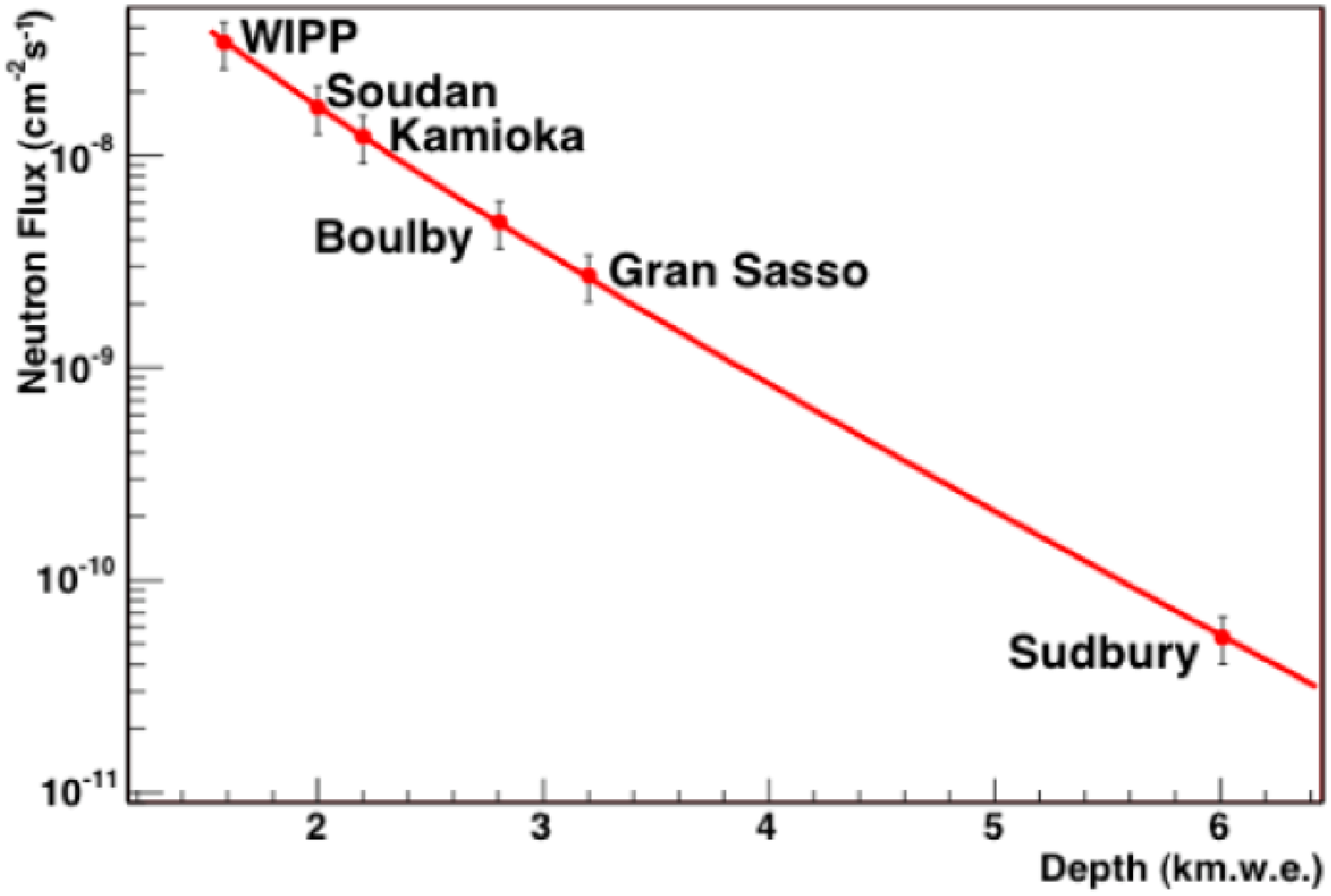}
\end{center}
\caption{\small (a) Left: The total muon flux measured for the various
underground sites as a function of the equivalent vertical depth
relative to a flat overburden. The smooth curve is a global fit
function to those data taken from sites with flat overburden.  (b) Right:
The total muon-induced neutron flux deduced for the various
underground sites displayed. Uncertainties on each point reflect those
added in quadrature from uncertainties in knowledge of the absolute
muon fluxes and neutron production rates based upon simulations
constrained by the available experimental data.  (All
from~\cite{mei_hime05}.)}
\label{fig:depth1}
\end{figure}

Figure~\ref{fig:depth1}a shows the total flux of cosmic-ray muons as a
function of overburden and for a selected set of underground
laboratories. Depth is defined in terms of the total muon flux that
has been experimentally determined and that would be present in a
laboratory with flat overburden. Figure~\ref{fig:depth1}b shows the
total neutron flux that is induced by the muons and that emerge at the
rock-cavern boundary of an underground site. Roughly speaking these
fluxes are attenuated by about one order of magnitude for every
increase in depth of 1500\,m.w.e.

\begin{figure}[!h]
\begin{center}
\includegraphics[width=3.0in]{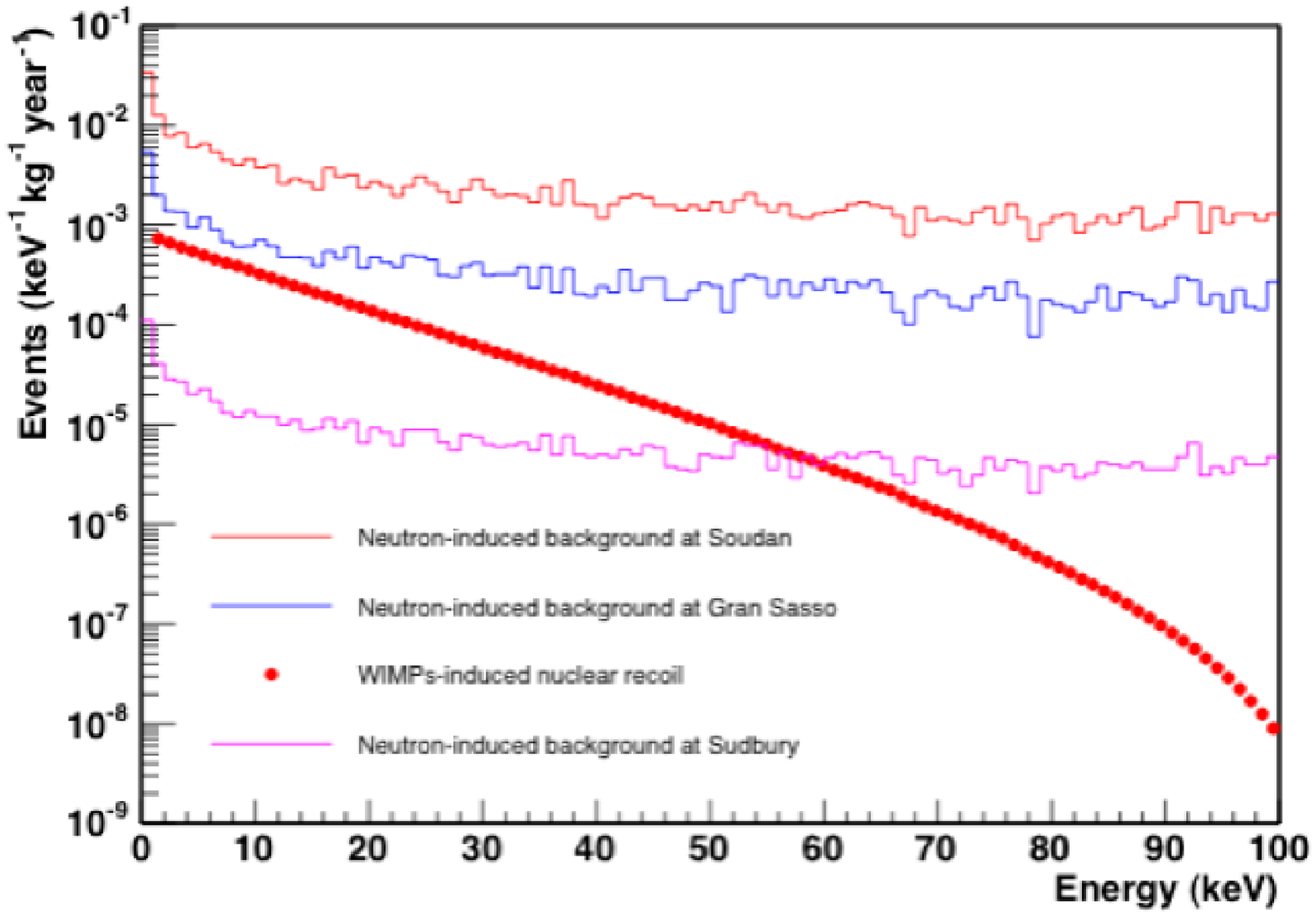}
\includegraphics[width=3.1in]{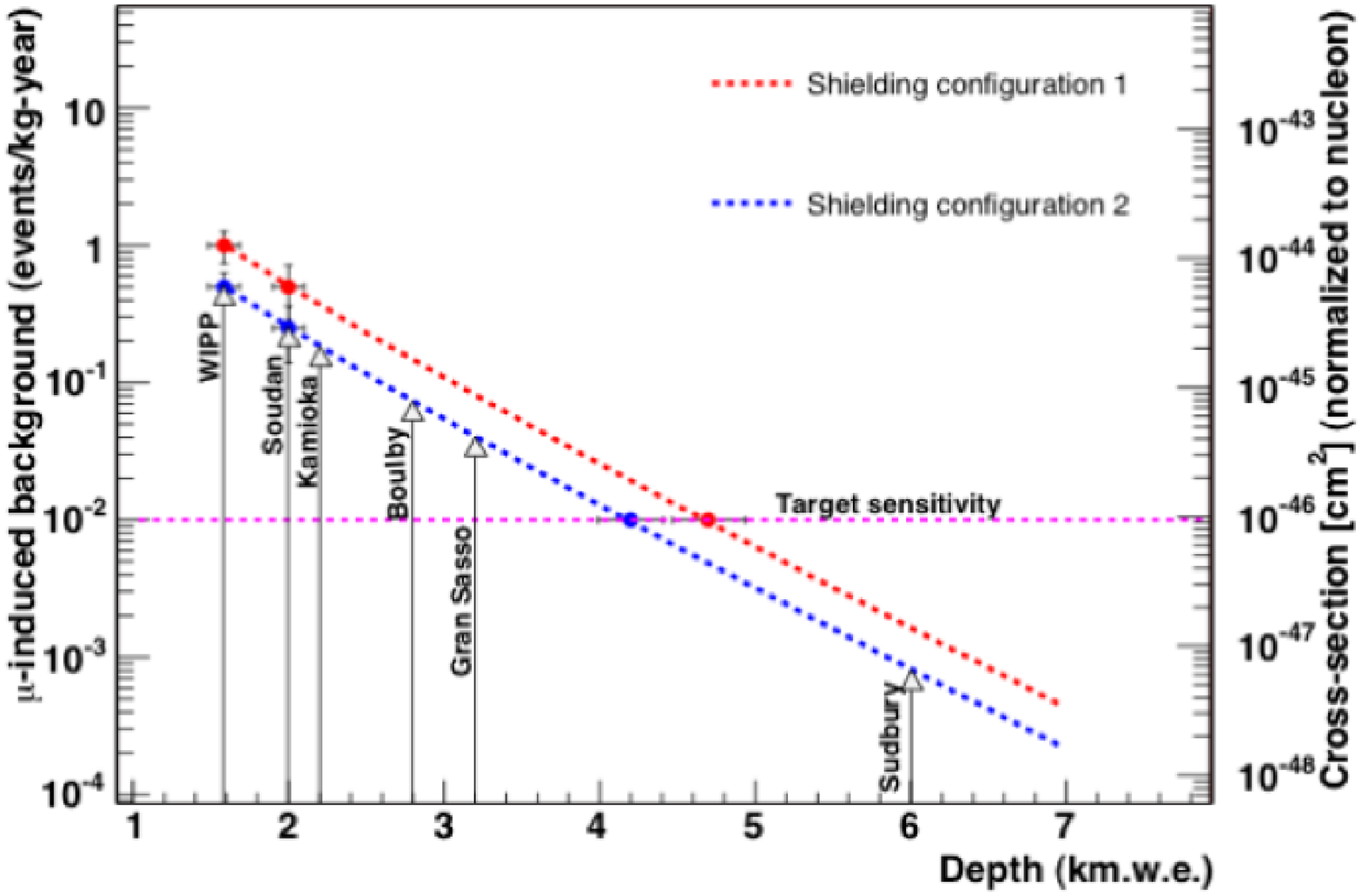}
\end{center}
\caption{\small (a) Left: The predicted event rates for spin-independent
WIMP-nucleon scattering (dotted-line) in Ge assuming a WIMP-nucleon
cross-section of $\sigma_p = 10^{-46}$\,cm$^2$ and a 100-GeV/c$^2$
WIMP mass. Muon-induced neutron backgrounds are also displayed for
comparison, indicating the need for greater and greater depth as
experiments evolve in scale and sensitivity.  (b) Right: The
Depth-Sensitivity-Relation (DSR) derived for the actual CDMS-II
detector geometry (upper curve) and for a reconfigured shield in which
all the lead gamma shield is exterior to the polyethylene moderator.
The muon-induced background is dominated by elastic scattering of
neutrons depositing visible recoil energy in a~10 to~100\,keV
window. Specific points are shown, for example, at the depth of the
Soudan mine where the CDMS-II detector has been
operating. Uncertainties reflect those present due to uncertainties in
the rock composition and in generating the muon-induced fast neutron
flux.  (All from~\cite{mei_hime05}.)}
\label{fig:depth2}
\end{figure}

\begin{figure}[!h]
\begin{center}
\bigskip
\medskip
\includegraphics[width=3.00in]{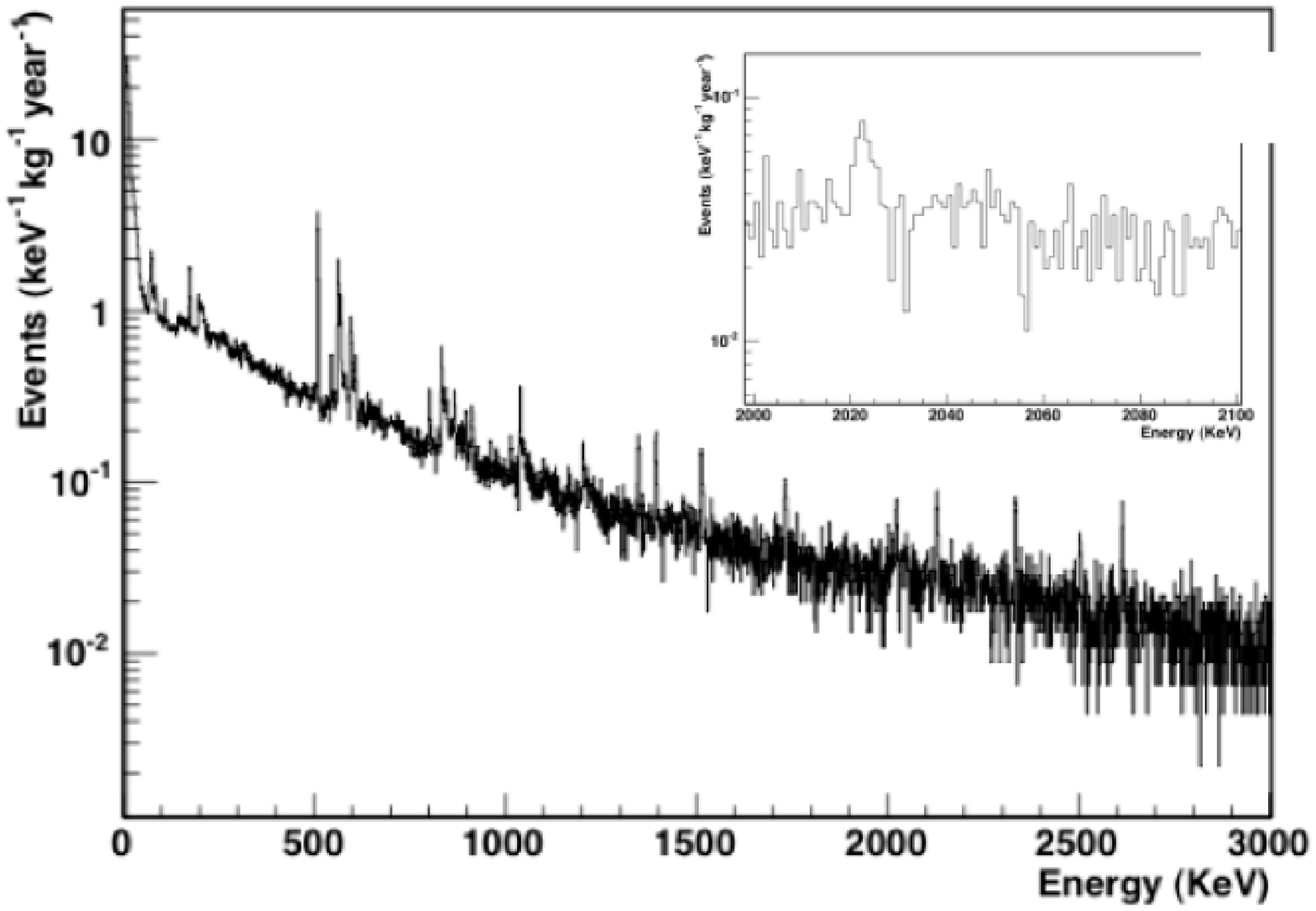}
\includegraphics[width=3.30in]{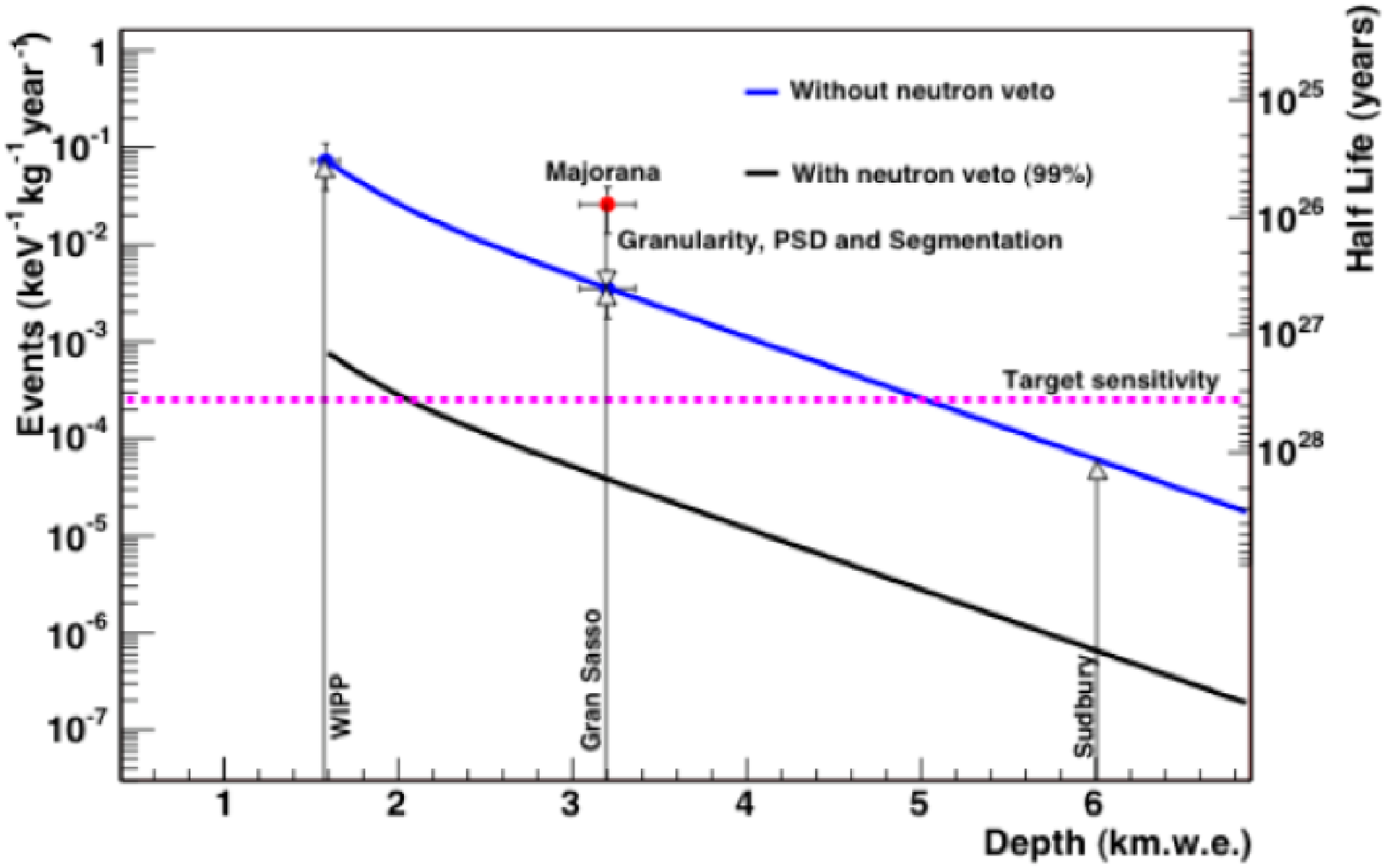}
\end{center}
\caption{\small
(a) Left: A simulation of the muon-induced background for a Majorana-like
experiment operating at an equivalent overburden provided by the Gran
Sasso Laboratory, showing the full spectrum and an expanded profile
(inset) spanning the Region-of-Interest (ROI) around the Q-value for
neutrinoless double beta decay at 2039\,keV. The peak at 2023\,keV is
characteristic of that produced via the $^{76}Ge(n,n^\prime \gamma)$
reaction.  (b) Right: The Depth-Sensitivity-Relation (DSR) applied to a
Majorana-like experiment. The raw event rate in the energy region of
interest of 0.026\,events/keV/kg/year can be reduced by a factor of
7.4 by exploiting the detector granularity, pulse-shape
discrimination, and detector segmentation. The upper curve displays
the background simulated in the case that no active neutron veto is
present and the lower curve indicates the reduction that would result
from an active neutron veto that is 99\% efficient.  (All
from~\cite{mei_hime05}.)}
\label{fig:depth3}
\end{figure}

The effect of this muon-induced background will depend upon
the details of a specific detector geometry and its scientific
goal. In ref.~\cite{mei_hime05}, the {\small DSR} was developed for
germanium detectors specific to the direct search for \WIMP\ dark
matter (\CDMSII) and that under development to search for neutrinoless
double beta decay (Majorana). As can be seen in Figure~\ref{fig:depth2},
elastic scattering of fast neutrons produced by cosmic-ray muons
represent an important background for direct dark matter searches,
while Figure~\ref{fig:depth3} demonstrates the sensitivity to these fast
neutrons in neutrinoless double-beta decay experiments owing
primarily to inelastic scattering processes. 

% dsa - substantial revision/additions through to end of section 
% in response to reviewer comment C.6

Next-generation dark matter experiments will either require depths in
the range of 4000--5000\,m.w.e. or greater, or significant steps to
actively shield or veto the fast neutrons produced through cosmic-ray
muon interactions to reach the desired sensitivity levels.  Which
option is more cost-effective will depend on the available of space at
depth, and the nature of the central detector hardware. For example,
some classes of experiments, such as noble liquids may lend themselves
to submersion in large water-filled cavities. Although the focus here
is on depth requirements to defeat fast neutron backgrounds, the
experiments will also need to address internal sources of neutrons
from ($\alpha$,n) reactions and nuclear fission in the residual
contamination in detector and shielding components, which should also
be sufficiently thick to shield against similar neutron sources in the
surrounding rock.

%\clearpage % clears figure queue and helps place fig here ([!h])

For example, simulations carried out by the \CDMSII\
collaboration~\cite{cdms119} at the Soudan Mine depth of
2080\,m.w.e. using Geant3 to propagate cosmogenic neutrons from the
rock predict that this Ge-based experiment would detect single-scatter
nuclear-recoil events from punch-through neutrons at a rate about one
per 400\,kg-d exposure. With an additional factor of two reduction
from a scintillator veto system, the estimated rate of unvetoed
neutrons becomes 0.5\,events/kg/year. An exposure of 3\,kg-years,
which corresponds to a \WIMP-nucleon sensitivity of $10^{-44}$\,cm$^2$
in the absence of background, would contain 1.5 background events at
this rate. More detailed simulatons that include correlations of the
parent muon, as well as all charged and neutral secondaries, were done
with {\small FLUKA} and {\small MCNPX} that simulate energy
depositions in the surrounding scintillator and in the dark matter
detectors~\cite{raul_cdms}.  These simulations predict a rate of
0.05\,events/kg/year\,$\pm$\,0.01\,(stat.)\,$\pm$\,0.03\,(sys.), where
the reduction compared with the previous simulations is in part
attributed to improved tracking of signals in the scintillator, and a
larger fraction of multiple-scatters (these latter simulations were
done for a 10-times larger array). At the higher rate, which is very
likely an overestimate, a factor of 300 improvement to reach
$10^{-46}$\,cm$^2$ sensitivity would require, according to the {\small
DSR} derived in~\cite{mei_hime05}, a depth of 6000\,m.w.e. in the
absence of additional shielding measures. At the lower rate a depth of
4200\,m.w.e. would suffice for a factor of 30 reduction.

% reduction factor r for increased depth x given by 
% r=1500mwe*log10(x) => r=3700 for x=300; depth = 2080 + 3700 ~ 6000mwe

While the absolute rate of unvetoed nuclear recoils estimated with
Monte Carlos lacks direct validation with experimental data for the
specific mechanisms that lead to background events in dark matter
experiments, detecting events correlated with unvetoed nuclear recoils
can be used to predict their rate. For example, in the CDMS
experiments performed at shallow depth, neutrons that multiple
scattered were combined with simulations to estimate the number of
single scatters associated with the same ambient neutron
population. New simulations underway based on {\small FLUKA} and
{\small MCNPX} show that an 60-cm-thick outer shield of liquid
scintillator loaded with 0.5\% Gd could efficiently detect a majority
of the spallation events that produce neutrons that could interact
with the dark matter detectors inside the shield~\cite{raul_cdms}. The
efficient detection of such events would simultaneously provide a
cross check on the background simulations and higher-statistics
measure of the in situ background compared with the multiple-scatter
technique.

Although depth is the simplest and most certain solution to the
fast-neutron background, several more ideas have been suggested that
could make sites shallower than about 4000\,m.w.e.
acceptable. Sophisticated shielding and vetoing could reduce this
background by one to two orders of magnitude. A thick (1--2\,m)
scintillator active veto around the detectors could tag high-energy
neutrons as they penetrate inward. If instrumented as a Gd-loaded
neutron multiplicity meter, as described above, a thickness of
60--80\,cm would provide an order of magnitude reduction as predicted
by simulations done for \CDMSII~\cite{raul_nim}. In addition the cavern
rock, or an outer heavy passive shield, could be instrumented with
additional veto detectors in order to catch some part of the shower
associated with the initial muon.  Work is also ongoing to investigate
very thick passive high-purity water shields, which could economically
shield both gammas and neutrons. Simulations of a 4-m-thick water
shield have shown a factor of 20 reduction in the recoil rate due to
punch-through neutron compared with the conventional external-lead
internal-polyethylene shield configuration referred to in
Figure~\ref{fig:depth2}~\cite{gaitskell_water}.

Increased granularity in detectors will enhance the multiple-scatter
rejection rate (primarily for inner detector at some cost of fiducial
mass), but would nonetheless serve as a cross check on
systematics. Along these lines, measuring the neutron multiplicity in
the energy range 100\,keV--10\,MeV in which neutrons produce
\WIMP-like recoils would also be helpful for benchmarking Monte Carlo
simulations and quantifying systematic errors. The focus here has been
on the depth-related fast-neutron background, but it is worth noting
that its suppression through the use of depth and passive and/or
active shielding must be accompanied by sufficient control of neutron
sources from radioactivity. Primary sources are ($\alpha$,n) processes from
uranium- and thorium-series contamination and from spontaneous fission
of uranium. Sufficiently thick moderator will adequately shield
($\alpha$,n) sources from the rock (e.g., 1--2-m thick), but radiopurity of
internal components will require continued vigilance. With regard to
fission, one source that has been considered is from Pb
shielding. G.~Heusser and co-workers have observed upper limits of
20\,ppt gram of uranium per gram of lead, with no positive
measurements in any samples counted~\cite{heusser_gempi}. That level
would produce on the order if 1 event per ton-year in a dark matter
detector, but as noted by Heusser it may be that the chemistry of
these heavy metals tends to make lead intrinsically pure.

In summary, next generation experiments, for example, those aiming for
sensitivity in the $10^{-45}$\,cm$^2$ range, could be performed at depths
on the order of 4000\,m.w.e. As experiments aim for greater
sensitivity, such as those that could occur in the first suite of
{\small DUSEL} experiments, depths of 6000\,m.w.e. become very
desirable. Ultimately, we find that these depths should be available
at a laboratory that is aimed at a long-duration program. Surely, if
\WIMPs\ are discovered with cross sections anywhere in the range
between present limits and $10^{-46}$\,cm$^2$, a {\small DUSEL} at
6000\,m.w.e. would be needed, possibly along with active vetoes, to
study \WIMPs\ with greater statistics and without contamination from
depth-related backgrounds.

\subsection{Materials Handling}
\label{handling}

Underground storage, handling, and fabrication facilities are
important for dark matter experiments. Materials exposed to cosmic
rays at the Earth's surface become radioactive through inelastic
interactions, primarily due to air-shower neutrons. Even a 30-day
exposure of an initially-pure sample of a medium-mass element will
result in long-lived low-energy activity due to $^3$H at several tens
of disintegrations per kg per year. Other radionuclides within 10--20
mass units lighter than the target are also produced, with specific
activities in some cases a hundred times higher than for $^3$H.  These
effects have been seen in experiments, and computer programs such as
{\small COSMO} and {\small SIGMA} have provided estimates of the activation
and decay processes, in reasonable agreement with measurements.

The activation rate is reduced by a factor $\sim$100 at a depth of 
just 10--20\,m.w.e.
Current and past experiments have taken effective
measures to shield or limit exposure times for critical components
during processing at the surface.  Underground storage for several
halflives of the most critical radionuclides is also effective at
reducing this background.  However, the event rates to be sought in
next-generation dark matter experiments are predicted by
Supersymmetry to be exceedingly low, of order 10~per ton per year.
Even experiments proposed to have excellent electron-recoil rejection
efficiency would benefit significantly from eliminating the
cosmogenics by conducting final purification and fabrication of
critical components at least in a shallow underground facility.

Exposure to radon during these final operations must also be limited,
to avoid deposition or incorporation of radon or its daughters
(including long-lived $^{210}$Pb) into critical components.  The
extent of radon mitigation required will be highly dependent on the
particular experiment and component involved.  It would be prudent to
provide at least a section of the facility with radon significantly
below the surface air level of $\sim$16\,Bq/m$^3$.  The Borexino
collaboration built an assembly area at Princeton with radon
concentration reduced by a factor of 100 from the local surface level.

Based on these considerations, {\small DUSEL} should have a
100--200\,m$^2$ clean shop and storage area at least
20\,m.w.e. underground, equipped with radon and particulate mitigation
infrastructure.  Cleaning stations, chemical hoods, and machine tools
should be present from opening day.  More materials processing
equipment such as copper electroforming and semiconductor crystal
growing may be incorporated as specific experiments are staged in, so
provision for these capabilities brings potential benefits.

\subsection{Facility Needs and Space}

Most dark matter experiments have modest-size footprints (typically
200\,m$^2$, or less) for the actual experiment.  Additional space is
of course needed for staging, {\small DAQ}, and control rooms.
The environments must be able to be 
maintained at modest, e.g., Class 1000 clean room conditions with some
mitigation of radon backgrounds. These needs are similar to those of
double-beta decay experiments. 

An initial suite of experiments would therefore be accommodated 
by one or two full-size caverns of
2000 cubic meters each, plus an additional 50 square meters for
equipment staging, storage, control rooms, and \RnD\
projects. The lab should also have the capability for further
expansions that can be excavated for new experiments with minimal
disruption to existing projects.

The more specific basic facility needs of dark matter
experiments  are itemized below:

\begin{itemize}

\item The space required to set up any one of the next-generation
detectors is typically a 10\,m high by 200\,m$^2$ footprint. Several 
such experiments would be mounted in a single large hall.  The hall 
should have crane(s) up to 30 tons with trained riggers available
by arrangement.  An overall muon veto and possibly water shielding
for the entire hall would be cost-effective solutions to the needs
of nearly all experiments. Water shields for individual experiments
might require a sunk-in pit of 10-m depth below floor level to
accommodate the shield base section.

\item Experimental rooms must be cleanable, that is, upgradeable to
clean-room standards during initial assembly and later operations.  
Soft-wall clean room systems are available commercially to allow 
flexible configuration of the underground space. 

\item Radon underground is a critical problem for sensitive low energy 
experiments.  Experimental requirements are typically specified
in the 10--100\,mBq/m$^3$ range.
Either each experiment will need its own radon-scrubbing
micro-environment, or the entire cavern
wall should be sealed and the hall continuously supplied with air 
scrubbed to reduce radon levels. 
Some experiments require constant radon purging for experiment
interiors, so a compressor system for old-air storage and filtration,
or cover gas derived from a liquid-nitrogen boiloff source
will be required. 

\item Each experiment requires additional underground laboratory floor
space to house data acquisition and experiment control rooms of
typically 50\,m$^2$. 

\item Because stability during extended periods of data-taking is
critical, temperature control to at least office-building standards
($\pm$1$^{\circ}$C) is necessary in all underground areas.

\item Average power requirements per experiment are typically 100\,kW
or less, with peak power no more than a factor of two higher. Critical
subsystems such as control computers and cryogenics will 
require uninterruptable power supplies 
which should be provided by the individual experiments.

\item The typical size of underground experimental crews during
installation and commissioning  is 10--20 people. 
Standard running requires only two to four
people accessing experimental areas, with larger numbers during
upgrade periods and run commissioning. The principal laboratory
requirement is 24/7 access in case of emergency. During commissioning
and detector start-up sequences, extended access (e.g., two shifts per
day) is beneficial. None of the experiments as envisioned requires
continuous underground presence.

\item Provision for radioactive calibration sources is required for
all experiments, including the necessary licensing, safety, and
storage protocols.

\item A 100--200\,m$^2$ clean shop, staging and storage area should be provided
at least 20\,m.w.e. underground, equipped with radon and particulate
mitigation infrastructure.  This facility could be shared with
non-dark matter users.  Cleaning stations, chemical hoods, and machine
tools should be present from opening day.  More elaborate materials
processing equipment such as copper electroforming and Ge detector
re-processing could benefit nearly all dark matter experiments.  The
infrastructure should allow these to be incorporated as specific
experiments are staged in.

\item A surface facility with laboratory space and experiment
control rooms providing computer links to the underground laboratory
is also needed.  This would amount to roughly 30\,m$^2$ per experiment.

\end{itemize}

Special facility needs (experiment specific): Each experiment has
specific additional needs not discussed above, but which are listed
in the Infrastructure Table.  These tend to fall into groupings as follows.

\begin{itemize}

\item Large inventories of liquid cryogen are used in several experiments.
Safety systems must be included to minimize the possibility of large
accidental releases, and to mitigate the oxygen-deficiency hazard
which could develop with accidental release or accumulation of boiloff
gas. Oxygen deficiency alarms, standby ventilation, and/or personnel
escape/refuge facilities would be needed.

\item {\small EURECA} is calling for a helium liquefier located near
their experiment.  This would have to be engineered to allow safe
operation or shutdown in case of a power outage.

\item As presently configured, the {\small TPC} 
experiments use flammable, toxic gases for the negative-ion drift.
The detectors themselves operate below atmospheric pressure so
catastrophic releases are unlikely.  However, appropriate procedures
must be in place for safe storage, transfer and disposal of these
materials.  {\small COUPP} presently plans to use several hundred
kilograms or more of a non-flammable heavy liquid (CF$_3$I) which is
much less toxic than the CS$_2$ negative-ion capture agent but is used
under several bar pressure.  Release prevention and management systems
will still be required for this material.

\item Water shielding tanks are planned for {\small COUPP}, {\small CLEAN}, 
and possibly others.  These require personnel safeguards to prevent
injury and equipment damage.

\item Power outages of duration longer than 30~minutes can be
disruptive for cryogenic experiments that plan to use cryocoolers,
such as {\small XENON} and \SuperCDMS. Since battery-type
Uninterruptable Power Supplies {\small UPSs} are not practical for
long outages, the provision for experiment-run backup generators may
be called for.

\end{itemize}

\section{International Context}
\label{international}

As we have discussed in section~\ref{depth}, it is best to
perform dark matter experiments as deep as possible. For ton-scale
experiments, the neutron background can be eliminated at depths of
6000\,m.w.e., or at shallower depths if adequate active shielding is
used and systematic effects are well studied. Several current efforts
that are building or performing experiments are developing
next-generation plans. These plans were quite evident through the
recent series of workshops held at {\small SNOLAB}, the site of the
Sudbury Neutrino Observatory, or {\small SNO}.

{\small SNOLAB} is located at a depth of 6000\,m.w.e. and, using funds
provided by the Canadian Fund for Innovation, is constructing a
laboratory to house a suite of four to six new experiments as early as
2007. A call to the physics community for Letters of Interest yielded
18 responses, 7 of which were for experiments aimed at detecting
\WIMPs. While it is unlikely that all 18 of these experiments will be
funded and built (e.g., some represent similar competing
technologies), it is likely that the capacity of {\small SNOLAB} will
be exceeded by approximately a factor of two, though such an estimate
is inherently uncertain and depends on many factors.

In addition to {\small SNOLAB}, other major laboratories that will
host next-generation dark matter experiments include Boulby, Gran
Sasso, Modane, Canfranc, and Kamioka. None of these is as deep as
{\small SNOLAB}, but several dark matter experiments which are already
operating at these locations are likely to remain there, to take
advantage of the existing infrastructure and proximity to home
institutions.  For example, the Italian-led {\small WARP} and the
Swiss-led {\small ArDM} liquid-argon experiments (which follow on the
developments for {\small ICARUS}) and the Japanese-led {\small XMASS}
liquid-xenon experiment, are likely to remain at Gran Sasso, Canfranc
and Kamioka, respectively. Strong candidates for {\small DUSEL} from
the international community are the {\small UK} program and Eureca, in
addition of course to the various {\small US} efforts and others
discussed in Section~\ref{experiments}, including {\small CLEAN},
Coupp, {\small DEAP}, {\small DRIFT}, \SuperCDMS, {\small XENON}, and
{\small ZEPLIN}. Given the strong demand for deep space for dark
matter experiments, the larger scale of ``next-next'' generation
experiments timed for, say, a 2012 opening of {\small DUSEL}, and the
capacity limitations of {\small SNOLAB}, it is clear that a robust and
exciting dark matter program will be part of the initial {\small
DUSEL} program.

\section{Summary and Outlook}
\label{summary}

In this report we have described the broad and compelling range of
astrophysical and cosmological evidence that defines the dark matter
problem, and the \WIMP\ hypothesis, which offers a solution rooted in
applying fundamental physics to the dynamics of the early
universe. The \WIMP\ hypothesis is being vigorously pursued, with a
steady march of sensitivity improvements coming both from
astrophysical searches and laboratory efforts. The connections between
these approaches are profound and will reveal new information from
physics at the smallest scales to the origin and workings of the
entire universe.

Direct searches for \WIMP\ dark matter require sensitive detectors
that have immunity to electromagnetic backgrounds, and are located in
deep underground laboratories to reduce the flux from fast
cosmic-ray-muon-induced neutrons which is a common background to all
detection methods. With {\small US} leadership in dark matter searches
and detector \RnD, a new national laboratory will lay the foundation
of technical support and facilities for the next generation of
scientists and experiments in this field, and act as magnet for
international cooperation and continued {\small US} leadership.

The requirements of depth, space and technical support for the
laboratory are fairly generic, regardless of the approach. Current
experiments and upgraded versions that run within the next few years
will probe cross sections on the $10^{-45}$--$10^{-44}$\,cm$^2$ scale,
where depths of 3000--4000\,m.w.e. are sufficient to suppress the
neutron background. On the longer term, greater depths on the
5000--6000 level are desirable as cross sections down to
$10^{-46}$\,cm$^2$ are probed, and of course, if \WIMPs\ are discovered
then building up a statistical sample free of neutron backgrounds will
be essential to extracting model parameters and providing a robust
solution to the dark matter problem. 

While most of the detector technologies are of comparable physical
scale, i.e., the various liquid and solid-state detector media under
consideration have comparable density, a notable exception is the
low-pressure gaseous detectors. These detectors are very likely to
play a critical role in establishing the galactic origin of a signal
if the remaining challenges of background rejection and low threshold
can be demonstrated, and so it is important to design the lab with
this capability in mind. For example, for a \WIMP-nucleon cross
section of $10^{-43}$\,cm$^2$ (just below the present
limit~\cite{cdms119}), 100 modules of the size and pressure currently
being investigated by the {\small DRIFT-II} collaboration (1\,m$^3$ at
40\,torr CS$_2$~\cite{drift}) would require a two-year
exposure~\cite{gaitskell04} to get the approximately 200
events~\cite{copi2005} required to establish the signal's galactic
origin. While detector improvements are under investigation, a simple
scaling for the bottom of the {\small MSSM} region at
$10^{-46}$\,cm$^2$ would require a 100,000\,m$^3$ detector volume. If
a factor of 10 reduction in required volume is achieved (e.g., higher
pressure operation, more detailed track reconstruction, etc.) then an
experimental hall of (50\,m)$^3$ could accommodate the experiment.

Because the \WIMP-nucleon cross section is unknown, it is impossible to make
a definitive statement as to the ultimate requirements for a
directional gaseous dark matter detector, or any other device, for
that matter. What is clear, however, is that whatever confidence one
gives to specific theoretical considerations, the foregoing
discussion clearly indicates the high scientific priority of, broad
intellectual interest in, and expanding technical capabilities for
increasing the ultimate reach of direct searches for \WIMP\ dark
matter. Upcoming experiments will advance into the low-mass
Supersymmetric region and explore the most favored
models in a complementary way to the \LHC, and on a similar time
scale. The combination of astrophysical searches and accelerator
experiments stands to check the consistency of the
solution to the dark matter problem and provide powerful constraints
on the model parameters. Knowledge of the particle properties from
laboratory measurements will help to isolate and reduce the
astrophysical uncertainties, which will allow a more complete picture of
the galactic halo and could eventually differentiate between, say,
infall versus isothermal models of galaxy formation. 

The scientific landscape of dark matter, which spans particle physics,
astrophysics and cosmology, is very rich and interwoven. Exploring
this exciting program following an initial detection will need many
observables and hence a range of capabilities for follow-up experiments
including different targets to sort out the mass and coupling of the
\WIMP, and directional sensitivity to confirm its galactic origin and
open the age of \WIMP\ astronomy. Clearly, this broad and fascinating
program is ideally suited to the multi-decade span of {\small DUSEL}.

\section{Acknowledgements}

This work was carried out, in part, at several workshops sponsored
with the support of the NSF {\small DUSEL} ``Solicitation 1''
grant. In particular, the Dark Matter working group held meetings
during the August 2004 workshop in Berkeley, the January 2005 workshop
at CU Boulder, and the July 2005 workshop at the University of
Minnesota. We also wish to thank the Institute for Nuclear Theory at
the University of Washington, Seattle, for hosting several members of
this working group for a week of discussions on dark matter and
underground science in August, 2005. We thank the following
participants at these various workshops for contributing to
discussions that have positively influenced this report: J.F.~Beacom,
D.~Berley, P.L.~Brink, J.I.~Collar, D.O.~Caldwell, P.~Cushman,
J.~Filippini, R.~Hennings-Yeomans, S.~Kamat, L.M.~Krauss, D.~McKinsey,
R.~Mahapatra, R.~Ragahvan, T.~Shutt, H.~Wang, and J.~White.  We thank
G.~Chardin, S.~Elliot, W.~Rau, D.~Snowden-Ifft, T.J.~Sumner, and
J.~White for helpful written communications on specific experiments
and research plans.  This report is in part based on material from the
Homestake-2003~\cite{homestake} and Cascade-2005~\cite{cascade}
Collaboration Science Books, which summarized the results of a series
of community workshops. We are grateful particularly to W.~Haxton,
L.~Baudis, R.J.~Gaitskell, and R.W.~Schnee for allowing us to
incorporate portions of these reports.

\section*{Contributors and Working Group Members}
D.S.~Akerib (Co-chair), Case Western Reserve University;
E.~Aprile (Co-chair), Columbia University;
E.A.~Baltz, Kavli Institute for Particle 
    Astrophysics and Cosmology, {\small SLAC}; 
M.R.~Dragowsky, Case Western Reserve University;
R.J.~Gaitskell, Brown University;
P.~Gondolo, University of Utah;
A.~Hime,  Los Alamos National Laboratory;
C.J.~Martoff, Temple University;
D.-M.~Mei, Los Alamos National Laboratory;
H.~Nelson, University of California, Santa Barbara;
B.~Sadoulet, University of California, Berkeley;
R.W.~Schnee, Case Western Reserve University;
A.H.~Sonnenschein, Fermi National Accelerator Laboratory;
and L.E.~Strigari, University of California, Irvine.

%%%%%%%%%%%%%%%%%%%%%%%%%%%%%%%%%%%%%%%%%%%%%%%%%%%%%%%%%%%%%%%%%%%%%%


\begin{thebibliography}{99}

\bibitem{turner} 
Committee on the Physics of the Universe, National Research Council,
M.S.~Turner, chair.
{\it Connecting Quarks with the Cosmos:
Eleven Science Questions for the New Century}
Washington, D.C.: The National Academies Press, 2003.
% http://www.nap.edu/catalog/10079.html

\bibitem{barish} 
Neutrino Facilities Assessment Committee, National Research Council,
B.~Barish, chair.
{\it Neutrinos and Beyond: New Windows on Nature}.
Washington, D.C.: The National Academies Press, 2003.
% http://www.nap.edu/catalog/10583.html

\bibitem{drell} http://www.science.doe.gov/hep/HEPAP/Quantum\_Universe\_GR.pdf

\bibitem{ostp} http://www.ostp.gov/html/physicsoftheuniverse2.pdf

\bibitem{epp2010}
Committee on Elementary Particle Physics in the 21st Century, National
Research Council, S.~Dawson and H.~Schapiro, chairs. 
{\it Revealing the Hidden Nature of Space and Time:
Charting the Course for Elementary Particle Physics}. 
Washington, D.C.: The National Academies Press, 2006.
% http://www.nap.edu/catalog/11641.html 

\bibitem{rubin} V.C. Rubin and W.K. Ford Jr., {\it Astrophys. J.} {\bf
159}, 379 (1970); Y. Sofue and V.C. Rubin, {\it Ann. Rev. Astron. \&
Astrophys.} {\bf 39}, 137-174 (2001).

\bibitem{zwicky} F. Zwicky, {\it Helv. Phys. Acta.} {\bf 6}, 110 (1933).

\bibitem{colley} Image credit: W.N.~Colley and E.~Turner (Princeton
University), J.A.~Tyson (Bell Labs, Lucent Technologies) and NASA.

\bibitem{tyson} J.A. Tyson, G.P. Kochanski and I.P. Dell'Antonio,
{\it Ap. J. Lett.} {\bf 498}, 107 (1998).
% astro-ph/9801193

\bibitem{allen2004} S.W. Allen, R.W. Schmidt, H. Ebeling, A.C. Fabian
and L.~van~Speybroeck,
{\it Mon. Not. Roy. Astron. Soc.} {\bf 353}, 457 (2004).
% see also older paper astro-ph/0205007 

\bibitem{spergel} D.N.~Spergel \etal., {\it Astrophys. J. Suppl.} {\bf 148}
175 (2003). 

\bibitem{tegmark} M. Tegmark \etal., 
{\it Phys. Rev.} {\bf D69}, 103501 (2004).
% astro-ph/0310723 

\bibitem{lee1977} B.W. Lee and S. Weinberg, {\it Phys. Rev. Lett.} 
{\bf 39}, 165 (1977).

\bibitem{jungman} G. Jungman, M. Kamionkowski and K. Griest, 
{\it Phys. Rep.} {\bf267}, 195 (1996).

\bibitem{axions} L.J. Rosenberg and K.A. van Bibber, 
{\it Phys. Rep.} {\bf 325}, 1-39 (2000).

\bibitem{wimpzillas} E.W. Kolb, D.J.H. Chung and A. Riotto,
hep-ph/9810361.

\bibitem{abbiendi2003} G. Abbiendi \etal. (the {\small ALEPH}
Collaboration, the {\small DELPHI} Collaboration, the {\small L3}
Collaboration, the {\small OPAL} Collaboration and the {\small LEP}
Working Group for Higgs Boson Searches),
{\it Phys. Lett.} {\bf B565}, 61 (2003).
% hep-ex/0306033

\bibitem{goodman} M.W. Goodman and E. Witten, 
{\it Phys. Rev.} {\bf D31}, 3059 (1985).

\bibitem{zeplin} G.J. Alner \etal. ({\small UK} Dark Matter Collaboration),
{\it Astroparticle Phys.} {\bf 23}, 444 (2005); see also a critique of
this result by
A.~Benoit \etal., {\it Phys. Lett.} {\bf B637}, 156 (2006),
and response by
N.J.T.~Smith, A.S.~Murphy, and T.J.~Sumner, 
{\it Phys. Lett.} {\bf B642}, 567 (2006).

\bibitem{edelweiss}
A.~Benoit \etal. ({\small EDELWEISS} Collaboration),
{\it Phys. Lett.} {\bf B545}, 43 (2002).

\bibitem{cresst2005} 
G.~Angloher \etal ({\small CRESST} Collaboration),
{\it Astroparticle Phys.} {\bf 23}, 325 (2005).

\bibitem{warp}
P.~Benetti \etal.  ({\small WARP} Collaboration),
astro-ph/0701286. 

\bibitem{cdms119} D.S. Akerib \etal. (CDMS Collaboration), 
{\it Phys. Rev. Lett.} {\bf 96}, 011302 (2006).
%astro-ph/0509259

\bibitem{lewin96}
J.D.~Lewin and P.F.~Smith, {\it Astropart. Phys.} {\bf 6}, 87 (1996).

\bibitem{dama} R.~Bernabei \etal., {\it
Phys. Lett.} {\bf B480}, 23-31 (2000); 
R. Bernabei \etal., astro-ph/0405282. 

\bibitem{Kim02}
Y.G. Kim, T.~Nihei, L.~Roszkowski and R.~Ruiz~de~Austri,
{\it J. of High Energy Phys.} {\bf 212}, 34 (2002).
% hep-ph/0208069

% trimmed to simplify figure 2
%\bibitem{Baltz03} E. Baltz and P. Gondolo,
%{\it Phys. Rev.} {\bf D67}, 063503 (2003).

\bibitem{Bottino03} A.~Bottino, F.~Donato, N.~Fornengo and S.~Scopel,
{\it Phys. Rev.} {\bf D69}, 037302 (2004).  
% hep-ph/0307303

\bibitem{Baltz04} E.A. Baltz and P. Gondolo, 
{\it J. of High Energy Phys.} {\bf 410}, 52 (2002).
% hep-ph/0407039.

\bibitem{Chattopadhyay04} U. Chattopadhyay, A. Corsetti and P. Nath,
{\it Phys. Atom. Nucl.} {\bf 67}, 1188 (2004).
% hep-ph/0310228

\bibitem{Baer03} H. Baer, C. Balazs, A. Belyaev and J. O'Farrill,
{\it J. of Cosmo. and Astropart. Phys.} {\bf 309}, 7 (2003).
% hep-ph/0305191

\bibitem{Giudice04} G.F. Giudice and A. Romanino, 
{\it Nucl. Phys.} {\bf B699}, 65-89 (2004); 
Erratum-ibid. {\bf B706}, 65-89 (2005).
% hep-ph/0406088.

\bibitem{Pierce04} A. Pierce, 
{\it Phys. Rev.} {\bf D70}, 075006 (2004).
% hep-ph/0406144.

\bibitem{Battaglia03}
M. Battaglia, A. De Roeck, J. Ellis, F. Gianotti, K. A. Olive and L. Pape, 
{\it Eur. Phys.~J.} {\bf C33}, 273-296 (2004).
% hep-ph/0306219

\bibitem{majewski} S.R. Majewski, M.F. Skrutskie, M.D. Weinberg, 
and J.C. Ostheimer,
{\it ApJ} {\bf 599}, 1081 (2003).
 
\bibitem{freese04} K. Freese, P. Gondolo, H.J. Newberg, and M. Lewis,
{\it Phys.\ Rev.\ Lett.} {\bf 92}, 111301 (2004).

\bibitem{freese05} K. Freese, P. Gondolo, and H.J. Newberg,
{\it Phys.\ Rev.} {\bf D71}, 043516 (2005).

\bibitem{diemand} J. Diemand, M. Kuhlen, and P. Madau, 
{\it ApJ} {\bf 649}, 1 (2006).

\bibitem{zhao} H. Zhao, D. Hooper, G.W. Angus, J.E. Taylor, and J. Silk, 
{\it ApJ} {\bf 654}, 697 (2007).

\bibitem{kamion1998} M. Kamionkowski and A. Kinkhabwala, 
{\it Phys. Rev.} {\bf D57}, 3256 (1998).

\bibitem{gates95} E.I. Gates, G. Gyuk, and M.S. Turner, 
{\it ApJ} {\bf 449}, L123 (1995).

\bibitem{pdg04}
S. Eidelman \etal. (Particle Data Group), 
{\it Phys. Lett.} {\bf B592}, 1 (2004).

\bibitem{tovey2000} D.R. Tovey \etal., 
{\it Phys. Lett.} {\bf B488}, 17 (2006).

\bibitem{heusser} G. Heusser,
{\it Ann. Rev. Nucl. Part. Sci.} {\bf 45}, 543-90 (1995).

\bibitem{Gondolo05} P. Gondolo and G. Gelmini, 
{\it Phys. Rev.} {\bf D71}, 123520 (2005).

\bibitem{SuperK} S. Desai \etal. (Super-Kamiokande Collaboration),
{\it Phys. Rev.} {\bf D70}, 083523 (2004).
% hep-ex/0404025

\bibitem{Bernabei03} R. Bernabei \etal.,
{\it Riv. Nuovo Cim.} {\bf 26N1}, 1-73 (2003).
% astro-ph/0307403

\bibitem{Savage04} C. Savage, P. Gondolo and K. Freese,
{\it Phys. Rev.} {\bf D70}, 123513 (2004).
% astro-ph/0408346

\bibitem{CRESSTI}
G. Angloher \etal. ({\small CRESST} Collaboration), 
{\it Astropart. Phys.} {\bf 18}, 43 (2002).

\bibitem{Picasso05} M. Barnabe-Heider \etal. ({\small PICASSO}
Collaboration), 
{\it Phys. Lett.} {\bf B624}, 186 (2005).

\bibitem{cdms_sd} D.S. Akerib \etal. (\CDMS\ Collaboration), 
{\it Phys. Rev.} {\bf D73}, 011102(R) (2006).

\bibitem{zeplin_sd} V.A. Kudryavstev for the {\small UKDM}
Collaboration, presented at the {\it Fifth International Workshop on
the Identification of Dark Matter}, Edinburgh, Scotland (2004).

\bibitem{naiad} G.J. Alner \etal. ({\small UKDM} Collaboration),
{\it Phys. Lett.} {\bf B616}, 17 (2005).

\bibitem{Ellis01} J.Ellis, A. Ferstl and K.A. Olive,
{\it Phys. Rev.} {\bf D63}, 065016 (2001).
% hep-ph/0007113 SD MSSM

\bibitem{Ellis00} J.Ellis, A. Ferstl and K.A. Olive,
{\it Phys. Lett.} {\bf B481}, 304 (2000).
% hep-ph/0001005 SD CMSSM

\bibitem{Silk:1985} J. Silk,  K.A. Olive, and M. Srednicki, 
{\it Phys. Rev. Lett.} {\bf 55}, 257 (1985).

\bibitem{Freese:1986} K. Freese, 
{\it Phys. Lett.} {\bf B167}, 295 (1986).

\bibitem{Krauss:1986} L.M. Krauss, M. Srednicki, and F. Wilczek, 
{\it Phys. Rev.} {\bf D33}, 2079 (1986).

\bibitem{Gunn:1978} J.E. Gunn, B.W. Lee, I. Lerche, D.N. Schramm, 
and G. Steigman, 
{\it ApJ} {\bf 223}, 1015 (1978).

\bibitem{Stecker:1978} F.W. Stecker, 
{\it ApJ} {\bf 223}, 1032 (1978).

\bibitem{Silk:1984} J. Silk and M. Srednicki, 
{\it Phys. Rev. Lett.} {\bf 53}, 624 (1984).

\bibitem{Donato:2000} F. Donato, N. Fornengo, and P. Salati, 
{\it Phys. Rev.} {\bf D62}, 043003 (2000).

\bibitem{Gondolo:1999} P. Gondolo and J. Silk, 
{\it Phys. Rev. Lett.} {\bf 83}, 1719 (1999).

\bibitem{Gondolo:1994} 
P. Gondolo, 
  {\it Nucl.\ Phys.\ Proc.\ Suppl.} {\bf 35}, 148 (1994); 
E.A. Baltz, C. Briot, P. Salati, R. Taillet, and J. Silk, 
  {\it Phys. Rev.} {\bf D61}, 3514 (2000); 
C. Tyler, 
  {\it Phys. Rev.} {\bf D66}, 3509 (2002); 
A. Falvard \etal., 
  {\it Astrop. Phys.} {\bf 20}, 467 (2004); 
A. Tasitsiomi, J. Gaskins, and A. Olinto, 
  {\it Astrop. Phys.} {\bf 21}, 637 (2004).

\bibitem{edsjo} J. Edsj\"o, in preparation (private communication).

\bibitem{antid1} 
H.~Baer and S.~Profumo,
{\it J. of Cosmo. and Astropart. Phys.} {\bf 0512}, 8 (2005).
% astro-ph/0510722

\bibitem{BESSlimit} H.~Fuke \etal. ({\small BESS} Collaboration), 
{\it Phys. Rev. Lett.} {\bf 95}, 081101 (2005).
% astro-ph/0504361

\bibitem{GAPS}
C.J.~Hailey \etal., 
{\it J. of Cosmo. and Astropart. Phys.} {\bf 601}, 7 (2006);
% astro-ph/0509587
K.~Mori \etal., {\it Astrophys. J.} {\bf 566}, 604 (2002).
% astro-ph/0109463

\bibitem{antid2} 
R.~Duperray \etal., {\it Phys. Rev.} {\bf D71}, 083013  (2005).
% astro-ph/0503544



\bibitem{heat} 
S.W. Barwick \etal., 
  {\it ApJ} {\bf 498}, 779 (1998); 
S.W. Barwick \etal., 
  {\it J. Geophys. Res.} {\bf 103}, 4817 (1998); 
J.J. Beatty \etal., 
  {\it Phys. Rev. Lett.} {\bf 93}, 241102 (2004).

\bibitem{Baltz:2002} E.A. Baltz,  J. Edsj\"o, K. Freese, and P. Gondolo, 
{\it Phys. Rev.} {\bf D65}, 63511 (2002).

\bibitem{Kane:2002} G.L. Kane, L. Wang, and J.D. Wells, 
{\it Phys. Rev.} {\bf D65}, 5770 (2002).

\bibitem{Cumberbatch:2007} D.T. Cumberbatch, and J. Silk, 
{\it MNRAS} {\bf 374}, 455 (2007).

\bibitem{deboer} W. de Boer, C. Sander, V. Zhukov, A.V. Gladyshev, 
and D.I. Kazakov, 
{\it Astron. Astrophys.} {\bf 444}, 51 (2005).

\bibitem{hess} F.A. Aharonian \etal., 
{\it Astron. Astrophys.} {\bf 425}, L13 (2004).

\bibitem{veritasgc} K. Kosack \etal., 
{\it ApJ} {\bf 608}, L97 (2004).

\bibitem{cangaroogc} K. Tsuchiya \etal., 
{\it ApJ} {\bf 606}, L115 (2004).

\bibitem{horns} D. Horns, 
{\it Phys. Lett.} {\bf B607}, 225 (2005); Erratum: {\bf B611}, 297 (2005).

\bibitem{bergstrom05} L. Bergstrom, T. Bringmann, M. Eriksson, 
and M. Gustafsson, 
{\it Phys. Rev. Lett.} {\bf 94}, 131301 (2005).

\bibitem{profumo} S. Profumo, {\it Phys. Rev. }{\bf D72}, 103521 (2005).

\bibitem{hall} J. Hall, and P. Gondolo, 
{\it Phys. Rev.} {\bf D74}, 063511 (2006).

\bibitem{Baltz01} E.A.~Baltz and P.~Gondolo, 
{\it Phys. Rev. Lett.} {\bf 86}, 5004 (2001).
% hep-ph/0102147

\bibitem{Ellis03}
J.~ Ellis, K. A.~Olive, Y.~Santoso and V.C.~Spanos,
{\it Phys. Lett.} {\bf B565}, 176-182 (2003).
% hep-ph/0303043

\bibitem{Arkani04}
N. Arkani-Hamed and S. Dimopoulos, 
{\it J. of High Energy Phys.} {\bf 506}, 73 (2005).
% hep-th/0405159.

\bibitem{hewett} J.L.~Hewett, B.~Lillie, M.~Masip and T.G.~Rizzo, 
{\it J. of High Energy Phys.} {\bf 409}, 70 (2004).
% hep-ph/0408248.

\bibitem{KKdarkmatter} 
K. Agashe and G.~Servant, 
{\it Phys. Rev. Lett.} {\bf 93}, 231805 (2004); % hep-ph/0403143
G.~Servant and T.M.P.~Tait, {\it Nucl. Phys.}, {\bf B650}, 391 (2003);
H.C.~Cheng, J.L.~Feng and K.T.~Matchev, {\it Phys. Rev. Lett.} 
{\bf 89}, 211301 (2002).

\bibitem{lcc} E.A. Baltz, M. Battaglia, M. Peskin and T. Wizansky, 
{\it Phys. Rev.} {\bf D74}, 103521 (2006); see also, 
% hep-ph/0602187
http://www.physics.syr.edu/$\sim$trodden/lc-cosmology.

\bibitem{horowitz03}
C.J.~Horowitz, K.J.~Coakley and D.N.~McKinsey,
{\it Phys. Rev.} {\bf D68}, 023005 (2003).
% astro-ph/0302071
 
\bibitem{beacom01}
J.F.~Beacom, W.M.~Farr and P.~Vogel,
{\it Phys. Rev.} {\bf D66}, 033001 (2002).
% hep-ph/0205220
 
\bibitem{strigari05}
L.E.~Strigari, J.F.~Beacom, T.P.~Walker and P.~Zhang,
{\it J. of Cosmo. and Astropart. Phys.} {\bf 0504}, 017 (2005).
% astro-ph/0502150

\bibitem{gaitskell04}
R.J.~Gaitskell, {\it Ann. Rev. Nucl. Part. Sci.} {\bf 54}, 315 (2004).

\bibitem{mei_hime05} D-M. Mei and A. Hime, 
{\it Phys. Rev.} {\bf D73}, 053004 (2006).
% astro-ph/0512125.

\bibitem{raul_cdms} R. Hennings-Yeomans \etal., 
private communication (in preparation).

\bibitem{raul_nim} R. Hennings-Yeomans and D.S.~Akerib,
astro-ph/0611371, 
accepted for publication in 
{\it Nucl. Instrum. Meth. in Phys. Res.}

\bibitem{gaitskell_water} R.J. Gaitskell, private communication.

\bibitem{heusser_gempi}
G.~Heusser, M. Laubenstein, and H. Nedera, in {\it 
Proc. of Intern. Conf. Isotop. Environm. Studies Aquatic Forum}, 
Monte Carlo, Monaco, October 2004.

\bibitem{drift} G.J. Alner \etal.,
{\it Nucl. Instrum. Meth. in Phys. Res.} {\bf A555}, 173 (2005).

\bibitem{copi2005} C.J. Copi, L.M. Krauss, D. Simmons-Duffin and
  S.R.~Stroiney, 
{\it Phys. Rev.} {\bf D75}, 023514 (2007).
% astro-ph/0508649

\bibitem{homestake} Homestake Collaboration, nucl-ex/0308018.

\bibitem{cascade} {\small DUSEL}-Cascades Collaborations, 
http://www.int.washington.edu/s2

\end{thebibliography}
\end{document}